\documentclass[manuscript]{acmart}

\usepackage{booktabs}
\usepackage{multirow}
\usepackage{multicol}
\usepackage[skip=0.5\baselineskip]{caption}
\usepackage{array,makecell}
\usepackage[export]{adjustbox}
\usepackage{arydshln}
\usepackage{graphicx}
\usepackage{subfig}
\usepackage{rotating}
\usepackage{wasysym}
\usepackage{soul}

\AtBeginDocument{%
  \providecommand\BibTeX{{%
    \normalfont B\kern-0.5em{\scshape i\kern-0.25em b}\kern-0.8em\TeX}}}

\setcopyright{acmcopyright}
\acmJournal{TOIS}
\acmYear{2022} \acmVolume{1} \acmNumber{1} \acmArticle{1} \acmMonth{1} \acmPrice{15.00}\acmDOI{10.xxx}

\begin{document}

\title[A Systematic Analysis on the Impact of Contextual Information on Point-of-Interest Recommendation]{A Systematic Analysis on the Impact of Contextual Information on Point-of-Interest Recommendation}

\author{Hossein A. Rahmani}
\authornote{Work done while Hossein A.~Rahmani was affiliated with Universit\`a della Svizzera italiana (USI), Switzerland and MASLab at University of Zanjan, Iran.}
\affiliation{%
  \institution{University College London}
  \country{United Kingdom}}
\email{h.rahmani@ucl.ac.uk}

\author{Mohammad Aliannejadi}
\affiliation{%
  \institution{University of Amsterdam}
  \country{The Netherlands}
}
\email{m.aliannejadi@uva.nl}

\author{Mitra Baratchi}
\affiliation{%
  \institution{Leiden University}
  \country{The Netherlands}
}
\email{m.baratchi@liacs.leidenuniv.nl}

\author{Fabio Crestani}
\affiliation{%
  \institution{Università della Svizzera italiana (USI)}
  \country{Switzerland}}
\email{fabio.crestani@usi.ch}

\renewcommand{\shortauthors}{H.~A.~Rahmani, M.~Aliannejadi, M.~Baratchi, and F.~Crestani}

\begin{abstract}
  As the popularity of Location-based Social Networks (LBSNs) increases, designing accurate models for Point-of-Interest (POI) recommendation receives more attention. POI recommendation is often performed by incorporating contextual information into previously designed recommendation algorithms. Some of the major contextual information that has been considered in POI recommendation are the location attributes (i.e., exact coordinates of a location, category, and check-in time), the user attributes (i.e., comments, reviews, tips, and check-in made to the locations), and other information, such as the distance of the POI from user's main activity location, and the social tie between users. The right selection of such factors can significantly impact the performance of the POI recommendation. However, previous research does not consider the impact of the combination of these different factors. In this paper, we propose different contextual models and analyze the fusion of different major contextual information in POI recommendation. The major contributions of this paper are: (i) providing an extensive survey of context-aware location recommendation (ii) quantifying and analyzing the impact of different contextual information (e.g., social, temporal, spatial, and categorical) in the POI recommendation on available baselines and two new linear and non-linear models, that can incorporate all the major contextual information into a single recommendation model, and (iii) evaluating the considered models using two well-known real-world datasets. Our results indicate that while modeling geographical and temporal influences can improve recommendation quality, fusing all other contextual information into a recommendation model is not always the best strategy.
\end{abstract}

\ccsdesc[500]{Information systems~Information retrieval}
\ccsdesc[500]{Information systems~Recommender systems}

\keywords{Contextual Information, Context-aware Recommendation, Point-of-Interest Recommendation, Location-based Social Networks}

\maketitle

\section{Introduction and Motivations}
\label{sec:intro}
With the ever-increasing number of smartphone users, Location-based Social Networks (LBSNs), such as Foursquare\footnote{https://www.foursquare.com}, Swarm\footnote{https://www.swarmapp.com}, and Yelp\footnote{https://www.yelp.com} are growing. LBSNs allow users to check-in through their smartphones at a location or Point-of-Interest (POI) such as restaurants, malls, or movie theaters. Users' check-in data often include geographical information, i.e., latitude and longitude, and the check-in timestamp. With thousands of potential POIs in the vicinity of each user, the process of choosing a POI by a user becomes overwhelming. POI recommendation algorithms strive to address this issue by filtering through a large variety of options available and returning those that are most likely to be of the user's interest. In recent years, Matrix Factorization (MF), as a linear technique, and Neural Network (NN), as a non-linear one, have proven to offer promising solutions to the problem of designing efficient filtering algorithms for recommendation systems~\cite{liu2017experimental,zhang2019deep,adomavicius2005toward,ricci2015recommender,ricci2011introduction}. The main difference between MF and NN lies in their approach of modeling the relation between users and POIs in a linear or non-linear way~\cite{he2017neural,rendle2020neural,anelli2021reenvisioning}. 
In general, POI recommendation suffers from two well-known problems, namely, data sparsity and cold start. Given a large amount of POIs and the user's ability to visit only a few of them, the user-POI interaction matrix becomes very sparse. This makes it very hard to model collaborative interactions between users and POIs. This problem is referred to as {\em data sparsity}. On the other hand, the {\em cold-start\/} problem refers to recommending POIs to those users who have very limited or no interaction records. Also, POIs with very limited or no interaction records are considered as cold items, and the models often fail to recommend these items effectively \cite{liu2017experimental,li2015rank,elahi2016survey,xiong2020go}. To address the data sparsity issue, more recently, Context-Aware Recommender Systems (CARSs) have gained popularity as many researchers in different disciplines such as recommender systems, information retrieval, and data mining have recognized the importance of contextual information \cite{liu2017experimental,stepan2016incorporating,baltrunas2008exploiting,rahmani2019regression,adomavicius2011context,chakraborty2020relevance}. A CARS should provide a user with recommendations taking into consideration the user's current context. The context of a user can be defined as a set of factors and limitations that impact users' perception and acceptance of a particular item. 
Various definitions of context exist in the literature~\cite{liu2017experimental,baral2018exploiting,sanchez2019exploiting,ricci2010mobile}, from which the most popular contextual factors for POI recommendation can be listed as follows: the time of check-in, weather, location, prices, or even the users' friendships.

\begin{figure}[h]
    \centering
    \includegraphics{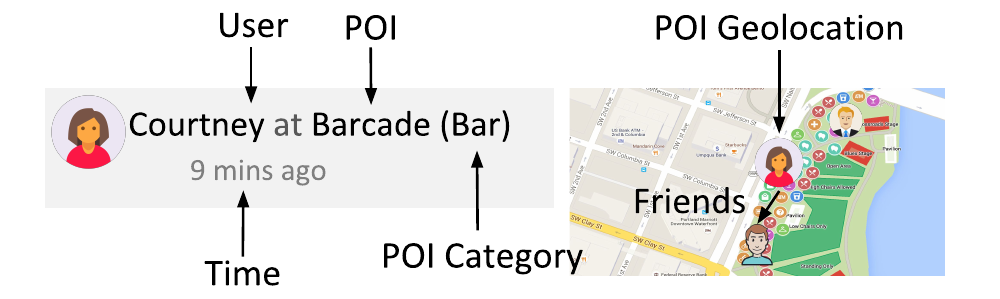}
    \caption{Illustration of a typical check-in in LBSN.}
    \label{fig:checkin}
\end{figure}

Fig.~\ref{fig:checkin} shows a typical check-in record in an LBSN. Based on such records, four important and effective contextual information are typically considered for POI recommendation, namely, geographical, temporal, social, and categorical information \cite{liu2017experimental,baral2018exploiting,ricci2002travel,manotumruksa2018contextual}. Among various contextual factors, finding the right combination of contextual information for a specific user is of great importance, as it affects both the effectiveness and efficiency of the models. Previous studies have shown that incorporating all available contextual factors is not always beneficial~\cite{liu2017experimental,baral2018exploiting}. In addition to the extra data processing load, using all contextual information will not necessarily lead to an improvement in recommendation accuracy~\cite{baral2018exploiting}. Furthermore, the effect of different contextual factors and their combinations on linear and non-linear algorithms performances has not been studied in depth \cite{liu2017experimental,baral2018exploiting,manotumruksa2019effective}. Yet, \citet{liu2017experimental} studied the different POI recommendation methods that have been proposed, and \citet{baral2018exploiting} exploited the impact of contextual information only on the PageRank algorithm.
Indeed, a careful selection of such factors, while taking into account the characteristics of the recommender model, can significantly impact the effectiveness of the system. Therefore, the main challenge is to identify which contextual information or which combination of them should be incorporated into a POI recommendation system to improve recommendation quality. To create an accurate POI recommender system, we need to be able to answer questions such as, \textit{what might be the impact of using temporal information instead of geographical information?} or \textit{what is the impact of using social and categorical information instead of geographical and temporal information?} Moreover, users in an LBSN have different behavior; for instance, a user might prefer to visit the same POIs again and again, while another user may prefer to discover new and unvisited locations. Therefore, analyzing the impact of users' behavioral biases on the performance of the models and the effectiveness of each contextual factor is of high importance. Such an analysis would enable systems to take different strategies for different groups of users. For instance, geographical context might be more effective for users who tend to regularly visit POIs, while social context might be more beneficial for users who tend to discover new POIs more often. With this knowledge, a recommender system would be able to employ separate strategies for each case.

In this work, we seek to provide a more comprehensive understanding of
contextual factors' impact on a set of representative POI recommendation models. Therefore, our main research questions can be summarized as follows:

\begin{itemize}
    \item \textbf{RQ1:} How effectively do different models incorporate multiple contextual factors in recommendation? (cf.~Section \ref{sec:results_contexts})
    \item \textbf{RQ2:} How can different evaluation metrics capture the effect of contextual information on various models? (cf.~Section \ref{sec:results_metrics})
    \item \textbf{RQ3:} How can we incorporate different contextual factors in linear and non-linear models? (cf.~Section \ref{sec:results_models})
    \item \textbf{RQ4:} How do models incorporating contextual information perform for users with different behavior? (cf.~Section \ref{sec:results_users})
\end{itemize}

To answer these research questions, we consider several baseline models previously proposed to prove the lack of deep consideration of contextual models. Also, we define different contextual models that can be used in a POI recommendation system and fuse them into the MF and NN recommendation approaches. Next, we analyze the impact of major influential contexts for POI recommendation by combining different contextual information on creating fused models. In particular, we are not aiming to justify the effect of any contextual information on any model. Instead, our goal is to characterize and provide a structured review of the impact of context on these two specific models. To the best of our knowledge, the impact of users' behavior on contextual models' performance in POI recommendation has not been extensively studied before. We intend to fill this gap via a detailed analysis defining different factors on users' behavior such as \textit{geographical distance}, \textit{check-ins density}, and \textit{exploration}. Our aim is to identify a selection of contextual factors that perform best and the impact of considering a combination of multiple factors (e.g., a combination of social and temporal factors or spatial and temporal factors). We analyze the results based on two different datasets to see how much the results can be generalized. This analysis can help select contextual factors that are suitable for implementation in real systems. Based on our findings, we see that neural networks-based models are more accurate than matrix factorization. Moreover, in most cases, temporal information has a greater effect than others. Also, the combination of temporal and geographical information helps the recommendation algorithms achieve better performance. Finally, to enable the results' reproducibility, we have made our codes, datasets, and analysis publicly available in open source.\footnote{\url{https://rahmanidashti.github.io/ContextsPOI/}}

The rest of this paper is organized as follows. We review the relevant studies on POI recommendations in Section \ref{sec:relatedwork}. Our proposed experimental framework and our analysis approach are presented in Section \ref{sec:methods}, followed by experimental evaluation in Section \ref{sec:evaluation}. Finally, Sections \ref{sec:discuss} and \ref{sec:conclusion} discuss and conclude this paper.

\section{Related Work}
\label{sec:relatedwork}
This section will discuss similar papers that tried to reproduce and examine multiple POI recommender models and analyze them. Then, we will review the different proposed contextual POI recommendations.

Previously some researchers reproduced and examined multiple POI recommendation models to analyze and discuss the impact of different contextual information in POI recommendation. In this regard, the first study was done by \citet{liu2017experimental} in 2017. The authors of this work reproduce and provide a comprehensive evaluation of 12 state-of-the-art POI recommendation models proposed by different researchers. Then, they compare them based on different evaluation metrics such as Precision, Recall, and nDCG (Sec.~\ref{sec:evalmetrics} shows the formula for each metric). In another work, \citet{stepan2016incorporating} incorporate spatial, temporal, and social information in their recommendation model. They analyze the impact of their contextual information on the proposed model by adding them separately. However, they do not consider the impact of the combination of contextual information. \citet{baral2018exploiting} exploited different contextual information in POI recommendation employing PageRank as a raking model. More recently, Sanchez \cite{sanchez2019exploiting} discusses the impact and the importance of evaluating contextual information in a POI recommendation model. They propose using evaluation metrics based on different contextual information to see the impact of contextual information on the performance of models. In addition, \citet{rendle2020neural} compare the performance of matrix factorization when it uses dot product and multi-layer perceptrons (MLP)\footnote{This approach is often referred to as neural collaborative filtering (NCF).}. They conclude that simple and traditional dot product archives better results, and MLP is too costly to use for recommendation in production environments. Our work is in line with the earlier mentioned studies performed by \citet{liu2017experimental}, \citet{stepan2016incorporating}, \citet{baral2018exploiting}, \citet{sanchez2019exploiting}. However, there are some major differences between our study and these works. The experimental research of \citet{liu2017experimental} studies several POI recommendation models and compares their performance in terms of precision, recall, and nDCG. In contrast, we propose two extensible models, one based on Matrix Factorization and the other based on Neural Networks, that can easily incorporate contextual information. By doing so, we are able to analyze the importance of  contextual information both on a linear and non-linear model. Furthermore, the study presented in \cite{liu2017experimental} only compares the individual, geographical and social components of the models (see Section 5.3 in \citet{liu2017experimental}). In this paper, we additionally study the categorical context. We also perform a much broader set of comparisons, going beyond comparing only social components or geographical components of different models, as done in \cite{liu2017experimental}, but also study the performance of models based on differences in user behaviors. The study of \citet{stepan2016incorporating} considers fusing all different contextual information into one single model. At the same time, we analyze the impact of having a different selection of contextual information in creating fused models. The study of \citet{baral2018exploiting} considers different fused models of contextual information. Still, they applied them into a single ranking model based on the PageRank algorithm that is a linear model. This study does not consider the analysis of performance by changing the underlying ranking model. We will, however, study the performance achieved by linear and non-linear models. The study of \citet{sanchez2019exploiting} implements some traditional POI recommendation systems by incorporating contextual information to show that the study and evaluation of different contextual information are important in the domain of POI recommendation without actually performing the analysis. In comparison, our paper provides a detailed analysis of the impact of different contextual factors in POI recommendation. In fact, to evaluate different combinations of the contextual information in different fused models, we analyze the impact of different contextual information in two commonly used models in POI recommendation, i.e., matrix factorization and neural networks, representing linear and non-linear models. To the best of our knowledge, none of the existing recommendation models incorporated these aspects in a linear and non-linear manner to the best of our knowledge.

Since different approaches use these two models in different settings, it is important to know which one is better and in which situation. In what follows, we will provide an overview of non-contextual and contextual-based POI recommendation systems and review relevant previous studies in each category. Table \ref{tbl:baselines} shows a summary of the related studies.

\begin{table*}[h]
\centering
\caption{Summary of POI recommendation papers in the related works in relation to the use of interaction (i.e., check-ins) and contextual information.}
\label{tbl:baselines}
\begin{tabular}{l|l|l|l|l|l}
\hline
Related work                                & \textbf{I}nteraction & \textbf{G}eographical & \textbf{T}emporal & \textbf{S}ocial & \textbf{C}ategorical \\ \hline
\citet{ye2011exploiting}           & $\checkmark$         & $\checkmark$          & $\times$          & $\checkmark$    &  $\times$ \\
\citet{ference2013location}   & $\checkmark$         & $\checkmark$          & $\times$          & $\checkmark$    &  $\times$ \\
\citet{cheng2012mgm}  & $\checkmark$         & $\checkmark$          & $\times$          & $\checkmark$    &  $\times$ \\
\citet{rahmani2019lglmf}   & $\checkmark$         & $\checkmark$          & $\times$          & $\times$    &  $\times$ \\
\citet{aliannejadi2019joint}   & $\checkmark$         & $\checkmark$          & $\times$          & $\times$    &  $\times$ \\
\citet{zhang2013igslr}   & $\times$         & $\checkmark$          & $\times$          & $\checkmark$    &  $\times$ \\
\citet{cheng2016unified}  & $\checkmark$         & $\checkmark$          & $\times$          & $\checkmark$    &  $\times$ \\
\citet{li2015rank}   & $\checkmark$         & $\checkmark$          & $\times$          & $\times$    &  $\times$ \\
\citet{guo2019location}   & $\checkmark$         & $\checkmark$          & $\times$          & $\times$    &  $\times$ \\
\citet{griesner2015poi}  & $\checkmark$         & $\times$          & $\checkmark$          & $\times$    &  $\times$ \\
\citet{gao2013exploring}   & $\checkmark$         & $\times$          & $\checkmark$          & $\times$    &  $\times$ \\
\citet{li2017time}   & $\checkmark$         & $\times$          & $\times$          & $\checkmark$    &  $\times$ \\
\citet{cho2011friendship} & $\times$         & $\times$          & $\checkmark$          & $\checkmark$    &  $\times$ \\
\citet{bao2012location} & $\checkmark$         & $\times$          & $\times$          & $\checkmark$    &  $\checkmark$ \\
\citet{gibson1998inferring} & $\checkmark$         & $\times$          & $\times$          & $\times$    &  $\checkmark$ \\
\citet{rahmani2019category} & $\checkmark$         &  $\times$         & $\times$          & $\times$    &  $\checkmark$ \\
\citet{baral2018exploiting} & $\checkmark$         & $\checkmark$          & $\checkmark$          & $\checkmark$    &  $\checkmark$ \\
\citet{stepan2016incorporating} & $\checkmark$         &  $\checkmark$         & $\checkmark$          & $\checkmark$    &  $\times$ \\
\citet{cheng2013you} & $\checkmark$         &  $\checkmark$         & $\checkmark$          & $\times$    &  $\times$ \\
\citet{baral2016maps} & $\checkmark$         &  $\checkmark$         & $\checkmark$         & $\checkmark$    &  $\checkmark$ \\
\citet{baral2016geotecs} & $\checkmark$         &  $\checkmark$         & $\checkmark$         & $\checkmark$    &  $\checkmark$ \\
\citet{liu2013point} & $\checkmark$         &  $\checkmark$         & $\checkmark$         & $\checkmark$    &  $\times$ \\
\citet{yin2013lcars} & $\checkmark$         &  $\times$         & $\times$         & $\checkmark$    &  $\times$ \\
\citet{hu2013spatial} & $\times$         &  $\checkmark$         & $\checkmark$         & $\times$    &  $\times$ \\
\citet{yin2015joint} & $\checkmark$         &  $\checkmark$         & $\checkmark$         & $\checkmark$    &  $\times$ \\
\citet{xie2016learning} & $\checkmark$         &  $\checkmark$         & $\checkmark$         & $\checkmark$    &  $\times$ \\
\citet{rahmani2020joint} & $\checkmark$         &  $\checkmark$         & $\checkmark$          & $\times$    &  $\times$ \\
\citet{chen2020learning} & $\checkmark$         &  $\checkmark$         & $\checkmark$          & $\times$    &  $\times$ \\
\citet{baral2019hirecs} & $\checkmark$         &  $\checkmark$         & $\times$          & $\times$    &  $\times$ \\
\citet{pan2019deep} & $\times$         &  $\checkmark$         & $\times$          & $\times$    &  $\times$ \\
\citet{zheng2020memory} & $\checkmark$         &  $\times$         & $\times$          & $\times$    &  $\times$ \\
\citet{lim2020stp} & $\checkmark$         &  $\checkmark$         & $\checkmark$          & $\times$    &  $\times$ \\
\citet{zhou2019adversarial} & $\checkmark$         &  $\times$         & $\times$          & $\times$    &  $\times$ \\
\citet{zhang2014lore} & $\times$         &  $\checkmark$         & $\checkmark$          & $\checkmark$    &  $\times$ \\
\citet{zhang2015geosoca} & $\times$         &  $\checkmark$         & $\times$          & $\checkmark$    &  $\checkmark$ \\
\citet{manotumruksa2017deep} & $\checkmark$         &  $\checkmark$         & $\times$          & $\times$    &  $\times$ \\
\citet{manotumruksa2018contextual} & $\checkmark$         &  $\times$         & $\checkmark$          & $\times$    &  $\times$ \\
\citet{manotumruksa2020contextual} & $\checkmark$         &  $\checkmark$         & $\checkmark$          & $\times$    &  $\times$ \\
\citet{chang2020learning} & $\checkmark$         &  $\checkmark$         & $\times$          & $\times$    &  $\times$ \\
\citet{lim2020stp} & $\checkmark$         &  $\checkmark$         & $\checkmark$          & $\times$    &  $\times$ \\
\citet{ma2018point} & $\checkmark$         &  $\checkmark$         & $\times$          & $\times$    &  $\times$ \\
\citet{zhou2019adversarial} & $\checkmark$         &  $\checkmark$         & $\times$          & $\checkmark$    &  $\times$ \\
\end{tabular}
\end{table*}

\subsection{Non-contextual Information}

\subsubsection{Interaction (I)}
Most traditional recommendation systems make recommendations for items such as movies or music using explicit ratings. In an LBSN, explicit ratings are rare, and usually, the check-in frequency (i.e.,~the interaction of users with POIs without any contexts such as geographical or temporal) implicitly reflects users' preferences for POIs. Hence, to produce POI recommendations, several earlier studies~\cite{Berjani2011,ye2011exploiting,mulligann2011analyzing,davidsson2011utilizing,manotumruksa2020contextual} adopted traditional recommendation models to infer users' personalized preference for POI by mining the check-in patterns of users. With the available check-in information, existing recommendation approaches (e.g., user-based and item-based Collaborative Filtering (CF)) can be employed for POI recommendation in LBSNs by treating POIs as items. By taking this approach, \citet{ye2010location} was the first research to provide location recommendations services in LBSNs that proposed user-based and item-based POI recommendation algorithms. The proposed approach assumes that similar users have similar tastes for locations and make POI recommendations based on most similar neighbors' opinions. On the other hand, this item-based POI recommendation approach assumes that users are interested in similar POIs.

\subsection{Contextual Information}

\subsubsection{Geographical information (G)}
Incorporating geographical information is one of the most important factors that distinguish a POI recommendation from a conventional item recommendation. Tobler's First Law of Geography (1970) states that ``everything is related to everything else, but near things are more related than distant things''. In fact, analysis of users' check-in data shows that a user's check-ins happen in geographically constrained areas \cite{ye2011exploiting,cheng2012mgm,sun2020go} and thus follow this general rule. This reflects the users' interest in visiting nearby POIs rather than distant ones. Several studies \cite{ye2011exploiting,cheng2012mgm,ference2013location,zhao2017geo} attempt to employ such geographical information to improve POI recommendation systems. \citet{ye2011exploiting} showed that users' check-in behavior follows a power-law distribution. They proposed a unified POI recommendation system by incorporating this geographical information to address the data sparsity problem. \citet{ference2013location} took into consideration user preference, geographical proximity, and social information for out-of-town POI recommendation. Cheng et al.~\cite{cheng2012mgm,cheng2016unified} proposed a Multi-center Gaussian model to capture users' movement patterns as they assumed users' check-ins happen around several centers. \citet{lian2014geomf} proposed a POI recommendation approach based on weighted matrix factorization. They explicitly model the so-called geographical users' ``activity area'' and ``the influence'' area of POIs. \citet{li2015rank} modeled the POI recommendation task as a ranking-based approach, where they incorporated the geographical information into the pair-wise ranking model. The geographical information is modeled by defining an extra factor matrix.

Conversely, \citet{zhang2013igslr} argued that geographical information should be considered for each user separately. To this end, a model was proposed based on kernel density estimation of the distance distributions between POIs checked-in by each user. \citet{rahmani2019lglmf} modeled this geographic information from two different perspectives: the user's and the location's. They showed that the recommendation model's performance could be improved by incorporating the impact of the neighboring POIs. Similarly, \citet{yuan2016joint} addressed the data sparsity problem, assuming that users tend to show more interest in POIs that are geographically closer to the one that they have already visited. \citet{guo2019location} proposed a location neighborhood-aware weighted matrix factorization model to exploit the location perspective, incorporating geographical distance among POIs. More recently, \citet{aliannejadi2019joint} proposed a two-phase collaborative ranking algorithm for POI recommendation that takes into account the geographical information of POIs located in the same neighborhood. \citet{manotumruksa2017deep} capture the complex relations between users and POIs using a deep recurrent collaborative filtering method. They apply a pairwise ranking function with the aim of capturing check-ins in the form of sequences of observed feedback using a multi-layer perceptron and a recurrent neural network architecture. In particular, their method can learn complex user and POI features using element-wise and dot products as well as the concatenation of latent factors. \citet{chang2020learning} proposed a graph neural network-based method inspired by the idea that consecutive check-ins at two POIs indicate a greater geographical influence between them. They designed a model to incorporate user preferences using a user-POI graph and geographical influences using a POI-POI graph in which edges of the POI-POI graph are weighted based on the frequency of users' consecutive visits. \citet{ma2018point} address the challenge of modeling more complex user-POI interactions from the sparse implicit feedback using an autoencoder-based approach named SAE-NAD. This network is a combination of a self-attentive encoder (SAE) and a neighbor-aware decoder (NAD). Their self-attentive encoder adopts a multi-dimensional attention mechanism to differentiate between user preference degrees. They also incorporate the geographical context information using the neighbor-aware decoder to make users’ reachability higher on the similar and nearby neighbors of checked-in POIs. This is achieved by the inner product of POI embeddings together with the radial basis function (RBF) kernel.

\subsubsection{Temporal information (T)}
Temporal constraints can result in specific user check-in patterns. Users' temporal check-in behaviors in LBSNs typically exhibit a periodic pattern. For instance, it is observed that users visit places around their office or home area on weekdays and spend time in shopping malls on weekends. There are hourly patterns observed in check-ins. For instance, user check-ins at restaurants typically happen during lunchtime, whereas check-ins at nightclubs happen, naturally, at night. Capturing such temporal information is of vital importance for POI recommendation. 

Many researchers have previously studied the effect of temporal information on users' preferences by proposing different models to improve the POI recommendation accuracy \cite{ding2005time,yuan2013time,zhao2017geo,liu2017experimental}. \citet{griesner2015poi} proposed an approach to integrate temporal information into weighted matrix factorization. They propose an approach to change the values of each POI's influence area through accounting for the time spent by a user to go from the current POI to the next. \citet{gao2013exploring} divided users' check-ins into different hourly time slots. Next, to train a user-based CF model, they compute the similarity between users based on their temporal overlap in visits to the same POIs. \citet{zhao2016stellar} proposed a latent ranking method that explicitly models the interactions between users, POIs, and time. In particular, they proposed to build upon a ranking-based pairwise tensor factorization framework. \citet{li2017time} proposed a time-aware personalized model adopting a fourth-order tensor factorization-based ranking, enabling the model to capture short-term and long-term preferences. \citet{yao2016poi} matched users' temporal regularity with the popularity of POIs to create a factorization-based algorithm. Also, \citet{yuan2013time} preserved the similarity of personal preferences in consecutive time slots by considering different latent variables at each time slot for each user. Moreover, \citet{manotumruksa2018contextual} proposed a contextual attention recurrent architecture model called CARA based on the success of recurrent neural network (RNN) models in modeling sequential patterns. Their model incorporates contextual information related to users' sequence of check-ins (e.g.,~time of the day) to effectively capture the users' dynamic preferences.

\subsubsection{Social Information (S)}
It has been observed that other users can socially influence a user's movements. The effect of social information has been studied to enhance POI recommendation based on the assumption that friends in LBSNs share more common interests than non-friends. Modeling social information has been explored in traditional recommendation systems \cite{tang2013social,jiang2012social}, and most of the existing work in POI recommendation is inspired by ideas taken from traditional recommendation systems. The analysis of \citet{cho2011friendship} shows that around 10–30\% of human movements can be socially influenced. Also, using the Gowalla dataset, they show an improvement in the accuracy of recommendation by considering the influence of friendships on users' mobility (estimated to be around 61\%) and the influence of mobility on new friendships (24\%). \citet{qiao2018socialmix} present SocialMix, a hybrid model that considers (i) user's familiarity and (ii) preference similarity for POI recommendation. To calculate users' familiarity score, they use three features: number of mutual friends, Jaccard similarity (based on user's friend list), and cosine similarity (based on user's check-in history). The weight of each feature is determined through maximum likelihood estimation. The preference similarity that shows users' similarity in terms of their preference in visited POIs is calculated based on the cosine similarity of user-location check-in data. \citet{zeng2018point} consider creating vectors representing user check-ins in 24-hour time-slots. They consecutively calculate the user similarities by measuring the cosine similarity of these vectors. Conversely, \citet{ye2010location} showed on a dataset of Foursquare check-ins that 96\% of users share less than 10\% of the commonly visited places, 87\% of people share nothing at all. However, to incorporate social information, they proposed a friend-based CF method to recommend POIs to users based on the commonly visited POIs. Moreover, \citet{gao2013exploring} assumed that people share their friends' check-in activities. Their model used the Hierarchical Pitman-Yor (HPY) language model to represent the check-in pattern with effective results.

\subsubsection{Categorical information (C)}
Category information of POIs provides a strong indication of the activities that can be performed in them. It is shown that users have distinct biases on the categories of POIs they check-in to. In LBSNs, POI categories are typically organized in a hierarchical category tree. Foursquare offers a 3-level category hierarchy. The top-most level consists of nightlife, food, while the lowest level, consisting of bars, pubs, Japanese food, or cafes. Considering such information in the analysis of check-in data can reflect users' preferences on the corresponding category. In the \cite{liu2013personalized}, the authors propose a category-aware recommendation to model the user's preference transition among POIs over their categories of each POI. Finally, they recommend POIs to a user based on the categories that the user prefers. \citet{bao2012location} model the preference of users based on their social opinions using Hypertext Induced Topic Search (HITS) (\citet{gibson1998inferring}). HITS regards an individual's visit to a POI as a directed link from the user to that POI. Each user has a hub score denoting their knowledge of a POI, and each location is associated with an authority score indicating its interest level. The target is to obtain a mutually reinforcing relationship between a user's knowledge and the interest level of a POI. The users' location history is categorized according to the POI's type (such as shopping or restaurants). A user-location matrix is used to identify the local experts who have a higher affinity towards a POI category, and such experts' social opinions were used in the recommendation. More recently, \citet{rahmani2019category} proposed a category-aware POI embedding model that considers both the users' sequence of check-ins and the category information of POIs. To this end, they made use of Word2Vec \cite{mikolov2013distributed} to generate a high-dimensional representation of the sequence of check-ins and POI categories.

\subsubsection{Fusion Models}
Contextual information has proven to be beneficial in improving the performance of POI recommendation models \cite{mazumdar2020cold,chang2020content,wang2020modeling,davtalab2020poi}. A class of studies tried to incorporate different contextual information in a single model \cite{liu2020exploiting,yu2020leveraging,ma2020exploring,si2019adaptive}. \citet{baral2016maps} proposed a multi-aspect POI recommendation system using geographical, temporal, categorical, and social information. They model a graph where each node corresponds to a location, and each user-time tuple is regarded as an attribute of the location node that shows the transition of a user between locations. For recommendation, they use the Top-Sensitive PageRank model \cite{haveliwala2003topic}, an extension of the PageRank algorithm, to rank each location for each user. Liu and Xiong \cite{liu2013point} propose a topic and location-aware POI recommendation system by exploiting textual and contextual information. They exploit an aggregated Latent Dirichlet Allocation (LDA) model to learn users' interest topics and find interesting POIs by mining textual information associated with them. Then, they utilized this contextual information for POI recommendation in combination with probabilistic matrix factorization. \citet{zhang2015geosoca} proposed GeoSoCa, which incorporates three different contextual factors, namely, geographical, social, and categorical information. GeoSoCa models geographical influence based on a check-in probability distribution over a two-dimensional space using a kernel density estimation (KDE) method. To model the social and categorical influences, GeoSoCa considers the check-in frequency of a user's friends on a POI and the power-law distribution of all users' categorical check-in frequency, respectively. \citet{baral2016geotecs} proposed a multi-aspect model based on the weighted matrix factorization. They incorporated geographical, temporal, categorical, and social aspects into the model. The intuition behind using the category information is that locations with the same category may have similar visits. For modeling the temporal information, they consider check-ins that happened simultaneously on POI as temporal popularity. Also, to model geographical information, they consider the user's activity area (or influence area), which is defined as the region where the user frequently appears. \citet{yin2013lcars} use POIs' content information such as item tags and category keywords to link the content-similar spatial items. Their model learns each user's interest and the local preference of each city by capturing item co-occurrence patterns and exploiting item contents to produce the top item recommendations. Hu and Ester \cite{hu2013spatial} used topic modeling to exploit user posts' spatial and textual aspects. \citet{cheng2013you} considered the users' movement constraints, i.e., moving around a localized region, and proposed a successive personalized POI recommendation model using a matrix factorization method that embedded personalized Markov chains and localized regions. \citet{yin2015joint} proposed a probabilistic generative model that exploited geographical, temporal, word-of-mouth, and semantic information. \citet{zhang2014lore} proposed a fusion method that considers geographical, temporal, and social influences. They model the geographical influence using a two-dimensional check-in probability distribution using KDE. The social influence is modeled based on the friendship relation of users. In contrast, the temporal influence is modeled by mining location sequences on location-location transition graph of all users and dividing the transition counts between every pair of POIs by the outgoing counts of each node. Also, \citet{xie2016learning} used geographical, temporal, and semantic aspects in their heterogeneous graph embedding model. They use a time decay method claimed to be an efficient predictor for the user's latest preferences. Moreover, \citet{rahmani2020joint} proposed a joint model of geographical and temporal information. To this end, they first find users' activity centers based on different temporal states and then recommend to users the unvisited POIs located in these areas to the users. \citet{lim2020stp} proposed a graph attention network using the spatial-temporal-preference data of users. Their method is able to exploit personalized user preferences and explore new POIs in global spatial-temporal-preference neighborhoods at the same time, while allowing users to learn from other users selectively. In addition, using a random walks approach, their method can mask a self-attention option to leverage the spatial-temporal-preference graph and find a new higher-order POI neighbor during exploration. \citet{zhou2019adversarial} proposed an adversarial model called APOIR to learn the distribution of user latent preferences of certain POIs to others. APOIR includes two major components: a recommender and a discriminator. The recommender module maximizes the probabilities of POIs that are unvisited based on the learned distribution to insert them in recommendation. The discriminator distinguishes the recommended POIs from the true check-ins and provides gradients to guide and improve the recommendation module in a reward framework.

\section{Evaluation Framework}
\label{sec:methods}
In this section, we describe how we designed our experiments to study contextual information for POI recommendation to answer the research questions posed in the introduction (Section \ref{sec:intro}). We outline what can be learned from each experiment, focusing on different contextual information and how their differences can be quantified. To this end, we propose an analytical and experimental evaluation framework in which we examine each contextual piece of information's effect. One of the important advantages of this framework is that it enables us to fuse every contextual model, either those proposed by ourselves or others.

To make use of contextual information, two tasks need to be performed. To show the ability to fuse previously proposed contextual information into our experimental framework, in the first task, we consider evaluating the contextual information that is incorporated into the previous POI recommendation systems\footnote{In this paper, we refer to them as baselines.}. Next, in the second task, to show the ability to incorporate any new contextual models, we model each piece of contextual information and fuse them in different ways into our analytical framework\footnote{In this paper, we refer to them as proposed models.}. Therefore, in this work, we consider two types of models: (i) we consider three state-of-the-art and well-known \textbf{context-aware POI recommendation algorithms} and analyze them within our proposed analytical and experimental evaluation framework, (ii) we also propose different \textbf{contextual models} and apply them into our proposed analytical framework. In our framework, to model (i), we have selected three state-of-the-art context-aware POI recommendation systems, namely, GeoSoCa \cite{zhang2015geosoca}, FCFKDEAMC (or LORE) \cite{zhang2014lore}, and PFMMGM \cite{cheng2012mgm}. These POI recommendation systems are well-known by the community for being able to make use of various contextual information. We will demonstrate how these contextual models will be incorporated into our proposed experimental framework. To show how new contextual information will be incorporated into the experimental framework, to model (ii), we consider matrix factorization (linear) and neural networks (non-linear) approaches to create contextual models for each piece of contextual information (geographical, social, temporal, and categorical information). Next, these contextual models will be fused by aggregation into matrix factorization and neural networks. Moreover, different users follow various behavior; for instance, some users prefer to visit nearby POIs rather than distant ones. Thus, different behavior impacts the modeling and the performance of different contextual information. Therefore, we conduct additional experiments to study the impact of users' behavior on contextual information. To this end, first, we categorize users based on three behavioral parameters in visiting different POIs, namely, geographical distance, check-ins density, and exploration. Furthermore, we study each model's performance by the different contextual models for each user category and compare the effect of contextual information for these user groups. In the following, we first introduce the baseline algorithms, contextual models, and then we present our analytical framework.

\subsection{Baseline Algorithm Analysis}
\label{sec:ex_baseline}

Here, we describe the three selected baseline algorithms that can incorporate different contextual information to improve the performance of the POI recommendation. These are GeoSoCa \cite{zhang2015geosoca}, FCFKDEAMC \cite{zhang2014lore}, and PFMMGM \cite{cheng2012mgm}.

GeoSoCa performs recommendation by exploiting geographical (Geo), social (So), and categorical (Ca) correlations among users and POIs. GeoSoCa models geographical influence based on a check-in probability distribution over a two-dimensional space using a kernel density estimation (KDE) method. To model the social influence, GeoSoCa considers the check-in frequency of users' friends on a POI to compute the social check-in frequency or rating between users and POIs. GeoSoCa computes a power-law distribution over users' categorical check-in frequency to incorporate the categorical information. FCFKDEAMC fuses temporal information based on sequential patterns (AMC), two-dimensional check-in probability distributions as geographical (KDE) information, and social (FCF) information. To model the geographical influence, similar to GeoSoCa, FCFKDEAMC considers a two-dimensional check-in probability distribution using KDE. It models the social influence using a friend-based collaborative filtering (FCF) approach that allows making POI recommendations based on similar friends. Moreover, FCFKDEAMC incrementally mines location sequences of all users and represents the sequential patterns as a dynamic location-location transition graph. Then, transition probabilities can be dynamically calculated by dividing the transition counts on the location-location transition graph by the outgoing counts of each node. PFMMGM fuses the probabilistic factorization model (PFM) and the multi-center Gaussian model (MGM), which models user check-ins based on a user's geographical centers to predict the probability of a user visiting a POI.

We have selected these methods for our experiments mainly for two reasons: (i) they have successfully modeled all the contextual factors that we plan to study in this work, and (ii) they have been recognized as state-of-the-art by a significant number of works in the literature~\cite{liu2017experimental}.

\begin{table*}[t]
\centering
\caption{Notation of proposed models. The \textit{Notation} column show the contextual information that are parts of the proposed models. The \textit{Information} column shows which pieces of information is captured by the notation. \textit{Interaction-based} means no context, when only check-ins information is considered.}
\begin{tabular}{ll}
\hline
\multicolumn{1}{c}{Notation} & \multirow{1}{*}{Information}   \\ \cline{1-2}
M & Matrix Factorization (\textit{interaction-based})                       \\
N & Neural Network (\textit{interaction-based})                             \\
G        & Geographical Information                         \\
T        & Temporal Information                             \\
S        & Social Information                               \\
C        & Categorical Information                        \\
MG        & Matrix Factorization and Geographical Information   \\
GT       & Geographical and Temporal Information            \\
GS        & Geographical and Social Information            \\
GC       & Geographical and Categorical Information         \\
TS       & Temporal and Social Information                  \\
TC       & Temporal and Categorical Information             \\
SC       & Social and Categorical Information               \\
GTS      & Geographical, Temporal, and Social Information   \\
GTC      & Geographical, Temporal, and Categorical Information   \\
GSC      & Geographical, Social, and Categorical Information   \\
TSC      & Temporal, Social, and Categorical Information   \\
GTSC     & Geographical, Temporal, Social, and Categorical Information   \\
\end{tabular}
\label{tbl:contextnotation}
\end{table*}

Our experiments on the baseline models are defined as follows. We carry out two different experiments on baseline models to consider both the impact of different contextual information and the evaluation metrics for analysis. Table \ref{tbl:contextnotation} shows the notations that we use to specify a different combination of baselines. The first column (i.e., Notation) shows the different contextual information parts of the proposed models (see Section \ref{sec:proposedmodels}). The second column (i.e., Information) shows which pieces of information are captured by the notations. To refer to the baseline models, we will use \textit{Baseline-(Contexts)} notations where \textit{Baseline} indicates the model name (i.e., GeoSoca, FCFKDEAMC, or PFMMGM), and \textit{Contexts} shows which contexts are included in the model. For instance, to refer to \textit{GeoSoCa}, when we only consider the geographical and categorical parts, we use the notation \textit{GeoSoCa-(GC)}. We will use these notations to refer to the combination of baselines later in the experiment (Section \ref{sec:discuss}).

\subsubsection{Experiment 1: Focus on Contexts}
In the first experiment, to answer \textbf{RQ1}, we study and analyze the impact of each contextual information separately. Also, we consider different combinations to evaluate the combined use of contextual information in the recommendation process.
To this end, we report each model's results while keeping only the mentioned contextual information and removing the ones that are not mentioned. This enables us to understand better how each of the contextual factors, as well their combinations, would affect each of the models. Given that a diverse set of combinations of contextual information is possible by selection from four pieces of information, this allows us to characterize the impact of different contextual factors more carefully.

\subsubsection{Experiment 2: Focus on Metrics}
In this experiment, we aim to answer \textbf{RQ2} by studying the results of different evaluation metrics that we use in our analytical framework to evaluate the proposed models and compare them on the use of contextual information. To this end, we consider the three most commonly used evaluation metrics in information retrieval and recommender system domains, namely Precision, Recall, and nDCG. Then, we evaluate the effect of each contextual information and of their different combinations using the evaluation metrics mentioned above. This experiment helps us see how different evaluation metrics can capture the effect of contextual information on various models. For example, if precision shows that a model using geographical information outperforms one using social information, is it the same based on Recall?

\begin{table*}[t]
\centering
\caption{Definition of terms used in the paper.}
\label{tbl:terms}
\begin{tabular}{ll}
\hline
Terms           & Definition \\ \hline
$U$             & The set of users in the dataset \\
$L$             & The set of POIs in the dataset \\
$CT$            & The set of all POI categories \\
$u$             & A sample user in $U$ \\
$l$             & A sample POI in $L$ \\
$\mathcal{K}$             & The dimension of latent features \\
$\mathcal{U}$   & The user latent-factor matrix of matrix factorization \\
$\mathcal{L}$   & The POI latent-factor matrix of matrix factorization \\
$\mathcal{E}$   & The user latent-factor matrix of neural network \\
$\mathcal{Q}$   & The POI latent-factor matrix of neural network \\
$\mathcal{R}$   & The user-POI frequency matrix \\
$\mathcal{S}$   & The social links matrix \\
$g_{u,l}$       & The categorical popularity for user $u$ on POI $l$ \\
$x_{u,l}$       & The number of check-ins of user $u$'s friends on POI $l$ \\
$\mathbf{B}_{u,c}$ & The frequency of user $u$ visiting the POIs that belong to category $c$ \\
$\mathbf{H}_{c,l}$ & The check-in frequency of all users on POI $l$ in category $c$ \\
$C_u$           & The number of check-ins by user $u$ \\
$Q_u$           & The total number of unique visited POIs by user $u$
\end{tabular}
\end{table*}

\subsection{Contextual Model Analysis}
\label{sec:proposedmodels}
We selected matrix factorization as the linear approach to model the relation between users and POIs. Neural networks are instead selected as the non-linear way of capturing such a relation. Different contextual information can be fused and used in combination with matrix factorization and neural networks. This section presents our proposed linear and non-linear models, followed by the contextual information models. Table~\ref{tbl:terms} presents the terms and notations used in the rest of this paper.

\subsubsection{Matrix Factorization}
To model the user's preferences based on check-in data, we apply matrix factorization, a popularly used linear model \cite{liu2017experimental,yuan2013time,rahmani2019lglmf}. Since user feedback in POI recommendation is implicit feedback, we use a probabilistic factor model to efficiently deal with check-in data by defining a Poisson distribution on the frequency \cite{cheng2012mgm,cheng2016unified}. 

\begin{figure}
    \centering
    \includegraphics[scale=0.8]{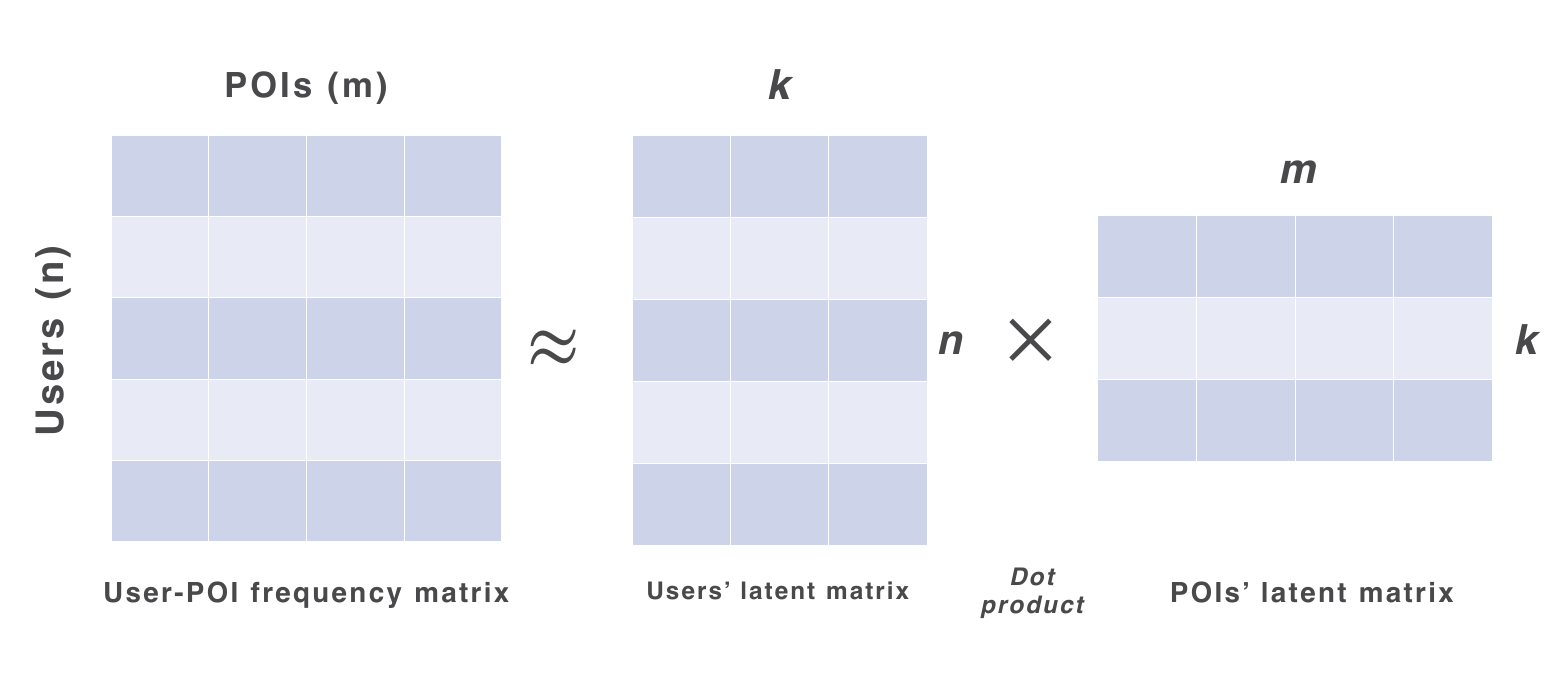}
    \caption{Matrix Factorization Model.}
    \label{fig:mfmodel}
\end{figure}

As shown in Fig.~\ref{fig:mfmodel}, the goal of matrix factorization is to find two low-rank matrices $\mathcal{U} \in \mathbb{R}^{\mathcal{K}\times|U|}$ and $\mathcal{L} \in \mathbb{R}^{\mathcal{K}\times|L|}$ based on the frequency matrix $\mathcal{R}$ such that $\mathcal{R} \approx \mathcal{U}^{T}\mathcal{L}$, where $\approx$ denotes ``approximately equal to''. The predicted probability of a user $u$, on a POI $l$, is determined by:
\begin{equation}
    P_{M_{u,l}} \propto \mathcal{U}_u^T\mathcal{L}_l
    \label{eq:m_score}
\end{equation}
\noindent
which can be obtained by solving the following optimization problem that places Beta distributions as priors on the latent matrices $\mathcal{U}$ and $\mathcal{L}$ while defining a Poisson distribution on the frequency:
\begin{equation}
\begin{split}
    min_{\{\mathcal{U},\mathcal{L}|\mathcal{R}\}} = \sum_{i=1}^{|U|}\sum_{k=1}^{\mathcal{K}}((\sigma_k-1)\ln(\mathcal{U}_{ik}/\rho_k)-\mathcal{U}_{ik}/\rho_k) \\ +
    \sum_{j=1}^{|L|}\sum_{k=1}^{\mathcal{K}}((\sigma_k-1)\ln(\mathcal{L}_{jk}/\rho_k)-\mathcal{L}_{jk}/\rho_k) \\ +
    \sum_{i=1}^{|U|}\sum_{j=1}^{|L|}((\mathcal{R}_{ij}\ln(\mathcal{U}^T\mathcal{L})_{ij}-(\mathcal{U}^T\mathcal{L})_{ij})+z)
\end{split}
\end{equation}
\noindent
where $\sigma=\{\sigma_1,...,\sigma_K\}$ and $ \rho=\{\rho_1,...,\rho_K\}$ are parameters for the Beta distributions, and $z$ is a constant.

\subsubsection{Neural Network}
\begin{figure}
    \centering
    \includegraphics[scale=0.6]{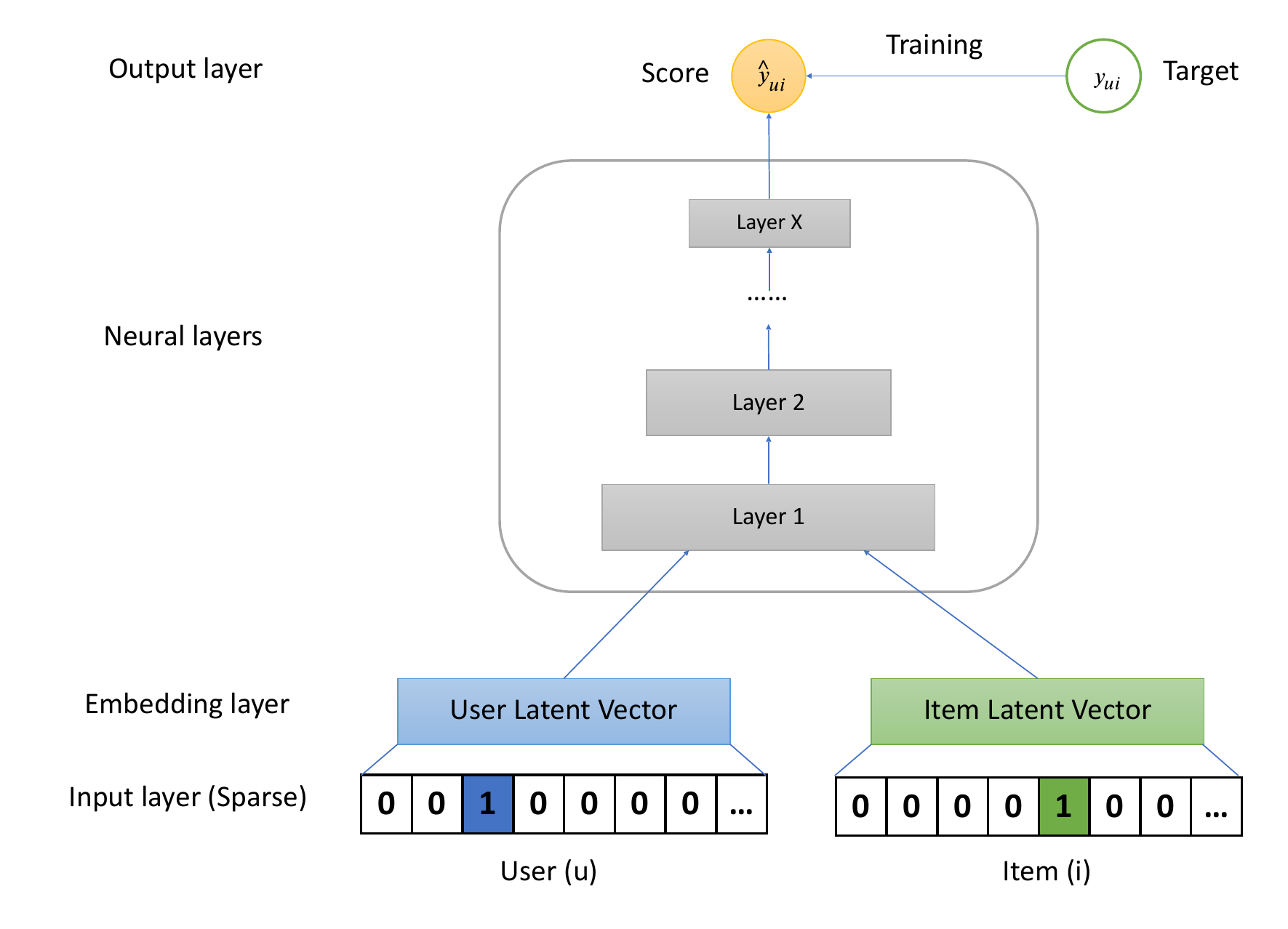}
    \caption{Neural Network Model.}
    \label{fig:nnmodel}
\end{figure}

To model a neural network of CF, as shown in Fig.~\ref{fig:nnmodel}, we consider a multi-layer representation of a user–POI interaction, similarly to \citet{he2017neural}.

The bottom input layer consists of user $v_u^U$ and POI $v_l^L$ feature vectors that describe user $u$ and POI $l$, respectively. Above the input layer is the embedding layer, a fully connected layer that projects the sparse representation to a dense vector. The user and POI embeddings are then fed into a multi-layer neural architecture to map the latent vectors to prediction scores. The final output layer is the predicted score $y$ that predicts if the user $u$ visit a POI $l$ or not:

\begin{equation}
    P_{N_{u,l}}=f(\mathcal{E}^Tv_u^U, \mathcal{Q}^Tv_l^L|\mathcal{E},\mathcal{Q},\theta_f)
    \label{eq:n_score}
\end{equation}
\noindent
where $\mathcal{E} \in \mathbb{R}^{\mathcal{K}\times|U|}$ and $\mathcal{Q} \in \mathbb{R}^{\mathcal{K}\times|L|}$, denote the latent factor matrix for users and POIs, respectively; while $\theta_f$ denotes the model parameters of the interaction function $f$. Since the function $f$ is defined as a multi-layer neural network, it can be formulated as:

\begin{equation}
    f(\mathcal{E}^Tv_u^U, \mathcal{Q}^Tv_l^L)=\diameter_{out}(\diameter_{X}(...\diameter_{2}(\diameter_{1}(\mathcal{E}^Tv_u^U, \mathcal{Q}^Tv_l^L))...))
\end{equation}
\noindent
where $\diameter_{out}$ and $\diameter_{X}$ respectively, denote the mapping function for the output layer and x-th neural layer when there are $X$ neural layers in total.

With the above two models, we can take into account all the different types of context that we have seen in Section~\ref{sec:intro}. In what follows, we provide a formal definition of these contextual factors.

\subsubsection{Geographical Information}
Different models use geographical information to generate a more accurate recommendation for each user based on the POI's and the user's location. Most of them are based on the distance between users and POIs. Some studies \cite{zhang2015geosoca,zhang2014lore} show that modeling geographical information leads to more recommendations as each user has different behavior. Here, we model a check-in probability distribution over a two-dimensional space using kernel density estimation \cite{zhang2015geosoca}. Then, we add a user-dependent variable for check-ins to make the geographical modeling more personalized. The geographical probability of visiting a new POI $l$ by a user $u$ is then estimated based on its location on the check-in probability distribution:

\begin{equation}
    P_{G_{u,l}}=\frac{1}{u_N}\sum_{i=1}^{n}(\mathcal{R}_{u,l_{i}}.Kh(l-l_{i}))
    \label{eq:geo_score}
\end{equation}
\noindent
where $u_N$ is the number of check-ins by user $u$, $\mathcal{R}$ shows the user-POI check-ins frequency matrix, and $Kh(l-l_{i})$ is the normal kernel function with a user-dependent variable.

\subsubsection{Social Information}
\citet{zhang2015geosoca} show that users' social check-ins follow a power-law distribution. We take this same approach to model social information. Therefore, in this case, the social link check-in frequency for user $u$ on POI $l$ is the check-in frequency on POI $l$ by user $u$'s friends. Then, we use the cumulative of the power-law distribution as the social information in recommendations as follows:

\begin{equation}
    P_{S_{u,l}}=1-(1+x_{u,l})^{1-\beta}
    \label{eq:so_score}
\end{equation}
\noindent
where $\beta$ is estimated by the check-in matrix $\mathcal{R}$ and social link matrix $\mathcal{S}$ as follows:

\begin{equation}
    \beta = 1+|U||L|[\sum_{u'\in{U}}\sum_{l'\in{L}}ln(1+\sum_{u''\in{U}}\mathcal{S}_{u',u''}.R_{u'',l'})]^{-1}
\end{equation}
\noindent
in which $\sum_{u''\in{U}}\mathcal{S}_{u',u''}.R_{u'',l'}$ is the social check-in frequency or rating of the friends $u''$ of user $u'$ on POI $l'$ and $x_{u,l}$ is the aggregation of the check-in frequency of $u$'s friends on the POI $l$, given by:

\begin{equation}
    x_{u,l}=\sum_{u'\in{U}}^{}\mathcal{S}_{u,u'}\mathcal{R}_{u',l}
\end{equation}
\noindent
where $\mathcal{S}_{u,u'}$ is a binary matrix that indicates whether there exists a social link between users $u$ and $u'$ or not. 

\subsubsection{Temporal information}
To model temporal information, we employ an Additive Markov chain \cite{zhang2014lore} that exploits the sequential transition pattern between users and POIs. The temporal probability of a user $u$ visiting a POI $l$ is based on the transition probability between all the user's visited POIs and the target POI, that is:

\begin{equation}
    P_{T_{u,l}}=\sum_{i=1}^{n}f(l_{n+1},l_i,n+1-i)
    \label{eq:temp_score}
\end{equation}
\noindent
where $n$ is the number of POIs and $f(l_{n+1},l_i,n+1-i)$ derive the sequential probability of visiting a new location conditioned on the sequence of visits based on the additive Markov chain that is calculated as:

\begin{equation}
    f(l_{n+1},l_i,n+1-i) = \frac{W(l_i).{TP(l_i\rightarrow{l_{n+1}})}}{\sum_{j=1}^{n}{W(l_j)}}
\end{equation}
\noindent
in which $W(l_i) = 2^{-\alpha.(n-i)}$ represents the sequence decay weight with the decay rate parameter $\alpha \geq 0$ and $TP(l_i\rightarrow{l_{n+1}}) = \frac{TCount(l_n, l_{n+1})}{OCount(l_n)}$ is the transition probability that $OCount(l_n)$ is the outgoing count of $l_i$ as a transition predecessor to other POIs. Also, $TCount(l_n, l_{n+1})$ indicates the number of transitions between $l_n$ and $l_{n+1}$.

\subsubsection{Categorical Information}
To model the categorical information, inspired by \citet{zhang2015geosoca}, we estimate a power-law distribution for users' categorical check-in frequency. This denotes the check-in frequency of user $u$ on all the POIs of category $c$. The cumulative distribution of the users' categorical check-in frequency is used as categorical information in recommendations as follow:

\begin{equation}
    P_{C_{u,l}}=1-(1+g_{u,l})^{1-\gamma},
    \label{eq:cat_score}
\end{equation}
\noindent
where $\gamma>1$ and $g_{u,l}=\sum_{c \in CT} \mathbf{B}_{u,c} \cdot \mathbf{H}_{c,l}$. Here, $CT$ is the set of all POI categories, $\mathbf{B}_{u, c}$ is the frequency of user $u$ visiting POIs that belong to the category $c$
and $\mathbf{H}_{c, l}$ is the check-in frequency of all users on POI $l$ in category $c$. Therefore, $g_{u,l}$ shows the categorical popularity of user $u$ on POI $l$ based on the set of categories and $\gamma$ is calculated as:

\begin{equation}
    \gamma = 1+|U||L|[\sum_{u'\in{U}}\sum_{l'\in{L}}ln(1+g_{u',l'})]^{-1}
\end{equation}

\subsubsection{Experiment 3: Focus on Contextual Models}
To answer \textbf{RQ1}, this experiment considers two well-known linear and non-linear methods: matrix factorization and neural networks. To answer \textbf{RQ2}, we evaluate the different combinations of our proposed contextual information on these two models. Finally, to integrate the proposed models with the contextual information given by Eqs.~(\ref{eq:m_score}), (\ref{eq:n_score}), (\ref{eq:geo_score}), (\ref{eq:temp_score}), (\ref{eq:so_score}), and (\ref{eq:cat_score}) into a unified preference score, $score(u,l)$, for user $u$ to unvisited POI $l$, we use the following sum rule:

\begin{equation}
    score(u, l) = \mathcal{M}_{u,l} + \mathcal{C}_{u,l}
    \label{eq:fusion}
\end{equation}
\noindent
where $\mathcal{M}_{u,l}$ is equal to $P_{M_{u,l}}$ for the matrix factorization based models or $P_{N_{u,l}}$ for the neural networks models. Also, $\mathcal{C}_{u,l} = [P_{G_{u,l}}, P_{T_{u,l}}, P_{S_{u,l}}, P_{C_{u,l}}]$ that is any sum combination of contextual models. The top-$n$ POIs $l$ with the highest score $score(u,l)$ are recommended to user $u$. It is worth mentioning that the sum rule has been widely used to fuse different factors for POI recommendations in previous works \cite{cheng2012mgm,zhang2015geosoca,zhang2013igslr,rahmani2019lglmf} and has shown high robustness.

As one can see in Table \ref{tbl:contextnotation}, each proposed model is denoted by a capital letter. To refer to the proposed models, we will use \textit{Model-(Contexts)} notations that \textit{Model} indicates the model name (i.e., M or N) and \textit{Contexts} shows which contexts are included in the model. For instance, to refer to the combination of matrix factorization with geographical and categorical context, we use the notation \textit{M-(GC)}. We will use these notations to refer to the combination of models later in the experiment (Section \ref{sec:discuss}).

Finally, to answer \textbf{RQ3}, we study the results of the proposed models based on the selected evaluation metrics, and we compare the results of different models on those metrics.

\subsection{Users Behavior Analysis}
Users have different behaviors based on the different contextual information. In this section, we define three different analyses to study the users' behavior. For instance, a user tends to visit the same POIs repeatedly, but another user might, differently, prefer to visit different POIs and never visit the same POI twice. Moreover, based on each context, users have different behavior. Therefore, we perform three experiments to consider user behavior, namely, (i) geographical distance, (ii) temporal check-in density, and (iii) exploration.

\subsubsection{User Behavior on Geographical Distance}
In order to model the users' behavior based on geographical information, we categorize users' geographical behavior. Users who tend to stay in a small neighborhood are different from those who move a lot. Thus, the geographical categorization distinguishes users by considering the range of consecutive check-ins' geographical distance.

\subsubsection{User Behavior on Temporal Check-in Density}
Similar to geographical distance, we investigate how users' density of check-ins impacts the performance of models. To this end, we consider the time between consecutive user check-ins at POIs. In this way, we distinguish between users who perform multiple check-ins on the same day for those who check in over several days.  We consider the temporal density of user check-ins and, therefore, the temporal distance of consecutive check-ins.

\subsubsection{User Behavior on Exploration}
We also define a new metric to measure how much a specific user wants to explore new POIs, instead of revisiting POIs. We define $Q_u$ as the total number of unique visited POIs by user $u$ and $C_u$ as the total number of check-ins by the user. As denoted in Eq.~\ref{eq:EF}, the user exploration factor is calculated simply by dividing the total number of unique visited POIs by the total number of check-ins. Therefore, if the exploration factor equals one, all the user's check-ins are to new POIs, and the user never visits a POI twice. Indeed, lower values mean that the user tends to visit the same POIs again and again. We categorize the users by the following exploration factor and analyze the performance of the models in relation to that.

\begin{equation}
    EF_u = \frac{Q_u}{C_u}
    \label{eq:EF}
\end{equation}

\begin{figure}[!tbp]
  \centering
  \subfloat[Exploration Geographical correlation.]
  {\includegraphics[width=0.32\textwidth]{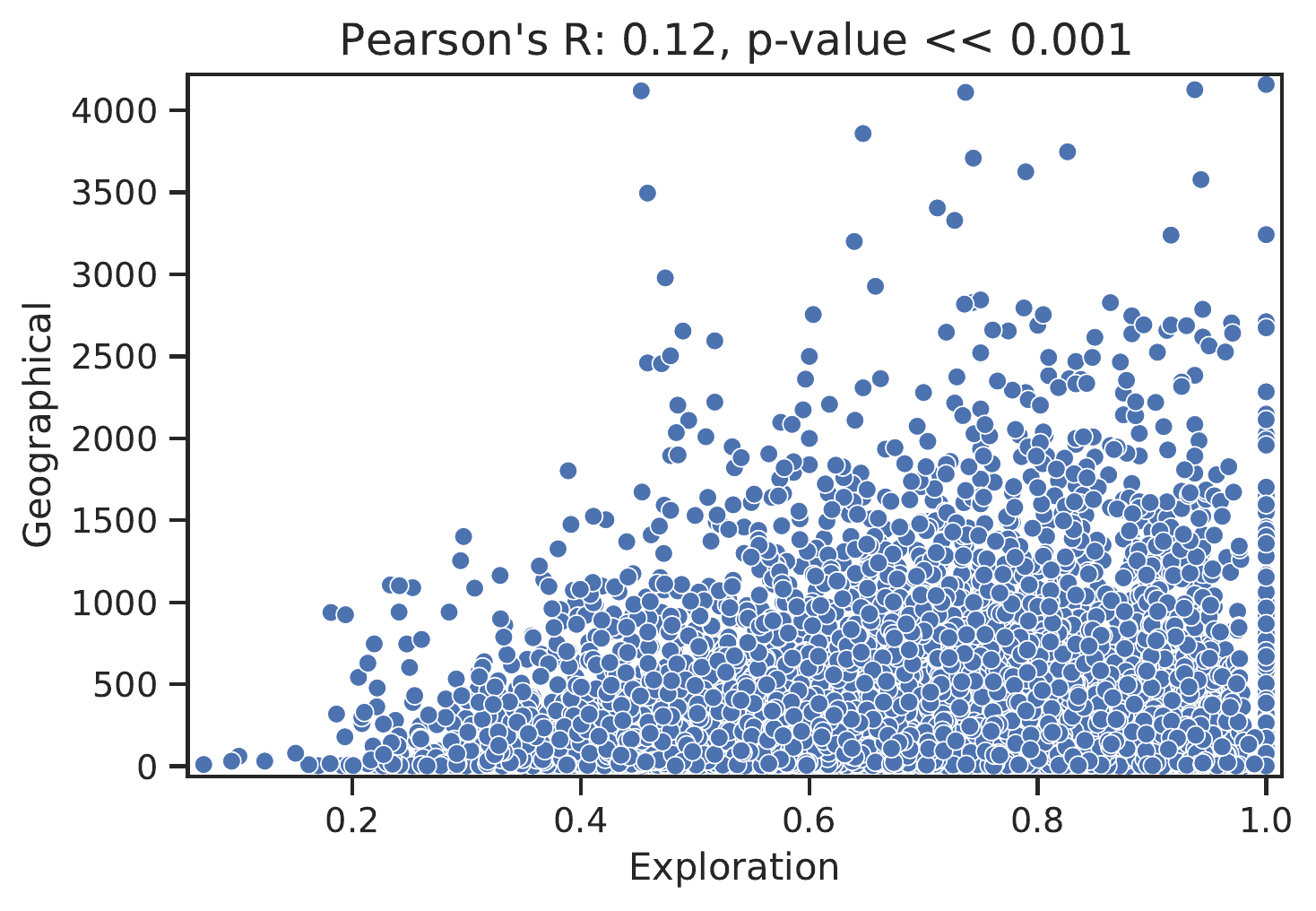}
  \label{fig:exp_geo_corr}}
  \hfill
  \subfloat[Exploration Temporal correlation.]
  {\includegraphics[width=0.32\textwidth]{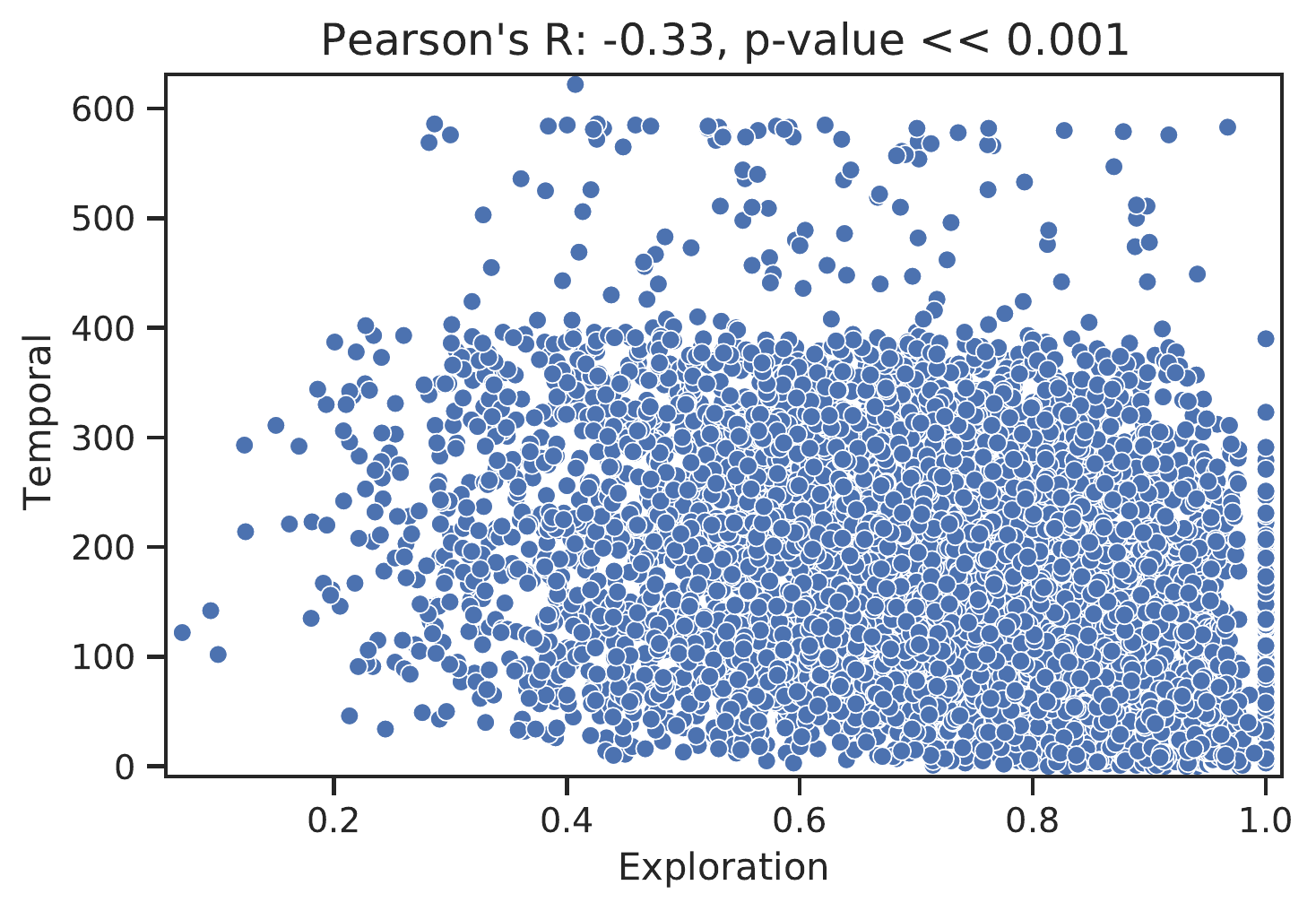}
  \label{fig:exp_temp_corr}}
  \hfill
  \subfloat[Temporal Geographical correlation.]
  {\includegraphics[width=0.32\textwidth]{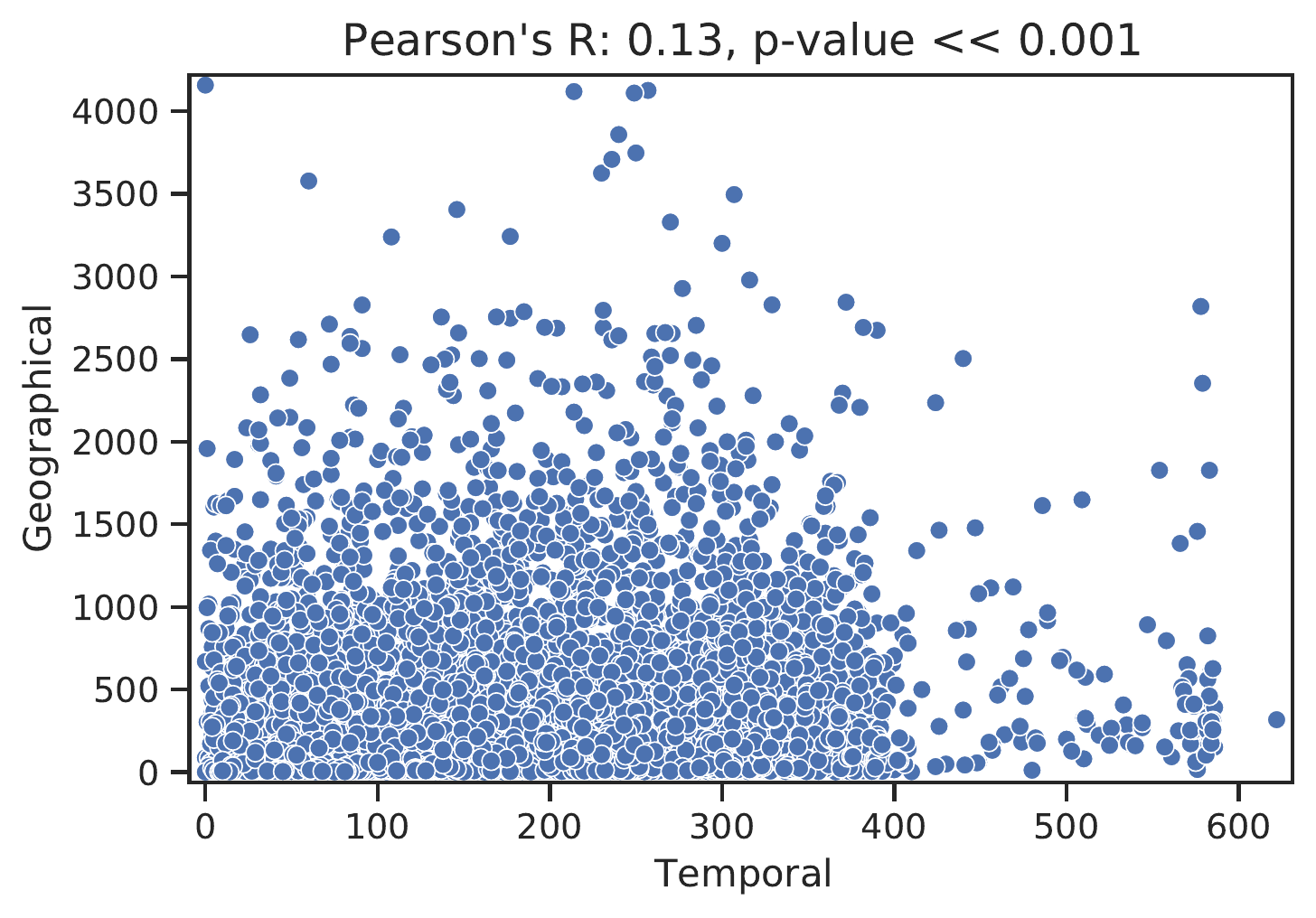}
  \label{fig:temp_geo_corr}}
  \caption{User Behavior Aspects Correlation}
  \label{fig:userAnalysisCorr}
\end{figure}

\subsection{Behavior Correlation}
In this section, we aim to show whether the three different aspects of user behavior that we study correlate with each other. One may argue that each of these variables depends on the others and, therefore, their values correlate. To test this, we compute the pair-wise correlations between these three factors based on \textit{Pearson's R} to analyze their relationship. These results are presented in Fig.~\ref{fig:userAnalysisCorr}. Overall, we observe a low correlation between these aspects/variables, supporting our assumption to consider them independently. Moreover, we see a \textit{negative correlation between the temporal and exploration aspects} (Pearson's R: -0.33). This suggests
that the more time between consecutive user check-ins, the less exploration the user carries out. Furthermore, from Fig.~\ref{fig:exp_geo_corr}, one can see that we found a \textit{low positive correlation between geographical distance and users' check-in exploration} (Pearson's R: 0.12). This indicates that the users that stay in the same neighborhoods tend to visit the same places, while when the geographical distance increases (e.g., visiting a new city or place which is far from their home location), they are more likely to go and explore more places. However, in none of the cases, we observe a considerable correlation between the variables. Therefore we conclude that even though there might be some dependencies among these variables (like those that we mentioned), the effect of those dependencies is negligible. 

\subsubsection{Experiment 4: Focus on Users Behavior}
In this experiment, we aim to answer \textbf{RQ4} by studying the users' behavior based on different contexts. We consider three different behavioral habit experiments based on geographical, temporal, and user preference information.

\section{Evaluation Methodology}
\label{sec:evaluation}
In this section, we present the experimental settings, including the datasets, the evaluation metrics, and the experimental setup we employed.

\subsection{Datasets}
We use two real-world check-in datasets from Yelp and Gowalla. They are commonly used by other papers as they have provided a lot of check-in information. Yelp dataset includes all the contextual information we needed (geographical coordinates, POI category, friendship information, and check-in timestamp). The Yelp dataset provided during the Yelp dataset challenge\footnote{https://www.yelp.com/dataset/challenge} round 7 (access date: Feb 2016) includes data from 10 metropolitan areas across two countries. From this dataset, we eliminate those users with fewer than 10 check-in POIs and those POIs with fewer than 10 visitors to consider enriched users and POIs and avoid the cold-start problem. The Gowalla dataset, on the other hand, does not include the POI category information. This dataset includes check-in data from February 2009 to October 2010. Before using this dataset, we filter out users with fewer than 15 check-ins and POIs with fewer than 10 visitors. The statistics of the datasets are provided in Table \ref{tbl:datasets}.

\begin{table*}[h]
\centering
\caption{Characteristics of the evaluation dataset used in the experiments. $\left| U \right|$ is the number of users, $\left| L \right|$ the number of POIs, $\left| C \right|$ is the number of check-ins, $\left| CU \right|$ the number of unique check-ins, $\left| CT \right|$ the number of category $\left| S \right|$ the number of social links.}
\label{tbl:datasets}
\begin{tabular}{lccccccccc}
\toprule
Datasets    & $|U|$ & $|L|$  & $|C|$ & $|CU|$ & $|CT|$ & $|S|$ & $\frac{|C|}{|U|}$ & $\frac{|C|}{|L|}$ & \%Sparsity   \\
\midrule
Yelp & 7,135 & 16,621 & 301,753 & 285,608 & 595 & 46,778 & 159.42 & 68.43 & 99.94\%      \\
Gowalla & 5,628 & 31,803 & 620,683 & 378,968 & - & 46,001 & 110.28 & 19.51 & 99.78\%      \\
\end{tabular}
\end{table*}

\subsection{Evaluation Metrics}
\label{sec:evalmetrics}
To evaluate the performance of the proposed experiments, we used three evaluation metrics commonly used for assessing the performance of location-based recommendation: \textit{Precision at K} (Pre@K), \textit{Recall at K} (Rec@K), and \textit{Normalized Discounted Cumulative Gain at N} (nDCG@K), with $K \in \{10, 20\}$. Given the top-K returned POIs for user $u$, Pre@K is defined as: 

\begin{equation}
    Pre@K = \frac{tp_u}{tp_u + fp_u}
\end{equation}
\noindent
while $Rec@K$ is defiend as:

\begin{equation}
    Rec@K = \frac{tp_u}{tp_u + tn_u}
\end{equation}
\noindent
where: $tp_u$ is the number of recommended POIs visited by $u$ (i.e., correct recommendations); $fp_u$ is the number of recommended POIs that are not visited by $u$ (i.e., incorrect recommendations); $tn_u$ is the number of POIs visited by $u$, that are not in the top-K recommendations.

Finally, we have the measure nDCG@K evaluating the ranking quality of the recommendation models. For each user, nDCG@K is defined as: 

\begin{equation}
    nDCG@K = \frac{DCG@K}{IDCG@K}
\end{equation}
\noindent
where $DCG@K=\sum_{i=1}^K\frac{2^{rel_i}-1}{log_2(i+1)}$. IDCG@K is the
DCG@N value when the recommended POIs are ideally ranked, and $rel_i$ refers to the graded relevance of the result ranked at position $i$. nDCG@K is in the range $0$ to $1$, and higher values mean better results. In the experiments, the average nDCG values of all users are reported.

\subsection{Experimental Setup}
In this paper, we use the critical difference diagram based on the Wilcoxon-Holm method to detect pairwise significance and to compare the ranking of contextual models where we evaluated them on all of the evaluation metrics. We use critical difference diagrams to present the overall ranking of methods. In this diagram, a thick horizontal line groups a set of models that are not significantly different. We partition each dataset into training, validation, and test sets. We first sort the check-ins of each user based on their timestamp, then for each user, we use the top 70\% of check-ins as training data, the following 20\% of check-ins as test data, and the remaining 10\% as validation data. To apply the significant difference, we first check the distribution of the datasets, and we observed that the datasets follow the normal distribution. Therefore, we determine the statistically significant differences between contextual models in terms of $Pre@K$, $Rec@K$, and $nDCG@K$ where $K \in \{10, 20\}$ using a two-tailed paired t-test at a 95\% confidence interval ($p < 0.05$). To evaluate the performance of neural network-based models, we need to add negative samples of check-ins, i.e., unvisited POIs, for each user into the datasets to prevent biasing of models. To this end, we add negative samples by first counting each user's unique visited POIs and next adding the same number of unvisited POIs for the same user \cite{he2017neural,he2018outer}. For the test set, we consider 1,000 negative samples for each user. We implement our methods with Python and Tensorflow\footnote{https://www.tensorflow.org/}. For the baseline algorithms, the parameters are initialized as reported in the corresponding papers \cite{zhang2015geosoca,zhang2014lore,cheng2012mgm,zhang2013igslr,liu2017experimental}. We set the latent factors parameter of K to $30$ for M and PFMMGM models. For the N model, we set user's and POI's embedding size to 30 and initialized them by latent factors that were extracted from matrix factorization. Finally, we optimized the model with mini-batch Adam optimizer \cite{kingma2014adam} and we employ two hidden layers for neural layers with size $128 \rightarrow 64$. Similarly, the choice of these parameters is based on literature \cite{he2017neural}.

\section{Results and Analysis}
\label{sec:discuss}
This section presents the results obtained from the experiments outlined in Section \ref{sec:methods}.

\begin{table*}[h]
\centering
\caption{Performance comparison in terms of Precision@$K$, Recall@$K$, and nDCG@$K$ for $K \in \{10,20\}$ on Yelp. The superscripts \textit{letters (a-q)} denote significant improvements compared to the other models ($p < 0.05$). The notation $\top$ shows the set of all the letters, and the best result of a different combination of contexts of baselines are shown in \textbf{bold}.}
\label{tbl:baseResultsYelp}
\begin{adjustbox}{max width=\textwidth}
\begin{tabular}{l|l|l|l|l|l|l|l}
\hline
Baselines               & Contexts & Pre@10 & Pre@20 & Rec@10 & Rec@20 & nDCG@10 & nDCG@20 \\ \hline
\multirow{7}{*}{GeoSoCa}& $^a$G & 0.0108$^{cop}$ & 0.0106$^{cop}$ & 0.0167$^{cop}$ & 0.0324$^{bcfop}$ & 0.0108$^{cop}$ & 0.0107$^{cop}$  \\ 
                         & $^b$S & 0.0127$^{acop}$ & 0.0113$^{cop}$ & 0.0154$^{cop}$ & 0.0267$^{cop}$ & 0.0130$^{acop}$ & 0.0120$^{acop}$  \\
                         & $^c$C & 0.0017 & 0.0017 & 0.0025 & 0.0047 & 0.0018 & 0.0018  \\ 
                         & $^d$GS & 0.0169$^{abcfop}$ & 0.0142$^{abcfop}$ & 0.0194$^{abcfop}$ & 0.0329$^{bcfop}$ & 0.0178$^{abcdfop}$ & 0.0157$^{abcfop}$  \\ 
                         & $^e$GC & 0.0156$^{abcfop}$ & 0.0142$^{abcfop}$ & \textbf{0.0230}$^{abcdfop}$ & \textbf{0.0424}$^{abcdfgop}$ & 0.0160$^{abcfop}$ & 0.0149$^{abcfop}$  \\ 
                         & $^f$SC & 0.0121$^{acop}$ & 0.0114$^{cop}$ & 0.0149$^{cop}$ & 0.0276$^{cop}$ & 0.0126$^{acop}$ & 0.0120$^{acop}$  \\ 
                         & $^g$GSC & \textbf{0.0183}$^{abcefop}$ & \textbf{0.0150}$^{abcfop}$ & 0.0214$^{abcfop}$ & 0.0340$^{bcfop}$ & \textbf{0.0195}$^{abcefop}$ & \textbf{0.0168}$^{abcefop}$  \\ \hdashline

\multirow{7}{*}{FCFKDEAMC} & $^h$G & 0.0108$^{cop}$ & 0.0103$^{cop}$ & 0.0163$^{cop}$ & 0.0311$^{cop}$ & 0.0108$^{cop}$  & 0.0104$^{cop}$  \\  
                         & $^i$S & 0.0156$^{abcfhop}$ & 0.0129$^{abcfhop}$ & 0.0177$^{abcfhop}$ & 0.0283$^{abcfhop}$ & 0.0162$^{abcfhop}$  & 0.0143$^{abcfhop}$  \\
                         & $^j$T & 0.0158$^{abcfhop}$ & 0.0145$^{abcfhiop}$ & 0.0229$^{abcfhiop}$ & 0.0406$^{abcfhiklnop}$ & 0.0165$^{abcfhlop}$  & 0.0154$^{abcfhop}$  \\ 
                         & $^k$SG & 0.0170$^{abcfhop}$ & 0.0146$^{abcfhiop}$ & 0.0192$^{abcfhopq}$ & 0.0329$^{abcfiopq}$ & 0.0179$^{abcfhop}$  & 0.0160$^{abcfhiop}$  \\ 
                         & $^l$ST & 0.0174$^{abcfh}$ & 0.0148$^{abcfhiopq}$ & 0.0195$^{abcfhopq}$ & 0.0321$^{abcfiopq}$ & 0.0185$^{abcfhiop}$  & 0.0164$^{abcfhiopq}$  \\ 
                         & $^m$GT & \textbf{0.0198}$^{\top-\{n\}}$ & \textbf{0.0173}$^{\top}$ & \textbf{0.0284}$^{\top}$ & \textbf{0.0487}$^{\top}$ & \textbf{0.0203}$^{{\top-\{ln\}}}$  & \textbf{0.0184}$^{{\top-\{n\}}}$  \\
                         
                         & $^n$SGT & 0.0182$^{\top-\{dgklm\}}$ & 0.0156$^{abcefhiopq}$ & 0.0203$^{abcefhiopq}$ & 0.0334$^{abcefhopq}$ & 0.0195$^{\top-\{dgklm\}}$ & 0.0173$^{\top-\{dgklm\}}$  \\ \hdashline

\multirow{3}{*}{PFMMGM}  & $^o$M & 0.0089$^{cp}$ & 0.0076$^{c}$ & 0.0104$^{c}$ & 0.0181$^{c}$ & 0.0086$^{cp}$ & 0.0078$^{c}$  \\ 
                         & $^p$G & 0.0075$^{c}$ & 0.0072$^{c}$ & 0.0113$^{c}$ & 0.0216$^{co}$ & 0.0074$^{c}$  & 0.0072$^{c}$  \\ 
                         & $^q$MG & \textbf{0.0160}$^{hop}$ & \textbf{0.0132}$^{hop}$ & \textbf{0.0241}$^{hiop}$ & \textbf{0.0398}$^{hiop}$ & \textbf{0.0172}$^{hop}$ & \textbf{0.0149}$^{hop}$  \\
\end{tabular}
\end{adjustbox}
\end{table*}

\begin{table*}[h]
\centering
\caption{Performance comparison in terms of Precision@$K$, Recall@$K$, and nDCG@$K$ for $K \in \{10,20\}$ on Gowalla. The superscripts \textit{letters (a-m)} denote significant improvements compared to the other models ($p < 0.05$). The notation $\top$ shows the set of all the letters and the best result of different combination of contexts of baselines are shown in \textbf{bold}.}
\label{tbl:baseResultsGowalla}
\begin{adjustbox}{max width=\textwidth}
\begin{tabular}{l|l|l|l|l|l|l|l}
\hline
Baselines  & Contexts & Pre@10 & Pre@20 & Rec@10 & Rec@20 & nDCG@10 & nDCG@20 \\ \hline

\multirow{3}{*}{GeoSoCa} & $^a$G & 0.0149 & 0.0151 & 0.0180$^{k}$ & 0.0363$^{k}$ & 0.0151 & 0.0151 \\ 
                         & $^b$S & 0.0207$^{adk}$ & 0.0181$^{adk}$ & 0.0212$^{ak}$ & 0.0355$^{k}$ & 0.0213$^{ad}$ & 0.0193$^{adk}$  \\
                         & $^c$GS &  \textbf{0.0215}$^{ad}$ & \textbf{0.0195}$^{ad}$ & \textbf{0.0253}$^{abd}$ & \textbf{0.0449}$^{abd}$ & \textbf{0.0222}$^{ad}$ & \textbf{0.0206}$^{ad}$ \\ \hdashline

\multirow{7}{*}{FCFKDEAMC}& $^d$G & 0.0159 & 0.0152 & 0.0191$^{k}$ & 0.0355$^{k}$ & 0.0159 & 0.0155  \\
                         & $^e$S & 0.0297$^{abcdklm}$ & 0.0242$^{abcdklm}$ & 0.0283$^{d}$ & 0.0441$^{d}$ & 0.0318$^{abcdklm}$ & 0.0274$^{abcdklm}$  \\ 
                         & $^f$T & \textbf{0.0502}$^{\top-\{i\}}$ & \textbf{0.0420}$^{\top-\{i\}}$ & \textbf{0.0470}$^{\top-\{i\}}$ & \textbf{0.0781}$^{\top-\{i\}}$ & \textbf{0.0536}$^{\top-\{hi\}}$ & \textbf{0.0469}$^{\top-\{i\}}$ \\ 
                         & $^g$SG & 0.0334$^{abcdeklm}$ & 0.0273$^{abcdeklm}$ & 0.0321$^{abcdeklm}$ & 0.0501$^{abcdeklm}$ & 0.0355$^{abcdeklm}$ & 0.0306$^{abcdeklm}$  \\ 
                         & $^h$ST & 0.0466$^{\top-\{fij\}}$ & 0.0360$^{\top-\{fij\}}$ & 0.0420$^{\top-\{fij\}}$ & 0.0612$^{\top-\{fij\}}$ & 0.0516$^{\top-\{fij\}}$ & 0.0428$^{\top-\{fij\}}$  \\ 
                         & $^i$GT & 0.0479$^{\top-\{fh\}}$ & 0.0404$^{\top-\{f\}}$ & 0.0459$^{\top-\{fj\}}$ & 0.0761$^{\top-\{f\}}$ & 0.0512$^{\top-\{fhj\}}$ & 0.0450$^{\top-\{fh\}}$ \\ 
                         & $^j$SGT & 0.0450$^{\top-\{fhi\}}$ & 0.0348$^{\top-\{fhi\}}$ & 0.0403$^{\top-\{fhi\}}$ & 0.0590$^{\top-\{fhi\}}$ & 0.0498$^{\top-\{fhi\}}$ & 0.0413$^{\top-\{fhi\}}$  \\ \hdashline
                         
\multirow{3}{*}{PFMMGM}  & $^k$M & 0.0168$^{a}$ & 0.0142$^{}$ & 0.0149$^{}$ & 0.0254$^{}$ & 0.0202$^{ad}$ & 0.0173$^{ad}$ \\ 
                         & $^l$G & 0.0187$^{a}$ & 0.0169$^{ak}$ & 0.0210$^{k}$ & 0.0364$^{k}$ & 0.0192$^{ad}$ & 0.0178$^{ad}$ \\ 
                         & $^m$MG & \textbf{0.0238}$^{abkl}$ & \textbf{0.0208}$^{abkl}$ & \textbf{0.0257}$^{abkl}$ & \textbf{0.0442}$^{abkl}$ & \textbf{0.0256}$^{abkl}$ & \textbf{0.0229}$^{abckl}$ \\
\end{tabular}
\end{adjustbox}
\end{table*}

\subsection{Experiment 1: Focus on Contexts (RQ1)}
\label{sec:results_contexts}
Here, our goal is to analyze the contribution of the contextual information and their different combinations on the POI recommendation's performance. In Tables \ref{tbl:baseResultsYelp} and \ref{tbl:baseResultsGowalla}, we show the results of a different combination of contextual models in baselines on the Gowalla and Yelp datasets. As the Gowalla dataset does not include categorical information, we cannot experiment with categorical information-based models on this dataset; thus Table \ref{tbl:baseResultsGowalla} has fewer number rows by removing the models with categorical context. The first column of Table \ref{tbl:contextnotation} shows the notation that we use to refer to different combinations of contextual information in baselines algorithms \cite{zhang2015geosoca,zhang2014lore,cheng2012mgm}. Below we summarise the conclusions we can draw from these tables. 

\textit{Temporal information plays a pivotal role among all combinations of contextual models.}
As seen, on both datasets displayed in Tables \ref{tbl:baseResultsYelp} and~\ref{tbl:baseResultsGowalla}, by using temporal context, \textit{FCFKDEAMC-(T)} is achieving higher performance against all models that include only a single contextual information (\textit{GeoSoCa-(G)}, \textit{GeoSoCa-(S)}, \textit{GeoSoCa-(C)}, \textit{FCFKDEAMC-(S)}, \textit{FCFKDEAMC-(G)}, \textit{FCFKDEAMC-(T)}, \textit{PFMMGM-(M)}, \textit{PFMMGM-(G)}). The location-based social networks capture the user's trajectory in the form of a sequence of visited locations. Our results show the importance of capturing such temporal information to model users' behavior. After \textit{FCFKDEAMC-(T)}, the results of \textit{GeoSoCa-(S)} and \textit{FCFKDEAMC-(S)} are close and, in some cases, even better than all geographical information, namely, \textit{GeoSoCa-(G)}, \textit{FCFKDEAMC-(G)}, and \textit{PFMMGM-(G)}. Moreover, temporal information is one of the most important contexts when combined with other contextual information. Table \ref{tbl:baseResultsYelp} shows \textit{FCFKDEAMC-(GT)} achieves the best results among all models, which include temporal information.

\textit{Considering geographical information per user on an individual basis is a more effective approach than creating universal distribution for all users.}
The results of \textit{GeoSoCa-(G)}, \textit{FCFKDEAMC-(G)}, and \textit{PFMMGM-(G)} show that among the models using geographical information (G), overall \textit{FCFKDEAMC-(G)} achieves the most stable results on both datasets. As we can see in Tables \ref{tbl:baseResultsYelp} and \ref{tbl:baseResultsGowalla}, \textit{PFMMGM-(G)} exhibits better results than \textit{GeoSoCa-(G)} and \textit{FCFKDEAMC-(G)} on the Gowalla dataset. However, on the Yelp dataset we see an opposite trend, where \textit{GeoSoCa-(G)} archives the best results. We see that on both datasets, even though less effective, \textit{FCFKDEAMC-(G)}'s performance is comparable with the other two models. Moreover, we observe a considerable difference in the performance of \textit{PFMMGM-(G)} and \textit{GeoSoCa-(G)} on the two datasets, where the models perform well on one dataset while failing on the other. 
Given that \textit{FCFKDEAMC-(G)} models the geographical component per user, while the two other models (i.e., \textit{PFMMGM-(G)} and \textit{GeoSoCa-(G)}) learn a universal geographical distribution, and \textit{FCFKDEAMC-(G)} exhibits a more robust performance on both datasets, we can conclude that utilizing a user-based individual geographical distribution leads to a more effective and robust geographical model.

\textit{Considering categorical information alone will not lead to improving the performance. However, its impact is significant when combined with geographical information.} Table \ref{tbl:baseResultsYelp} shows that the \textit{GeoSoCa-(C)} model is performing worse than all 16 other models in terms of all performance metrics. This could be due to the lack of consideration of relevant geographical context. For example, consider a situation of having POIs with the same category in different locations; users would prefer to visit nearby POIs rather than distant ones. Moreover, the result in Table \ref{tbl:baseResultsYelp} on \textit{GeoSoCa-(GC)} shows that a combination of \textit{GeoSoCa-(G)} and \textit{GeoSoCa-(C)} is significantly more effective than either of them separately.

\textit{Among different combinations of contextual information, the importance of geographical and temporal information is obvious.} As shown in Tables \ref{tbl:baseResultsYelp} and \ref{tbl:baseResultsGowalla}, \textit{FCFKDEAMC-(GT)} achieves the best performance in comparison with the other combinations of contextual information across $16$ out of $17$ models on the Yelp dataset and $11$ out of $13$ models on the Gowalla dataset. Moreover, one can see, when we add geographical information or temporal information to the models, they improve the performance of the models.

\textit{Considering all contextual information in a model will not necessarily improve the performance.} The comparison of \textit{FCFKDEAMC-(GT)} with \textit{FCFKDEAMC-(SGT)} shows that across 12 comparisons based on 6 evaluation metrics on two datasets, \textit{FCFKDEAMC-(GT)} beats the \textit{FCFKDEAMC-(SGT)} even when using fewer combinations of contextual information. 
It is worth noting that \textit{FCFKDEAMC-(SGT)} is the final model that \citeauthor{zhang2014lore} proposed~\cite{zhang2014lore}, but \textit{FCFKDEAMC-(ST)}, even with fewer contexts, outperforms \textit{FCFKDEAMC-(SGT)}. 

In summary, our main findings related to RQ1 are as follows:
\begin{itemize}
    \item Temporal contextual information plays a pivotal role among all combinations of contextual models.
    \item Categorical information alone will not lead to improvement in performance.
    \item Considering all the available contextual information in a model will not necessarily improve the performance.
    \item Among different combinations of contextual information, the importance of geographical and temporal information is clear.
\end{itemize}

\subsection{Experiment 2: Focus on Metrics (RQ2)}
\label{sec:results_metrics}
As seen in Tables \ref{tbl:baseResultsYelp} and \ref{tbl:baseResultsGowalla}, the performance of models varies when compared based on different evaluation metrics. For example, in terms of precision on both datasets \textit{GeoSoCa-(S)} performs more effectively compared with \textit{GeoSoCa-(G)}. However, the results on recall exhibit different behavior. We see that on recall, \textit{GeoSoCa-(G)} performs better than \textit{GeoSoCa-(S)}, suggesting that geographical information has more impact when focusing on recall. When we compare the results of \textit{PFMMGM-(M)} and \textit{PFMMGM-(G)}, which are respectively interaction-based and geographical-based models, we observe that \textit{PFMMGM-(G)} significantly outperforms \textit{PFMMGM-(M)} on recall. On the contrary, \textit{PFMMGM-(M)} achieves better results than \textit{PFMMGM-(G)} on the two other metrics, precision and nDCG. 

\begin{figure}[!t]
  \centering
  \subfloat[Performance on Rec@20]{\includegraphics[width=0.5\textwidth]{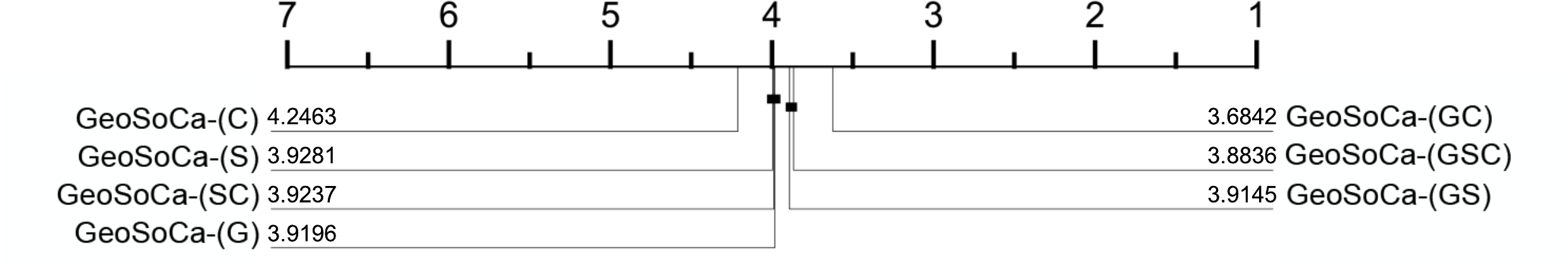}\label{fig:cd_base_rec}}
  \hfill
  \subfloat[Performance on nDCG@20]{\includegraphics[width=0.5\textwidth]{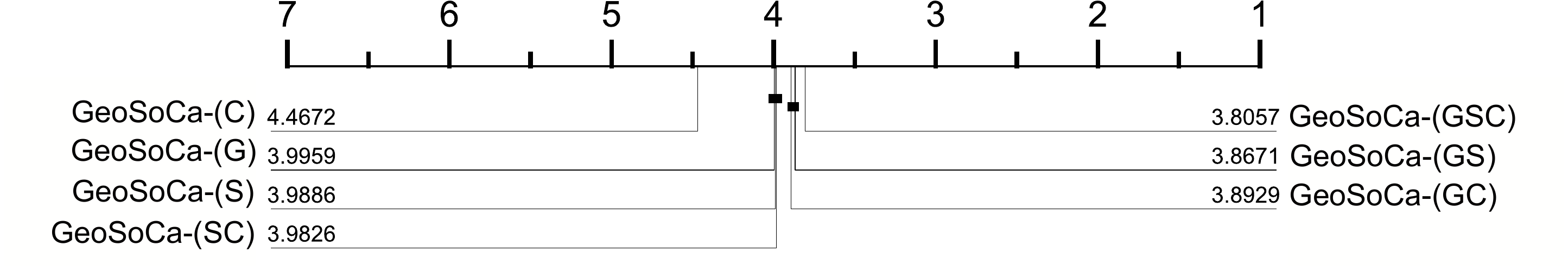}\label{fig:cd_base_ndcg}}
  \caption{Ranking of different combination of contextual information of GeoSoCa model on (a) Rec@20 and (b) nDCG@20.}
  \label{fig:cd_base}
\end{figure}

Considering these differences, in order to have an idea of the overall ranking of the models based on each performance metric, in Fig.~\ref{fig:cd_base} we present a critical difference diagram that allows comparing the statistical ranking of different models. In these graphs, the horizontal lines annotated by numbers show the ranking of models and, the thick horizontal lines group a set of models that are not significantly different. In Fig.~\ref{fig:cd_base}, we plot the ranking of a different combination of contextual models proposed in GeoSoCa \cite{zhang2015geosoca} based on two different evaluation metrics. As one can see in Fig.~\ref{fig:cd_base_rec}, the combination of geographical and categorical information achieve the best result based on the recall, while in Fig.~\ref{fig:cd_base_ndcg}, that shows the performance on nDCG, \textit{GeoSoCa-(GC)} ranked third and \textit{GeoSoCa-(SG)} is better. The difference is explained by the fact that these metrics measure different things. nDCG considers the ranking of the POIs when we recommend a list of POIs based on users' interests and behaviors, while the precision and recall only compare the list of recommended POIs and test set of users. One important fact to consider here is that these metrics are traditional metrics that are not designed for evaluating the quality of contextual information captured fully. There are currently no metrics that can assess the performance of recommendation with respect to the consistency of temporal, social, or geographical information. For instance, in evaluating the geographical context, we would ideally want to check if the recommended POIs are actually located near the user. Similarly, for evaluating a model incorporating temporal context, the sequence of visits or time of check-ins should be considered in the evaluation process and based on the users in the test set.

In summary, in relation to RQ2, we can observe different contrasting effects on traditional evaluation metrics of the use of context in POI recommendation models. For instance, one model might outperform other models on one metric like nDCG, while on other metrics such as precision and recall, the results might be completely different. The reasons for such behavior might be related to the different nature of these evaluation metrics. Thus, we need to propose new evaluation metrics based on contextual information.

\subsection{Experiment 3: Focus on Contextual Models (RQ3)}
\label{sec:results_models}
\begin{table*}[!t]
\centering
\caption{Performance comparison on matrix factorization based POI recommendation in terms of Precision@$K$, Recall@$K$, and nDCG@$K$ for $K \in \{10,20\}$ on Yelp. The superscripts \textit{letters (a-p)} denote significant improvements compared to the other models ($p < 0.05$). The notation $\top$ shows the set of all the letters. The best result between all models and the best result between different numbers of combinations are shown in \textbf{bold} and \textit{italic}, respectively.}
\label{tbl:MFYelp}
\begin{adjustbox}{max width=\textwidth}
\begin{tabular}{l|l|l|l|l|l|l|l}
\hline
Model & Contexts          & Pre@10 & Pre@20 & Rec@10 & Rec@20 & nDCG@10 & nDCG@20 \\ \hline
\multirow{16}{*}{M}&$^a$No Context & 0.0080$^{}$ & 0.0066$^{}$ & 0.0081$^{}$ & 0.0133$^{}$ & 0.0077$^{}$  & 0.0068$^{}$  \\ \cdashline{2-8}
&$^b$G       & \textit{0.0284}$^{\top-\{h\}}$ & \textit{0.0241}$^{\top-\{h\}}$ & \textit{0.0411}$^{\top-\{h\}}$ & \textit{0.0690}$^{\top-\{h\}}$ & \textit{0.0297}$^{\top-\{h\}}$  & \textit{0.0263}$^{\top-\{h\}}$ \\
&$^c$S       & 0.0178$^{ae}$ & 0.0160$^{ae}$ & 0.0218$^{ae}$ & 0.0380$^{ae}$ & 0.0185$^{ae}$  & 0.0170$^{ae}$ \\
&$^d$T       & 0.0221$^{ace}$ & 0.0195$^{ae}$ & 0.0308$^{ae}$ & 0.0537$^{afijop}$ & 0.0232$^{ae}$  & 0.0211$^{ajlo}$ \\
&$^e$C       & 0.0075$^{}$ & 0.0072$^{}$ & 0.0071$^{}$ & 0.0148$^{}$ & 0.0074$^{}$  & 0.0072$^{}$ \\ \cdashline{2-8}
&$^f$GS      & 0.0214$^{ace}$ & 0.0181$^{ace}$ & 0.0248$^{ace}$ & 0.0412$^{ace}$ & 0.0229$^{ace}$  & 0.0201$^{ace}$ \\
&$^g$GT      & 0.0253$^{\top-\{fhin\}}$ & 0.0219$^{\top-\{fhin\}}$ & 0.0371$^{\top-\{fhin\}}$ & 0.0617$^{\top-\{fhin\}}$ & 0.0264$^{\top-\{fhin\}}$  & 0.0237$^{\top-\{fhin\}}$ \\
&$^h$GC      & \textbf{0.0289}$^{\top-\{f\}}$ & \textbf{0.0246}$^{\top-\{f\}}$ & \textbf{0.0418}$^{\top-\{f\}}$ & \textbf{0.0701}$^{\top-\{f\}}$ & \textbf{0.0297}$^{\top-\{f\}}$  & \textbf{0.0265}$^{\top-\{f\}}$ \\ 
&$^i$ST      & 0.0195$^{ae}$ & 0.0166$^{ae}$ & 0.0224$^{ae}$ & 0.0369$^{ae}$ & 0.0203$^{ae}$  & 0.0181$^{ae}$ \\
&$^j$SC      & 0.0184$^{ae}$ & 0.0165$^{ae}$ & 0.0221$^{ae}$ & 0.0382$^{ae}$ & 0.0192$^{ae}$  & 0.0176$^{ae}$ \\
&$^k$TC      & 0.0231$^{ae}$ & 0.0202$^{ae}$ & 0.0322$^{ae}$ & 0.0564$^{ae}$ & 0.0243$^{ae}$  & 0.0220$^{ae}$ \\  \cdashline{2-8}
&$^l$GST     & 0.0204$^{ae}$ & 0.0172$^{ae}$ & 0.0239$^{ae}$ & 0.0382$^{ae}$ & 0.0213$^{ae}$  & 0.0188$^{ae}$ \\
&$^m$GSC     & 0.0219$^{ae}$ & 0.0186$^{ae}$ & 0.0253$^{ae}$ & 0.0417$^{ae}$ & 0.0232$^{ae}$  & 0.0205$^{ae}$ \\
&$^n$GTC     & \textit{0.0262}$^{\top-\{bhg\}}$ & \textit{0.0225}$^{\top-\{bhg\}}$ & \textit{0.0382}$^{\top-\{bhg\}}$ & \textit{0.0640}$^{\top-\{bhg\}}$ & \textit{0.0273}$^{\top-\{bhg\}}$  & \textit{0.0245}$^{\top-\{bhg\}}$ \\
&$^o$STC     & 0.0197$^{ae}$ & 0.0170$^{ae}$ & 0.0225$^{ae}$ & 0.0378$^{ae}$ & 0.0205$^{ae}$  & 0.0185$^{ae}$ \\ \cdashline{2-8}
&$^p$GSTC    & 0.0207$^{ae}$ & 0.0175$^{ae}$ & 0.0240$^{ae}$ & 0.0389$^{ae}$ & 0.0216$^{ae}$  & 0.0191$^{ae}$ \\
\end{tabular}
\end{adjustbox}
\end{table*}

\begin{table*}[!t]
\centering
\caption{Performance comparison on matrix factorization based POI recommendation in terms of Precision@$K$, Recall@$K$, and nDCG@$K$ for $K \in \{10,20\}$ on Gowalla. The superscripts \textit{letters (a-h)} denote significant improvements compared to the other models ($p < 0.05$). The notation $\top$ shows the set of all the letters. The best result between all models and the best result between different numbers of combinations are shown in \textbf{bold} and \textit{italic}, respectively.}
\label{tbl:MFGowalla}
\begin{adjustbox}{max width=\textwidth}
\begin{tabular}{l|l|l|l|l|l|l|l}
\hline
Model & Contexts   & Pre@10 & Pre@20 & Rec@10 & Rec@20 & nDCG@10 & nDCG@20 \\ \hline
\multirow{8}{*}{M} &$^a$No Context        & 0.0168 & 0.0142 & 0.0149 & 0.0254 & 0.0202 & 0.0173 \\ \cdashline{2-8}
&$^b$G       & 0.0377$^{ac}$ & 0.0326$^{ace}$ & 0.0404$^{ace}$ & 0.0680$^{ace}$ & 0.0402$^{ac}$ & 0.0359$^{ac}$ \\
&$^c$S       & 0.0261$^{a}$ & 0.0219$^{a}$ & 0.0242$^{a}$ & 0.0399$^{a}$ & 0.0290$^{a}$ & 0.0252$^{a}$ \\ 
&$^d$T       & \textbf{0.0542}$^{\top-\{d\}}$ & \textbf{0.0442}$^{\top-\{d\}}$ & \textbf{0.0517}$^{\top-\{df\}}$ & \textbf{0.0826}$^{\top-\{df\}}$ & \textbf{0.0583}$^{\top-\{df\}}$ & \textbf{0.0501}$^{\top-\{d\}}$ \\ \cdashline{2-8}
&$^e$GS      & 0.0372$^{ac}$ & 0.0295$^{ac}$ & 0.0360$^{ac}$ & 0.0543$^{ac}$ & 0.0407$^{ac}$ & 0.0343$^{ac}$ \\
&$^f$GT      & \textit{0.0503}$^{abcegh}$ & \textit{0.0410}$^{abcegh}$ & \textit{0.0493}$^{abcegh}$ & \textit{0.0789}$^{abcegh}$ & \textit{0.0538}$^{abceh}$ & \textit{0.0463}$^{abcegh}$ \\
&$^{g}$ST   & 0.0471$^{abcegh}$ & 0.0366$^{abcegh}$ & 0.0413$^{abce}$ & 0.0617$^{abce}$ & 0.0518$^{abceh}$ & 0.0431$^{abceh}$ \\ \cdashline{2-8}
&$^{h}$GST  & 0.0440$^{abce}$ & 0.0338$^{abce}$ & 0.0396$^{abce}$ & 0.0579$^{abce}$ & 0.0482$^{abce}$ & 0.0398$^{abce}$ \\
\end{tabular}
\end{adjustbox}
\end{table*}

\begin{table*}[!t]
\small
\centering
\caption{Performance comparison on neural network-based POI recommendation in terms of Precision@$K$, Recall@$K$, and nDCG@$K$ for $K \in \{10,20\}$ on Yelp. The superscripts \textit{letters (a-p)} denote significant improvements compared to the other models ($p < 0.05$). The notation $\top$ shows the set of all the letters. The best result between all models and the best result between different numbers of combinations are shown in \textbf{bold} and \textit{italic}, respectively.}
\label{tbl:NNYelp}
\begin{adjustbox}{max width=\textwidth}
\begin{tabular}{l|l|l|l|l|l|l|l}
\hline
Model&Contexts & Pre@10 & Pre@20 & Rec@10 & Rec@20 & nDCG@10 & nDCG@20 \\ \hline
\multirow{16}{*}{N}&$^a$No Context        & 0.066$^{bfghnlmnp}$ & 0.0551$^{bfghnlmnp}$ & 0.0849$^{bfghnlmnp}$ & 0.1405$^{bfghnlmnp}$ & 0.0687$^{bfghnlmnp}$ & 0.0604$^{bfghnlmnp}$ \\ \cdashline{2-8}
&$^b$G       & 0.0526$^{}$ & 0.042$^{}$ & 0.0599$^{}$ & 0.0934$^{}$ & 0.059$^{}$ & 0.0497$^{}$ \\
&$^c$S       & \textit{0.1104}$^{\top-\{cijo\}}$ & \textit{0.0863}$^{\top-\{cijo\}}$ & \textit{0.1412}$^{\top-\{cijo\}}$ & \textit{0.2123}$^{\top-\{cijo\}}$ & \textit{0.1205}$^{\top-\{cijo\}}$ & \textit{0.1008}$^{\top-\{cijo\}}$ \\ 
&$^d$T       & 0.0683$^{befghnlmnp}$ & 0.0565$^{bfghnlmnp}$ & 0.0886$^{befghnlmnp}$ & 0.1447$^{bfghnlmnp}$ & 0.0742$^{\top-\{cdijko\}}$  & 0.0642$^{\top-\{cdijko\}}$ \\
&$^e$C       & 0.0643$^{bhnfglmnp}$ & 0.0548$^{bhnfglmnp}$ & 0.0825$^{bhnfglmnp}$ & 0.1419$^{bhnfglmnp}$ & 0.0665$^{bhnfg}$  & 0.0594$^{bhnfglmnp}$ \\ \cdashline{2-8}
&$^f$GS      & 0.0563$^{hn}$ & 0.0449$^{hn}$ & 0.0644$^{bhn}$ & 0.1$^{bhn}$ & 0.0623$^{hn}$  & 0.0525$^{ghn}$ \\
&$^g$GT      & 0.0528 & 0.0421 & 0.0602 & 0.0936$^{}$ & 0.0597$^{}$ & 0.0501 \\
&$^h$GC      & 0.0521$^{}$ & 0.0418 & 0.0588$^{}$ & 0.0916 & 0.0576$^{}$ & 0.0487$^{}$ \\ 
&$^i$ST      & \textit{0.1115}$^{\top-\{cijo\}}$ & \textbf{0.0868}$^{\top-\{cijo\}}$ & \textit{0.1433}$^{\top-\{cijo\}}$ & \textit{0.2144}$^{\top-\{cijo\}}$ & \textit{0.1222}$^{\top-\{cijo\}}$ & \textbf{0.1019}$^{\top-\{cijo\}}$ \\
&$^{j}$SC   & \textit{0.1115}$^{\top-\{cijo\}}$ & 0.0863$^{\top-\{cijo\}}$ & 0.1422$^{\top-\{cijo\}}$ & 0.2128$^{\top-\{cijo\}}$ & 0.1213$^{\top-\{cijo\}}$ & 0.1009$^{\top-\{cijo\}}$ \\
&$^{k}$TC   & 0.0664$^{bfghlmnp}$ & 0.0558$^{bfghlmnp}$ & 0.0857$^{bfghlmnp}$ & 0.1451$^{bfghlmnp}$ & 0.0698$^{bfghlmnp}$ & 0.0615$^{bfghlmnp}$ \\ \cdashline{2-8}
&$^{l}$GST     & 0.0564$^{hn}$ & 0.045$^{hn}$ & 0.0646$^{bghn}$ & 0.1003$^{bghn}$ & 0.0625$^{hn}$ & 0.0527$^{hn}$ \\
&$^{m}$GSC     & 0.0569$^{bhn}$ & 0.0451$^{bhn}$ & 0.065$^{bhn}$ & 0.0994$^{bhn}$ & 0.0627$^{hn}$ & 0.0527$^{hn}$ \\
&$^{n}$GTC     & 0.0523$^{}$ & 0.0419 & 0.0592$^{}$ & 0.0919 & 0.058 & 0.049 \\
&$^{o}$STC     & \textbf{0.1124}$^{\top-\{cijk\}}$ & \textit{0.0867}$^{\top-\{cijk\}}$ & \textbf{0.1439}$^{\top-\{cijk\}}$ & \textbf{0.2146}$^{\top-\{cijk\}}$ & \textbf{0.1225}$^{\top-\{cijk\}}$ & \textit{0.1017}$^{\top-\{cijk\}}$ \\ \cdashline{2-8}

&$^{p}$GSTC    & 0.057$^{bhn}$ & 0.0451$^{bhn}$ & 0.0651$^{bhn}$ & 0.0998$^{bhn}$ & 0.0628$^{bhn}$ & 0.0528$^{hn}$ \\
\end{tabular}
\end{adjustbox}
\end{table*}

\begin{table*}[!h]
\centering
\caption{Performance comparison on neural network-based POI recommendation in terms of Precision@$K$, Recall@$K$, and nDCG@$K$ for $K \in \{10,20\}$ on Gowalla. The superscripts \textit{letters (a-h)} denote significant improvements compared to the other models ($p < 0.05$). The notation $\top$ shows the set of all the letters. The best result between all models and the best result between different numbers of combinations are shown in \textbf{bold} and \textit{italic}, respectively.}
\label{tbl:NNGowalla}
\begin{adjustbox}{max width=\textwidth}
\begin{tabular}{l|l|l|l|l|l|l|l}
\hline
Model & Contexts    & Pre@10 & Pre@20 & Rec@10 & Rec@20 & nDCG@10 & nDCG@20 \\ \hline
\multirow{8}{*}{N}&$^a$No Context      & 0.1142$^{}$ & 0.0874$^{}$ & 0.1$^{}$ & 0.1461$^{}$ & 0.1312  & 0.1074$^{}$  \\ \cdashline{2-8}
&$^b$G     & 0.1366$^{ad}$ & 0.1028$^{ad}$ & 0.1154$^{ad}$ & 0.1643$^{ad}$ & 0.1551$^{ad}$ & 0.1259$^{ad}$ \\
&$^c$S     & \textit{0.1767}$^{abdefh}$ & \textit{0.13}$^{abdefh}$ & \textit{0.1558}$^{abdefh}$ & \textit{0.2127}$^{abdefh}$ & \textit{0.1979}$^{abdefh}$ & \textit{0.159}$^{abdehf}$ \\ 
&$^d$T     & 0.1174$^{}$ & 0.0899$^{}$ & 0.1032$^{}$ & 0.1497$^{}$ & 0.1351$^{}$ & 0.1105$^{}$ \\ \cdashline{2-8}
&$^e$GS    & 0.1569$^{abdf}$ & 0.1145$^{abdf}$ & 0.1302$^{abdf}$ & 0.1774$^{abdf}$ & 0.1765$^{abdf}$ & 0.1412$^{abdf}$ \\
&$^f$GT    & 0.1373$^{ad}$ & 0.1032$^{ad}$ & 0.1161$^{ad}$ & 0.1649$^{ad}$ & 0.1564$^{ad}$ & 0.1269$^{ad}$ \\
&$^g$ST    & \textbf{0.1773}$^{abdefh}$ & \textbf{0.1303}$^{abdefh}$ & \textbf{0.1566}$^{abdefh}$ & \textbf{0.2133}$^{abdefh}$ & \textbf{0.199}$^{abdefh}$ & \textbf{0.1598}$^{abdefh}$ \\ \cdashline{2-8}
&$^h$GST   & 0.1571$^{abdf}$ & 0.1146$^{abdf}$ & 0.1306$^{abdf}$ & 0.1776$^{abdf}$ & 0.1773$^{abdf}$ & 0.1417$^{abdf}$ \\
\end{tabular}
\end{adjustbox}
\end{table*}

\begin{figure}[t]
\centering
\includegraphics[scale=0.6]{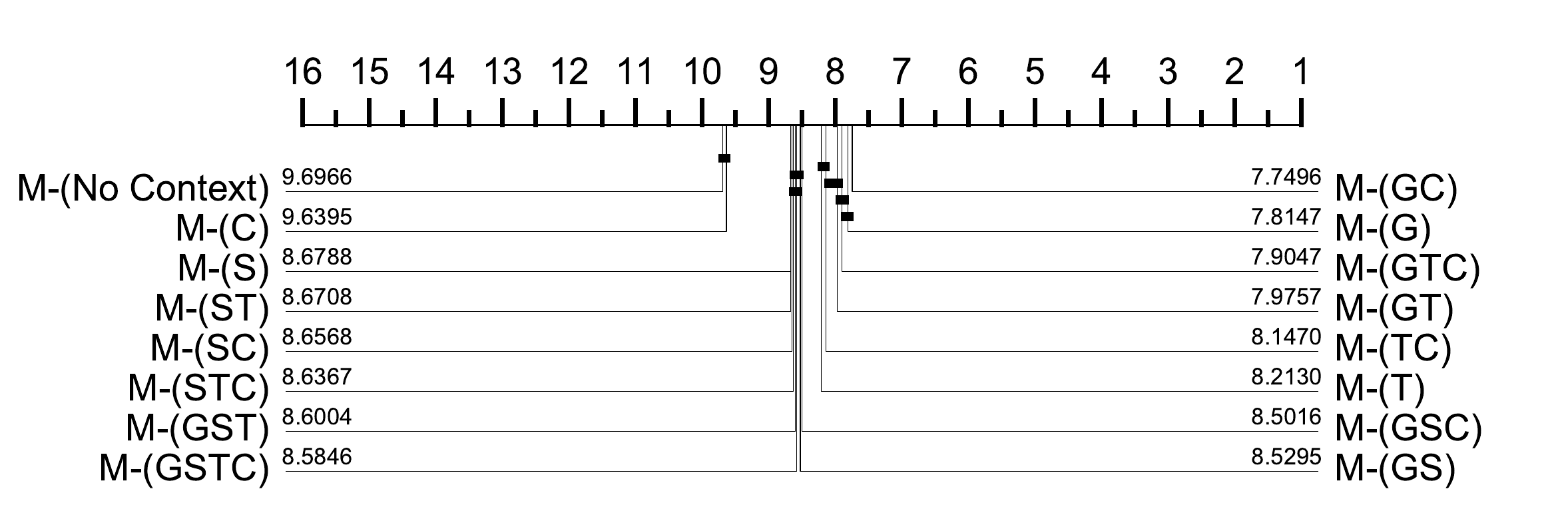}
\caption{Ranking of the contextual information in combination with matrix factorization.}
\label{fig:CD_MF}
\end{figure}

\begin{figure}[t]
\centering
\includegraphics[scale=0.6]{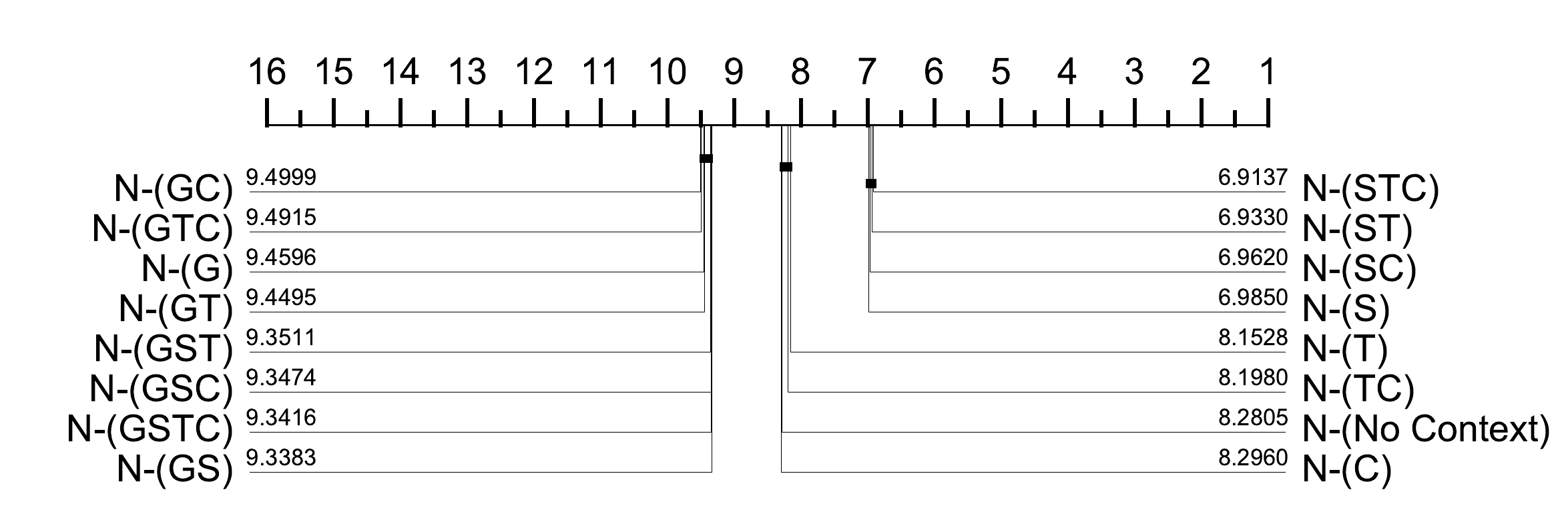}
\caption{Ranking of the contextual information in combination with neural network.}
\label{fig:CD_NN}
\end{figure}

After comparing how contextual information is modeled in baselines, we turn to compare our proposed models (based on matrix factorization and neural networks) using a similar combination of contextual information. In the rest of this section, we will refer to these models like the M and N models, respectively. We compare the performance of the models earlier presented in Table \ref{tbl:contextnotation}, and we report detailed results in terms of $Pre@K$, $Rec@K$, and $nDCG@K$ where $K \in \{10, 20\}$ in Tables \ref{tbl:MFYelp}, \ref{tbl:MFGowalla}, \ref{tbl:NNYelp}, and \ref{tbl:NNGowalla}. Also, in Figs.~\ref{fig:CD_MF} and \ref{fig:CD_NN}, we plot the ranking of the contextual information in combination with matrix factorization and neural networks, respectively. We can summarise the following observations using these comparisons:

\textit{The \textit{M} and \textit{N} models are more effective in all evaluation metrics on the Gowalla dataset while performing poorly on the Yelp dataset}. Tables \ref{tbl:MFYelp}, \ref{tbl:MFGowalla}, \ref{tbl:NNYelp}, and \ref{tbl:NNGowalla} show the results of the \textit{M} and \textit{N} models are much better on the Gowalla dataset. As seen in Table \ref{tbl:datasets}, while the sparsity of the two datasets is the same, there is a major difference in the ratio between the number of users and check-ins. In Yelp, there are far fewer check-ins per user. We also observed the same results on the \textit{Baselines} models in Tables \ref{tbl:baseResultsYelp} and \ref{tbl:baseResultsGowalla}.

\textit{The \textit{N} model achieves better performance compared to the \textit{M} model.} This is supported by the results on Tables \ref{tbl:NNYelp} and \ref{tbl:NNGowalla}. This shows that by modeling non-linear relations, the neural network can capture the latent relations between users and POIs better. We also see that all models have lower performance on Yelp than on Gowalla. This can be due to the fact that Gowalla has more check-ins, which is essential when you want to train a neural network-based model.

\textit{Fused models exhibit better performance compared with the simple check-in frequency-based models on M and N models.} In Tables \ref{tbl:MFYelp} and \ref{tbl:NNYelp}, we can see that the quality of POI recommendation relies on the number of contextual information fused. As one can see from Figs.~\ref{fig:CD_MF} and \ref{fig:CD_NN}, all of the contextual information and their combination have an impact on the performance of matrix factorization-based methods (by having a higher rank than the model with no context). However, for neural network-based methods, this depends on the incorporated contextual information: there are combinations of contextual information that can degrade the performance of recommendation (e.g., \textit{GC} and \textit{GST}).

\textit{Consideration of all available contextual information is not the best option for neither of the \textit{M} and \textit{N} models.} In both Figs.~\ref{fig:CD_MF} and \ref{fig:CD_NN}, the combination of all contextual information does not produce accurate results and can adversely lead to poor performance. The reason can be the operator rule (shown in Eq.~\ref{eq:fusion}) that is used to fuse them. For example, in this paper, we consider the sum rule based on previous research \cite{cheng2012mgm,zhang2015geosoca,zhang2013igslr}, but there are some other aggregation operators, such as the product rule, which can help to improve the different combinations of contextual information. We leave the further investigation of the effect of different ways of fusion (i.e., product, sum, mean rule, etc.) for future work.

\textit{The geographical or temporal information is more effective than other contextual information in improving the \textit{M} model.} The results of the \textit{M} model in Tables \ref{tbl:MFYelp} and \ref{tbl:MFGowalla} on both datasets show that if an additional context to interaction or check-in frequency is needed, it will be better to select geographical or temporal information.

\textit{Social information has a high impact on improving the \textit{N} model.} The combination of \textit{N} and \textit{S} is overall better than other combinations of different contextual information. Fig.~\ref{fig:CD_NN} shows the results of \textit{S} model where we see that social information improves the performance of the POI recommendation model. Therefore, it can help different models to provide better and more accurate results.

\textit{Temporal and categorical information appears to be the best contextual information in improving performance among the \textit{M} and \textit{N} models.}
Figs. \ref{fig:CD_MF} and \ref{fig:CD_NN} show that on the \textit{N} and \textit{M} models, the temporal and categorical information aspects have more impact on the user's behavior and are better incorporated. This can be because of the importance of considering users' sequential transition between the POIs in different temporal states and the category of POIs at the same time.

\textit{The combination of geographical and categorical information is the best case to model users' behavior on the \textit{M} model.} Fig.~\ref{fig:CD_MF} shows the combination of the \textit{G} and \textit{C} models with \textit{M} outperforming other combinations and archives the highest rank between the different contextual information. It shows that the synergy of geographical and categorical information on the user's preference works. The second best combination involves geographical and temporal information, which is the best combination on the Gowalla dataset since we do not have categorical information for that dataset.

\textit{The categorical information has the worst results among contextual models on the \textit{M} model.} This might be due to the fact that the \textit{M-(C)} model is ignorant of geographical context. Two POIs may have the same category, but distant POIs would be less likely to be visited. Although categorical information does not perform well alone, it seems to complement other information very well.

\textit{On datasets with a few check-ins per user, the \textit{M} model in combination with geographical information works best.} This conclusion is based on our analysis of two datasets with different levels of check-ins (see Table \ref{tbl:datasets}). It can be observed (in Table \ref{tbl:MFYelp}) that the \textit{M-(T)} model does not perform well, and its performance is lower than the \textit{M-(G)} model. However, the \textit{M-(T)} model performs better without the combination with other contextual information on the Gowalla dataset (see Table \ref{tbl:MFGowalla}).

In summary, our findings related to RQ3 are:
\begin{itemize}
    \item The ratio of the number of users and check-ins in datasets has a significant impact on the performance of the models.
    \item Neural network-based models overall perform better than  matrix factorization based models. This can be due to the fact that non-linear models achieve a higher-order representation of users and POIs relation.
    \item In both matrix factorization and neural network approaches, the fusion of contextual information exhibits better performance.
    \item The combination of all contextual information does not necessarily produce accurate results and can adversely lead to poor performance. 
    \item On datasets with a few check-ins per user, the matrix factorization model in combination with geographical information works best.
    \item Among both matrix factorization and neural networks, temporal and geographical information appears to be the best contextual information in improving the performance of models.
    \item Overall, the categorical information has the worst results among contextual models on models.
\end{itemize}

\subsection{Experiment 4: Focus on Users Behavior (RQ4)}
\label{sec:results_users}

\begin{figure}[!tbp]
  \centering
  \subfloat[Top baseline models from Table \ref{tbl:baseResultsGowalla}]{\includegraphics[scale=0.4]{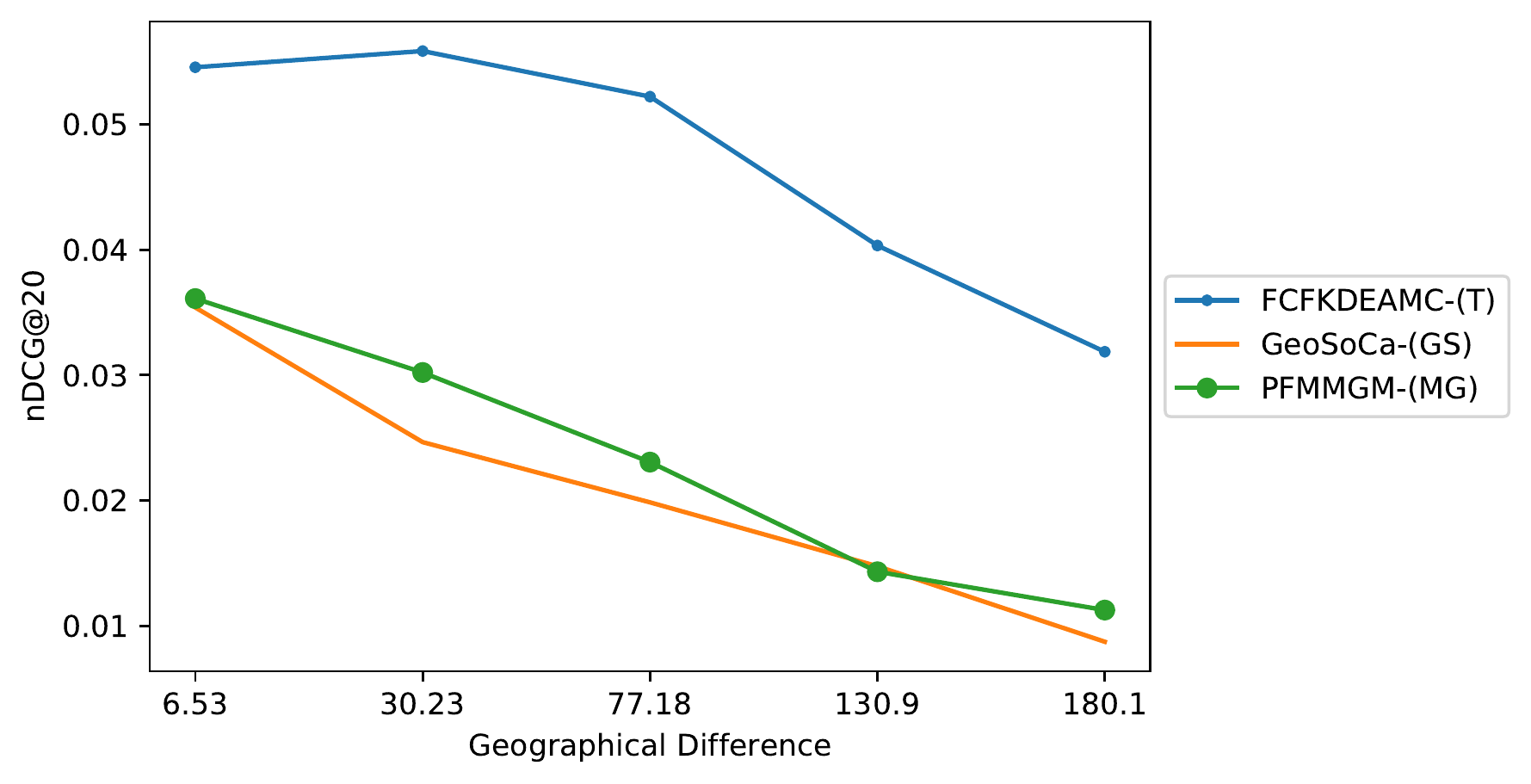}\label{fig:geoFactor_baselines_gowalla}}
  \hfill
  \subfloat[Top M models from Table \ref{tbl:MFGowalla}]{\includegraphics[scale=0.4]{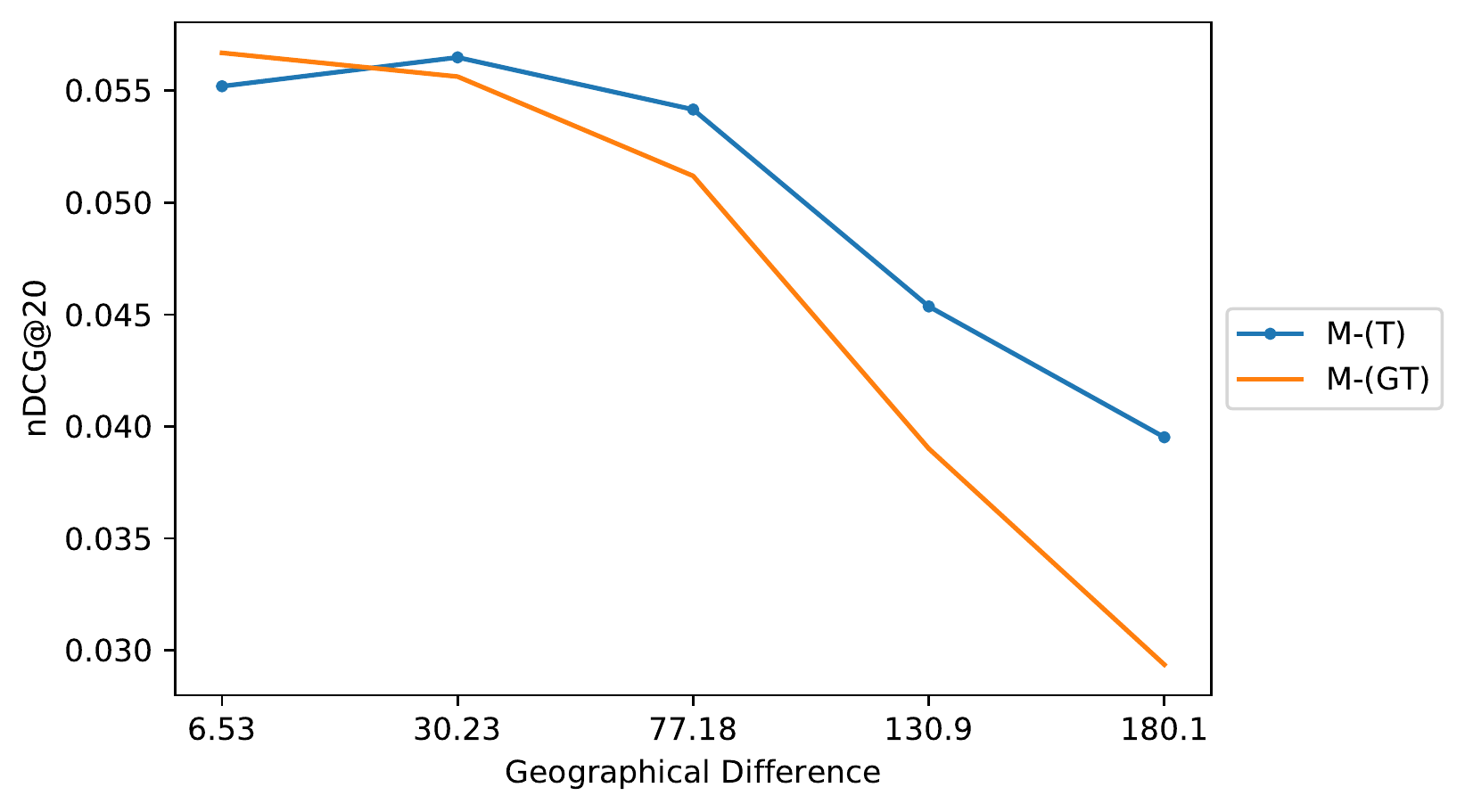}\label{fig:geoFactor_MFs_gowalla}}
  \hfill
  \subfloat[Top N models from Table \ref{tbl:NNGowalla}]{\includegraphics[scale=0.4]{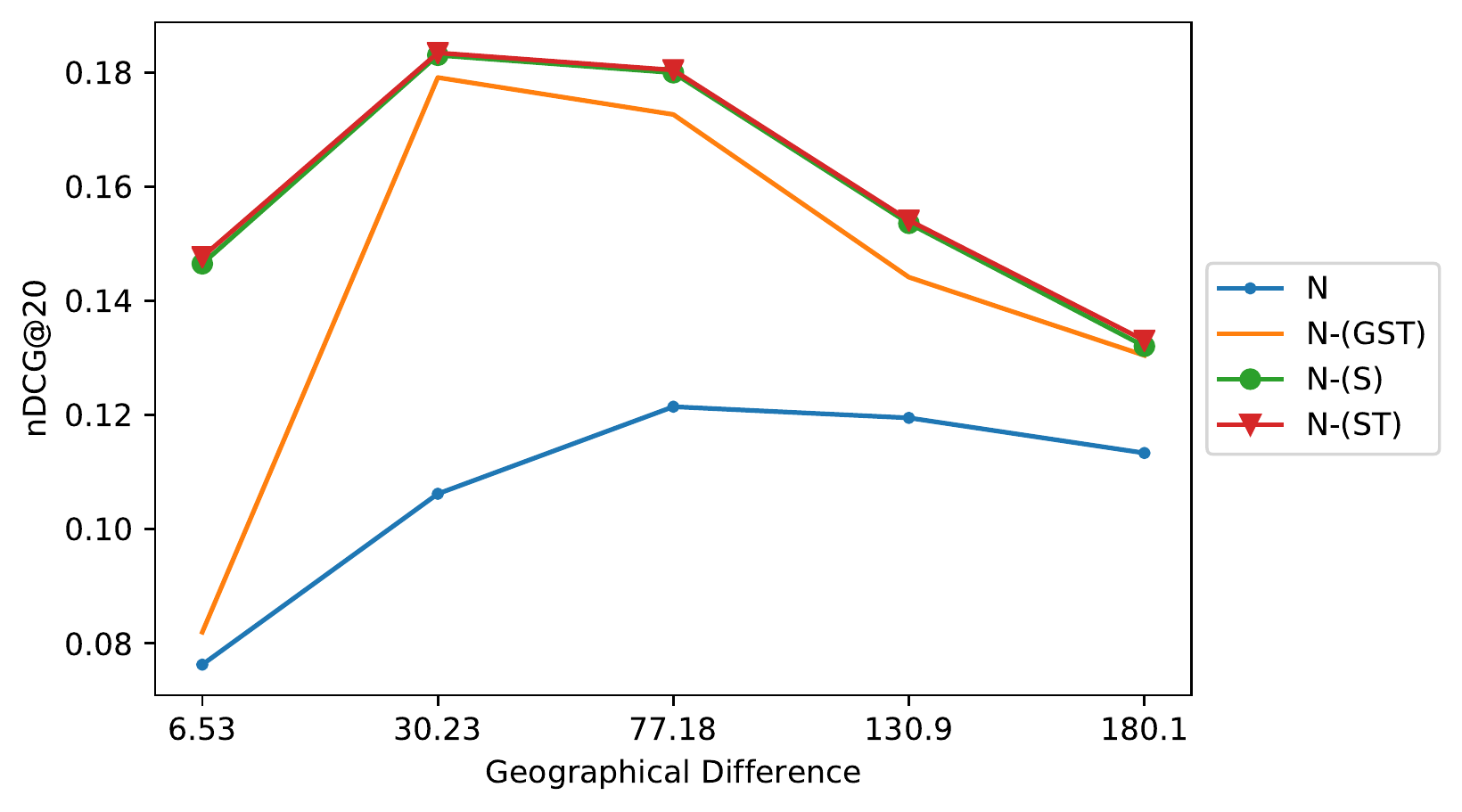}\label{fig:geoFactor_NNs_gowalla}}
  \hfill
  \subfloat[Top models with geographical information]{\includegraphics[scale=0.4]{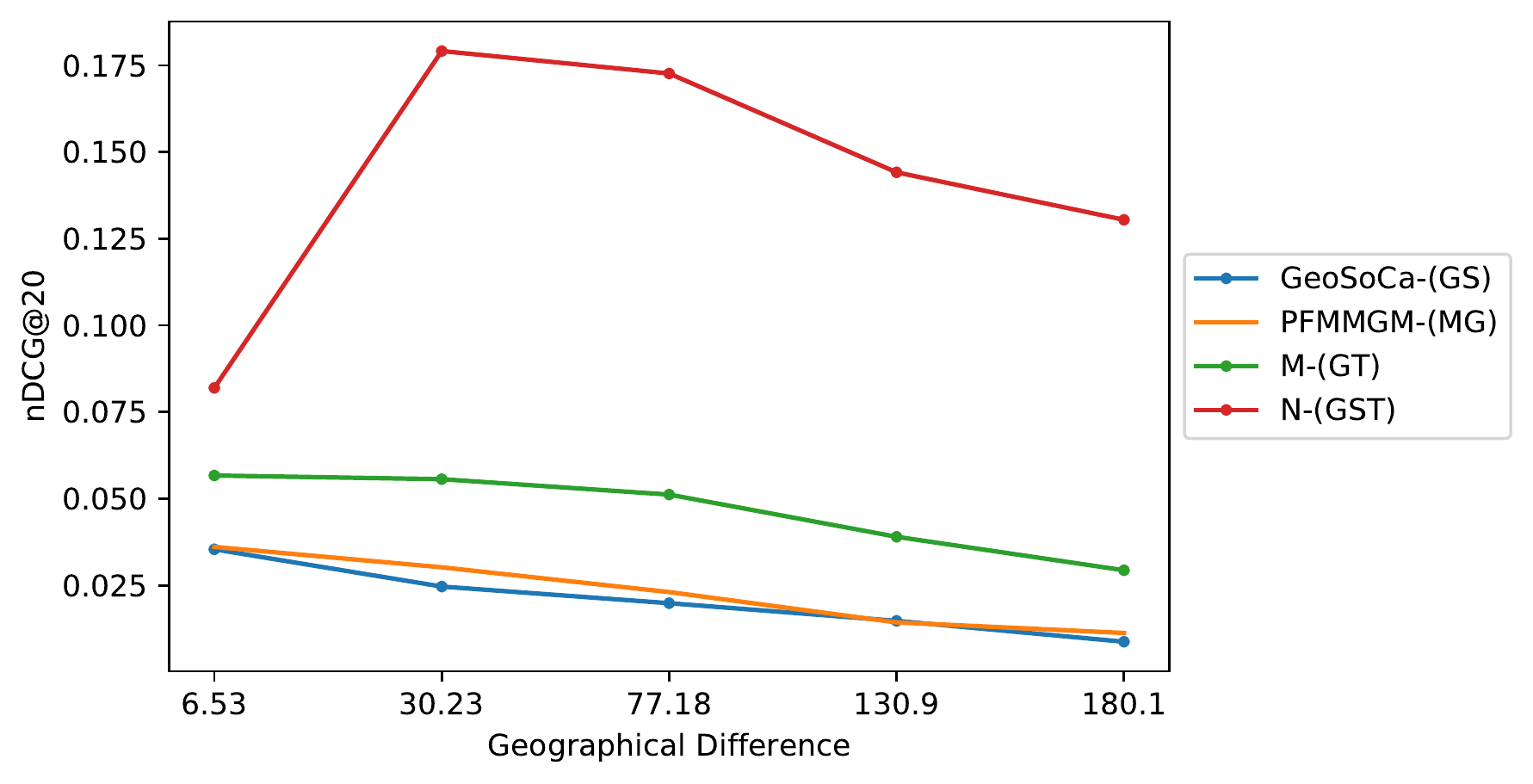}\label{fig:geoFactor_Geo_gowalla}}
  \hfill
  \subfloat[Top models with temporal information]{\includegraphics[scale=0.4]{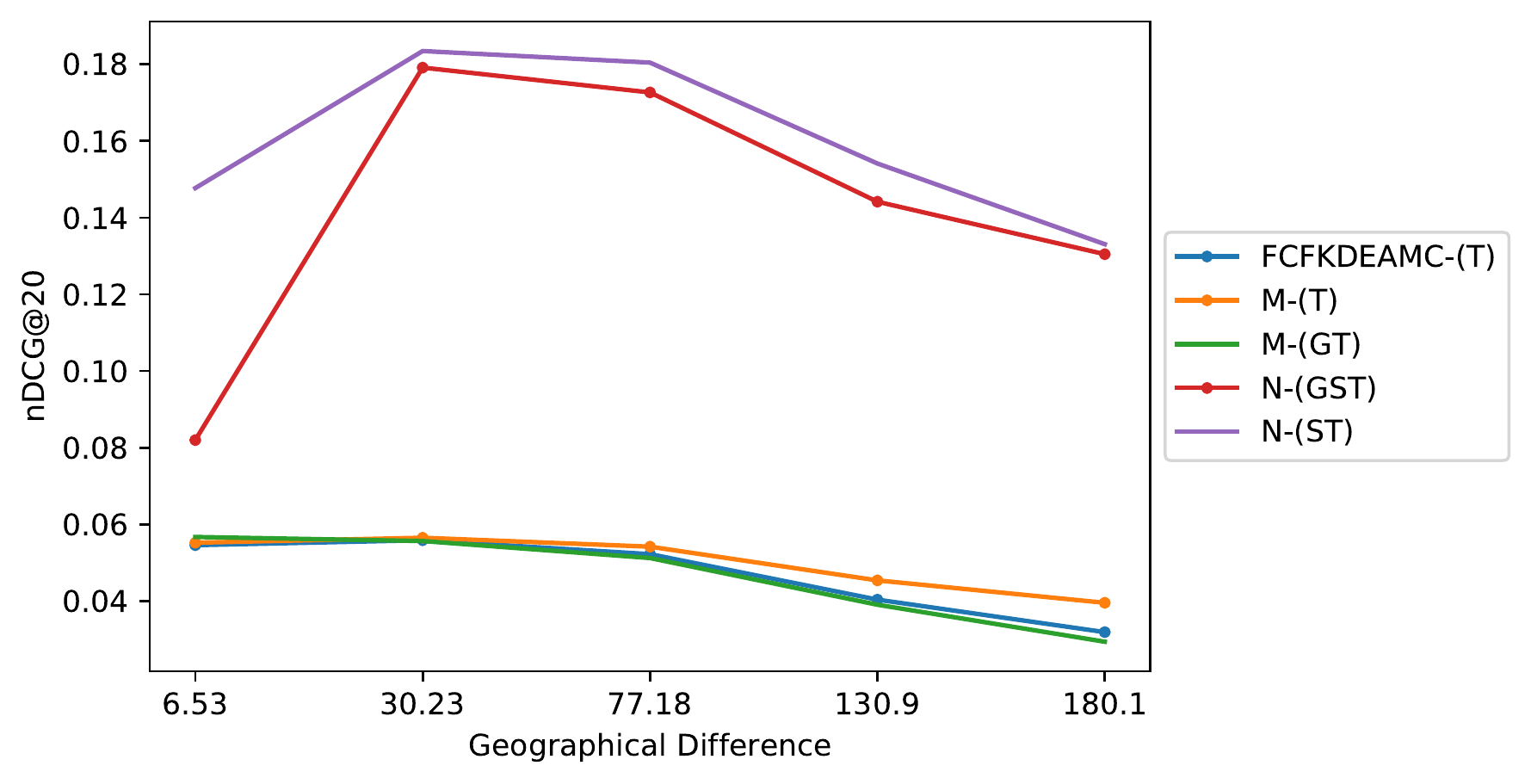}\label{fig:geoFactor_Temp_gowalla}}
  \hfill
  \subfloat[Top models with social information]{\includegraphics[scale=0.4]{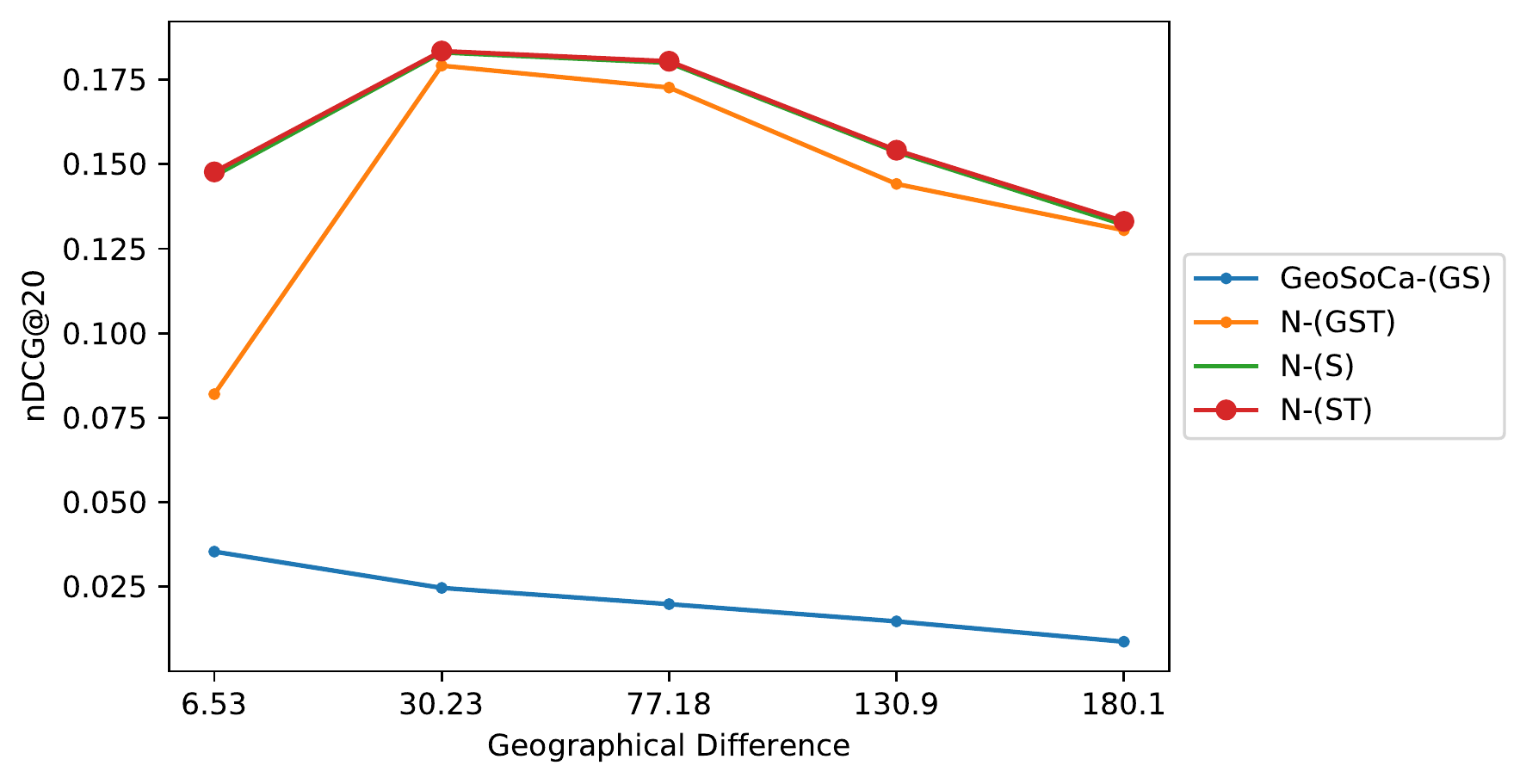}\label{fig:geoFactor_So_gowalla}}
  \caption{Analysis of recommendation performance as a function of geographical behavior of users on Gowalla.}
  \label{fig:geoFactor_gowalla}
\end{figure}

In this section, we report and analyze the results of geographical behavior, temporal behavior, and exploration factor based on the nDCG@20 evaluation measure. The objective is to answer \textbf{RQ4}, studying how different models perform for recommending POI to users with different behavior. We plot the results according to different categories of models, including baselines, matrix factorization, and neural network-based models, as well as contextual-based models (i.e., geographical, temporal, social). Notice that due to space considerations, we only report the results on the Gowalla dataset since we observed similar results on the Yelp dataset. Also, for better visibility and to avoid over-crowding the plot, we select and present only the models based on their performance on different combinations of contextual information from Tables \ref{tbl:MFGowalla}, and \ref{tbl:NNGowalla}. For example, in Table \ref{tbl:baseResultsGowalla}, between the different combinations of the \textit{GeoSoCa}, the \textit{GeoSoCa-(GS)} achieves the best performance on the Gowalla dataset thus, we select it for visualization. Also, in Fig.~\ref{fig:geoFactor_baselines_gowalla} only the best baselines with respect to nDCG@20 from Table \ref{fig:geoFactor_baselines_gowalla} are shown. We plot the rest of the graphs on the same basis.

\subsubsection{Impact of Geographical Distance}
Fig.~\ref{fig:geoFactor_gowalla} show the results of geographical behavior. In the graphs, on the x-axis, we present the geographical distance in kilometers between consecutive visits. The five values presented on the x-axis are extracted from the distribution of check-ins in the dataset. Here, we analyze the impact of users' behavior on the accuracy of recommendation. As seen in Figs.~\ref{fig:geoFactor_baselines_gowalla}, \ref{fig:geoFactor_MFs_gowalla}, \ref{fig:geoFactor_NNs_gowalla}, the accuracy of recommendation is higher for users that stay in smaller neighborhoods. This could be due to the fact that a user with a higher movement range will have more options compared to users with limited movement range and thus the task of recommendation is more difficult for them.

This agrees with the conclusions of previous studies that characterized users' geographical behavior \cite{ye2011exploiting,rahmani2019lglmf}. We see that for \textit{M-(GT)}, \textit{PFMMGM-(MG)}, \textit{M-(T)}, \textit{FCFKDEAMC-(T)}, and \textit{GeoSoCa-(GS)}, larger distances between consecutive check-ins lead to poorer performance of the models. Fig.~\ref{fig:geoFactor_MFs_gowalla} shows on matrix factorization-based methods, large distances between consecutive check-ins cause a sparser user-POI matrix; thus, the recommendation accuracy decreases. This effect is more evident for non-neural models where results heavily depend on geographical information. On the other hand, neural models seem to cope better with different geographical distances because of their capacity to model non-linearity. As one can see in Fig.~\ref{fig:geoFactor_NNs_gowalla}, there is a peak performance around a certain distance (i.e., around 30) for neural networks-based models. This happens for models that incorporate social information and indicates that the optimal performance is achieved for users with a consecutive check-in distance of 30 Km. Finally, Fig.~\ref{fig:geoFactor_NNs_gowalla} shows the results of neural network-based models. Neural network models that only use check-in frequency (i.e., \textit{N}) perform slightly better than models based on matrix factorization. In neural network models, we consider just users' check-in frequency without any contextual information. Moreover, neural network-based methods show that there are differences in incorporating linearity and non-linearity when considering consecutive check-ins of users' behavior.

\begin{figure}[!tbp]
  \centering
  \subfloat[Top baseline models from Table \ref{tbl:baseResultsGowalla}]{\includegraphics[scale=0.4]{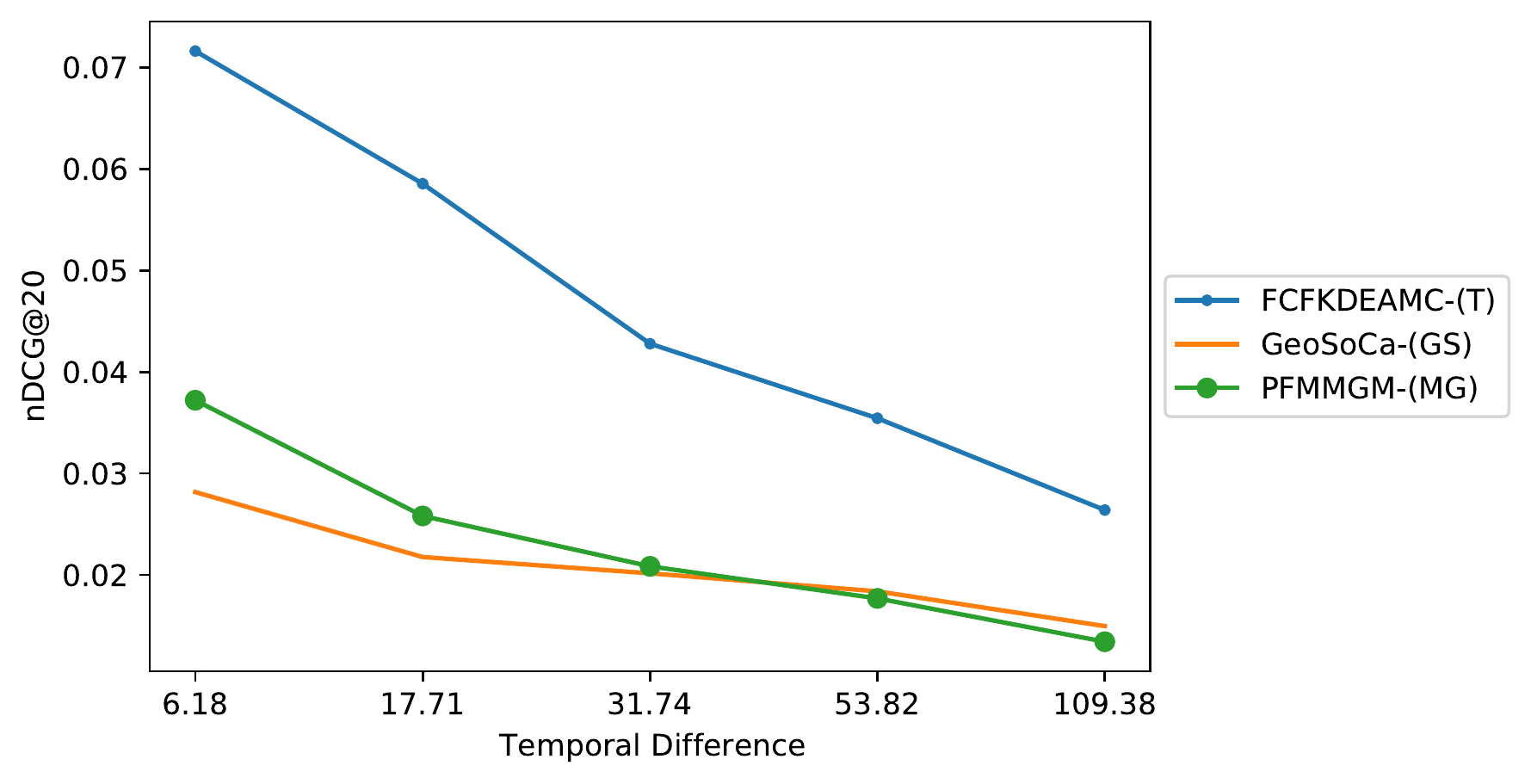}\label{fig:tempFactor_baselines_gowalla}}
  \hfill
  \subfloat[Top M models from Table \ref{tbl:MFGowalla}]{\includegraphics[scale=0.4]{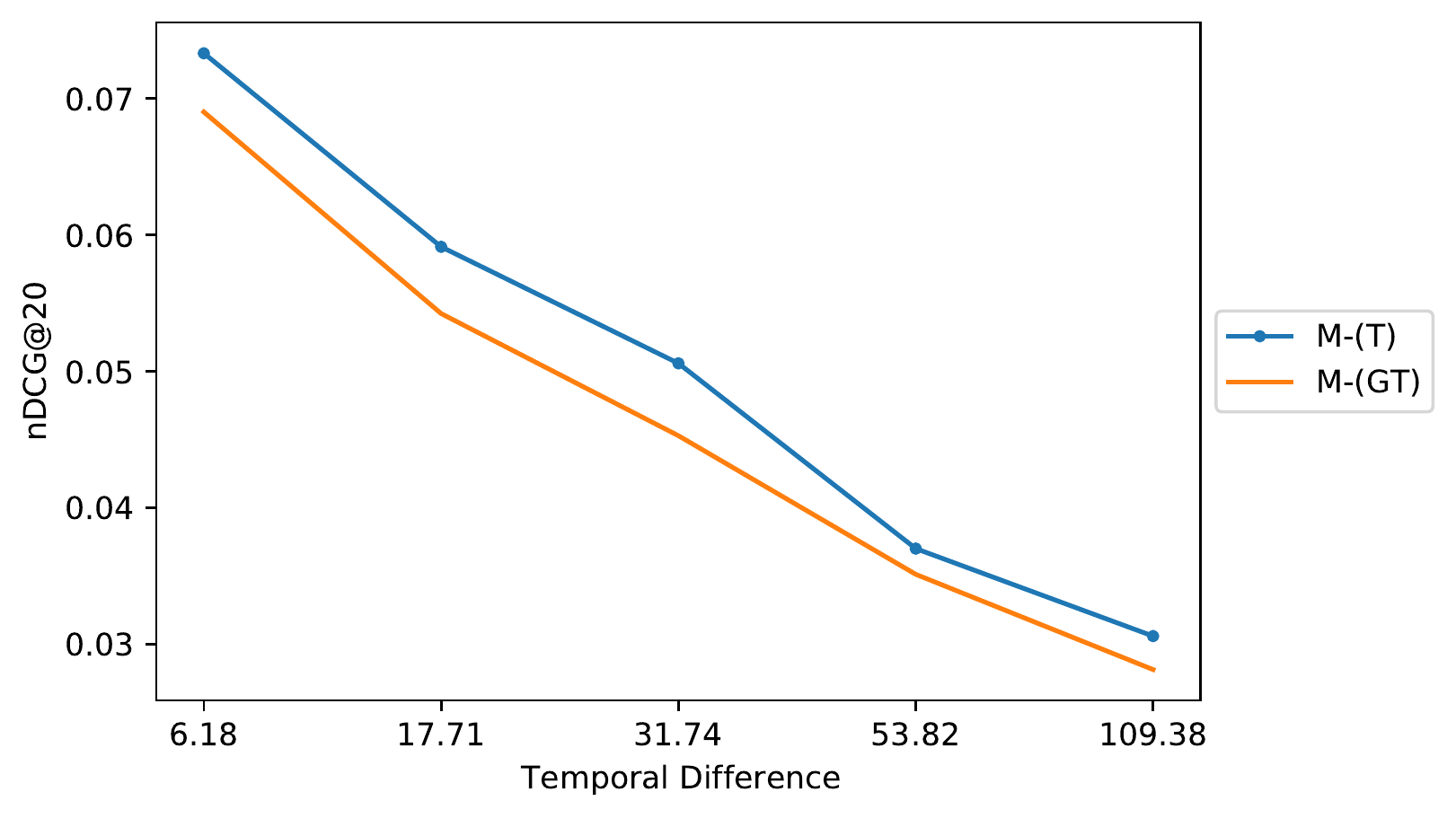}\label{fig:tempFactor_MFs_gowalla}}
  \hfill
  \subfloat[Top N models from Table \ref{tbl:NNGowalla}]{\includegraphics[scale=0.4]{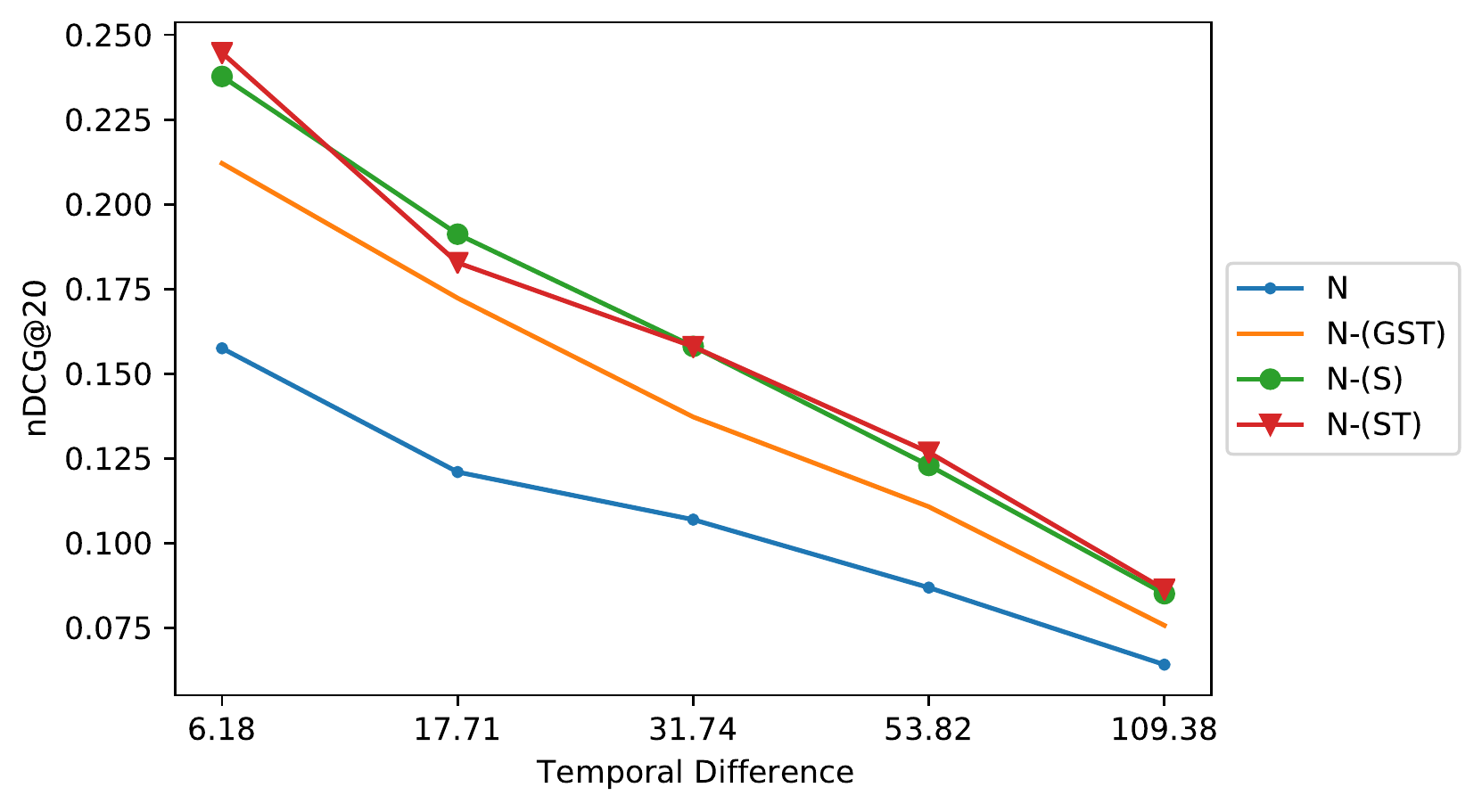}\label{fig:tempFactor_NNs_gowalla}}
  \hfill
  \subfloat[Top models with geographical information]{\includegraphics[scale=0.4]{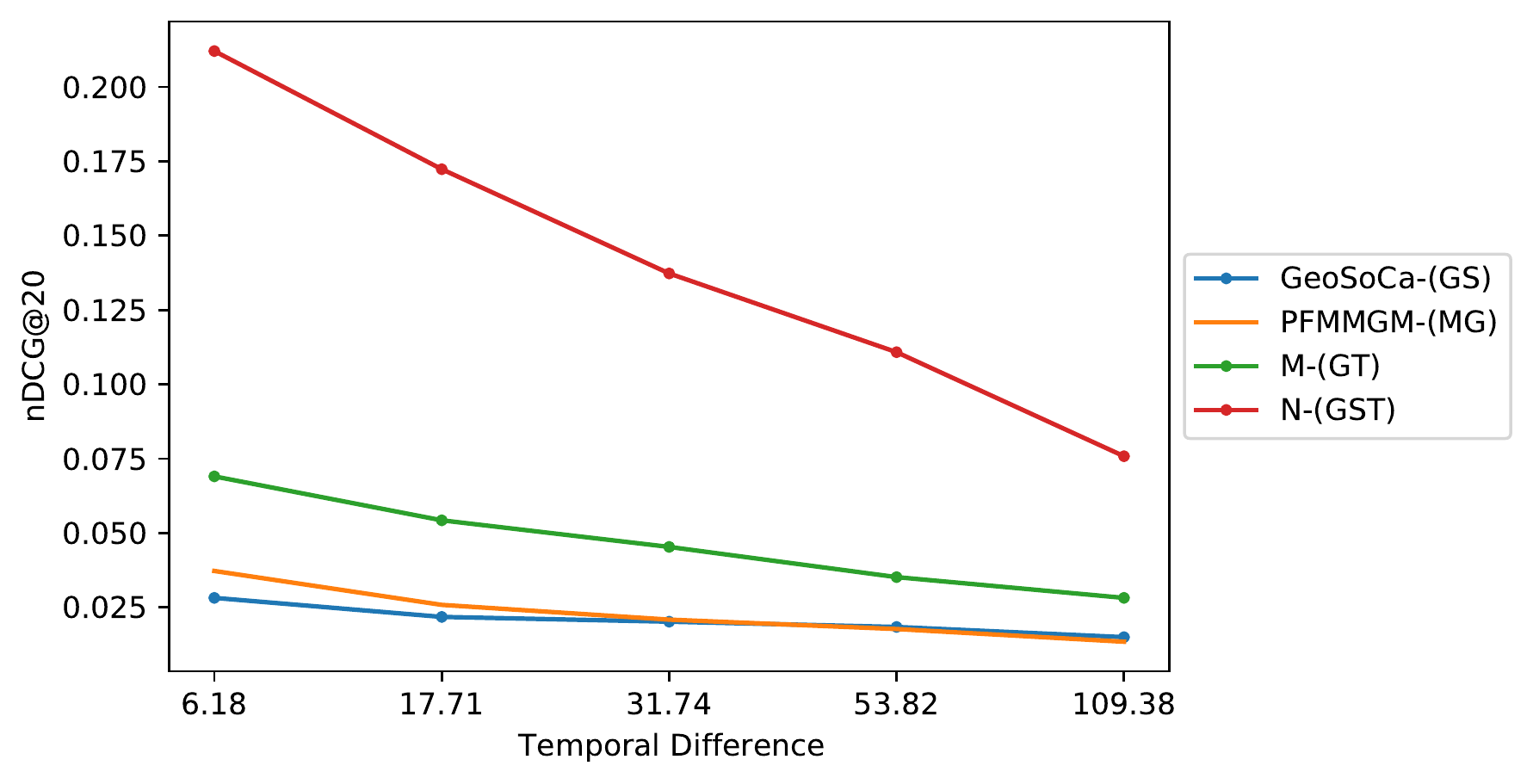}\label{fig:tempFactor_Geo_gowalla}}
  \hfill
  \subfloat[Top models with temporal information]{\includegraphics[scale=0.4]{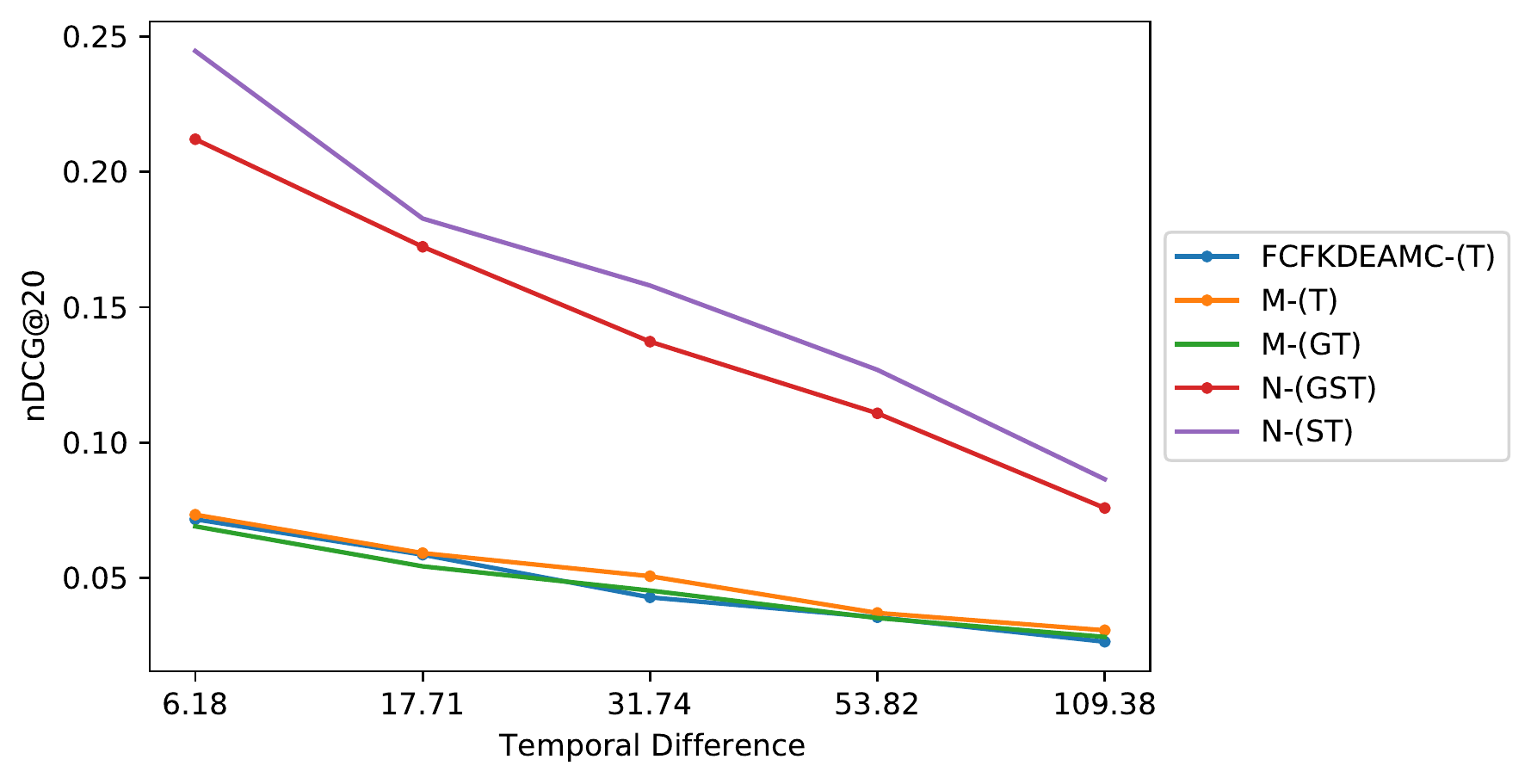}\label{fig:tempFactor_Temp_gowalla}}
  \hfill
  \subfloat[Top models with social information]{\includegraphics[scale=0.4]{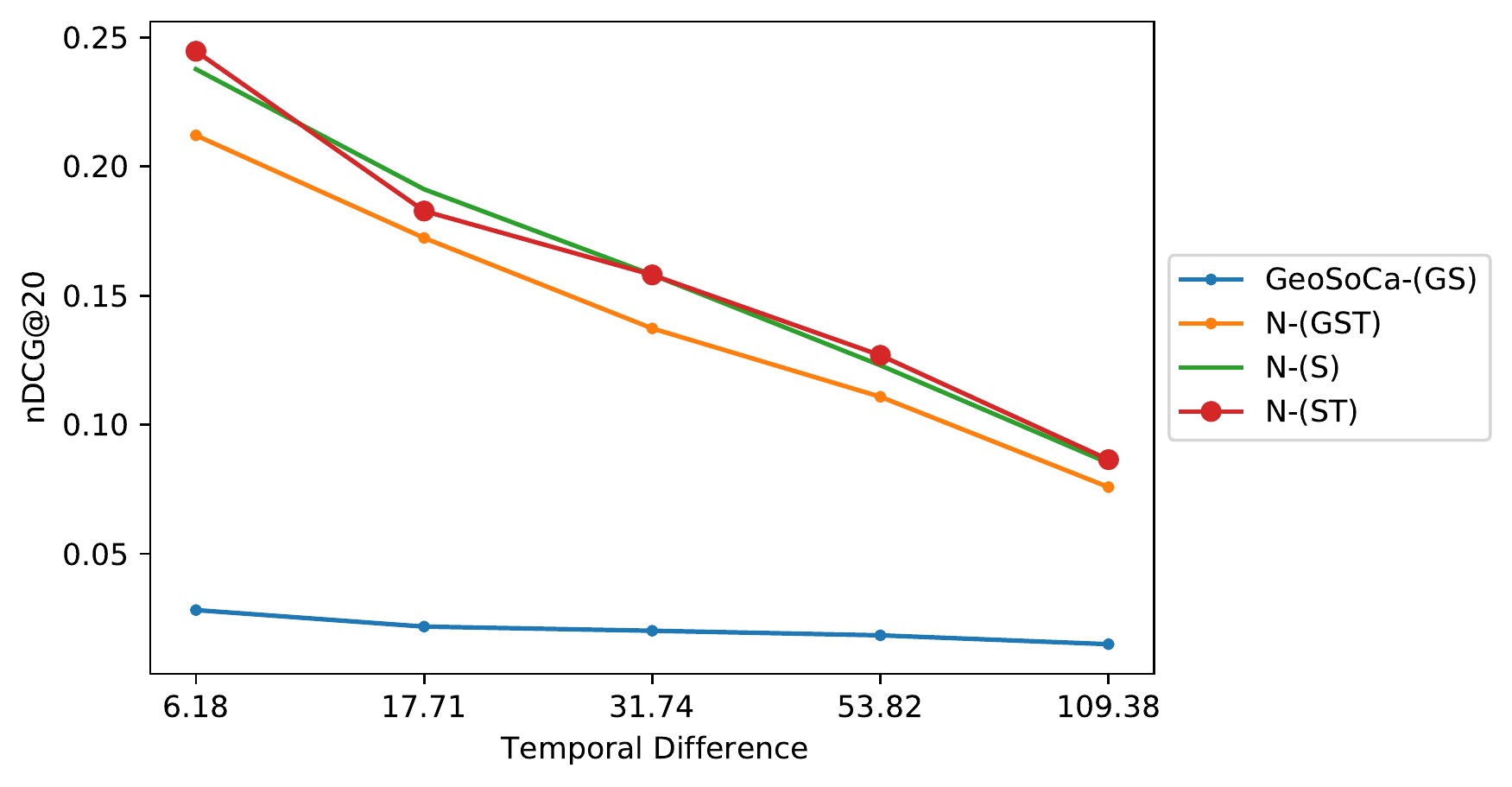}\label{fig:tempFactor_So_gowalla}}
  \caption{Analysis of recommendation performance as a function of temporal behavior of users on Gowalla.}
  \label{fig:tempFactor_gowalla}
\end{figure}

\subsubsection{Impact of Temporal Check-in Density}
In Fig.~\ref{fig:tempFactor_gowalla} we show the results of temporal behavior. We plot the results based on nDCG@20 and the distribution of the temporal distance between consecutive check-ins. In these figures, on the x-axis, five values are presented, dividing users based on the distribution of check-in timestamps. These figures show that the performance of recommendation decreases as the users' consecutive check-ins become more distant in time. This is consistently observed for all models, and temporal models are no exception either. This can be due to the lack of check-in information for users whose check-ins have higher temporal differences. Furthermore, as the time between check-ins increases, the age of recommendation information increases, making recommendations less relevant. 

\begin{figure}[!tbp]
  \centering
  \subfloat[Top baseline models from Table \ref{tbl:baseResultsGowalla}]{\includegraphics[scale=0.4]{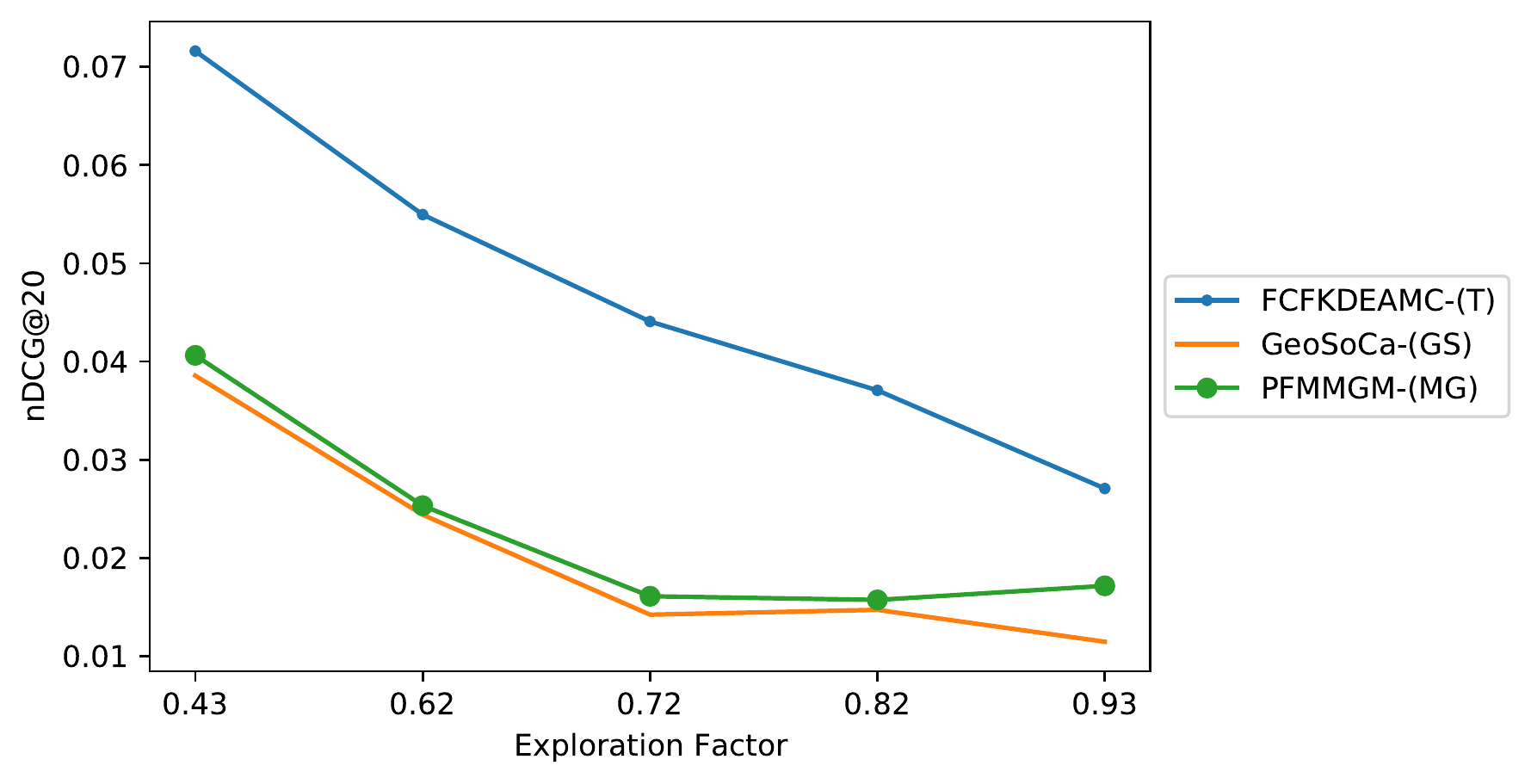}\label{fig:exFactor_baselines_gowalla}}
  \hfill
  \subfloat[Top M models from Table \ref{tbl:MFGowalla}]{\includegraphics[scale=0.4]{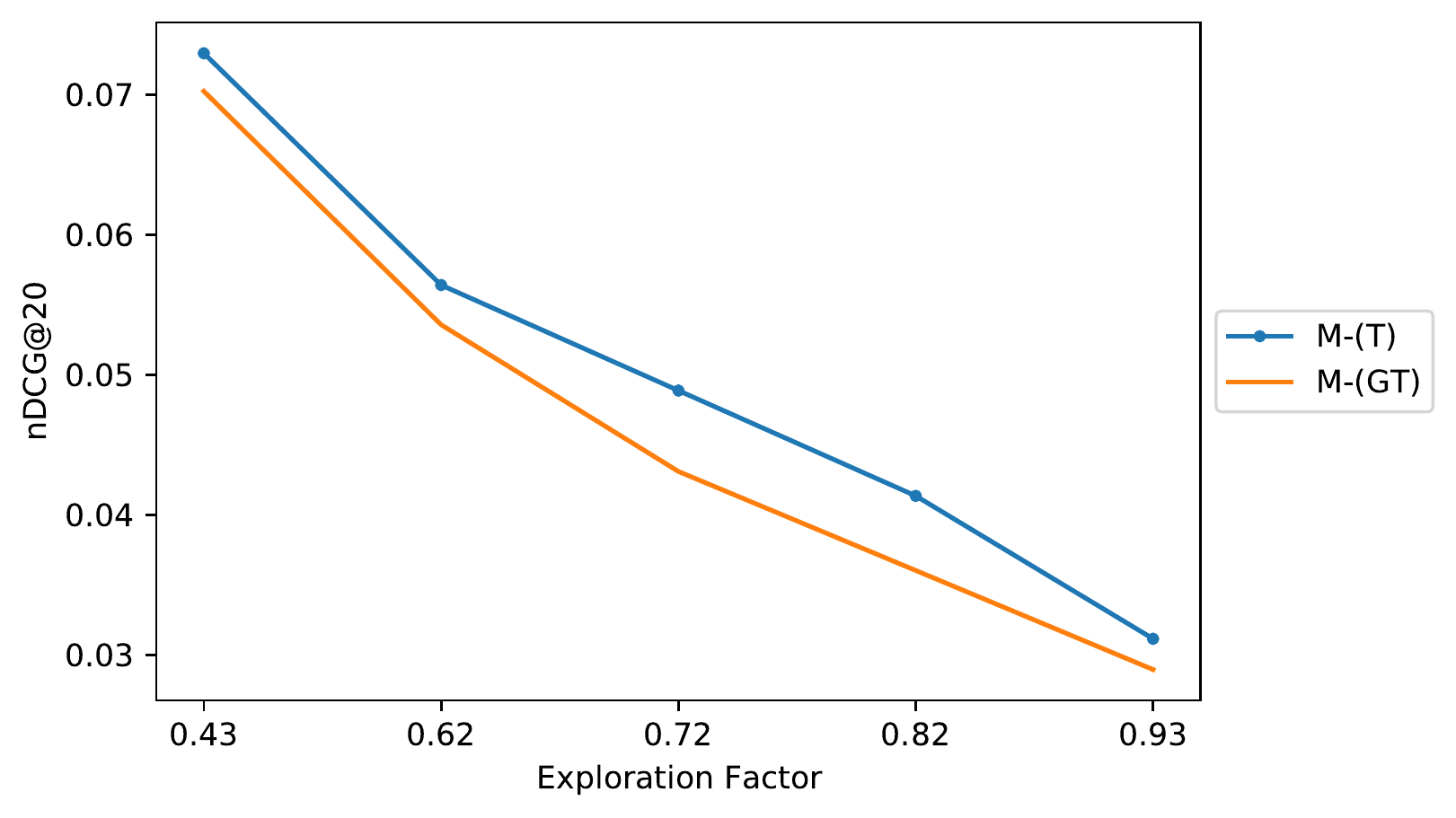}\label{fig:exFactor_MFs_gowalla}}
  \hfill
  \subfloat[Top N models from Table \ref{tbl:NNGowalla}]{\includegraphics[scale=0.4]{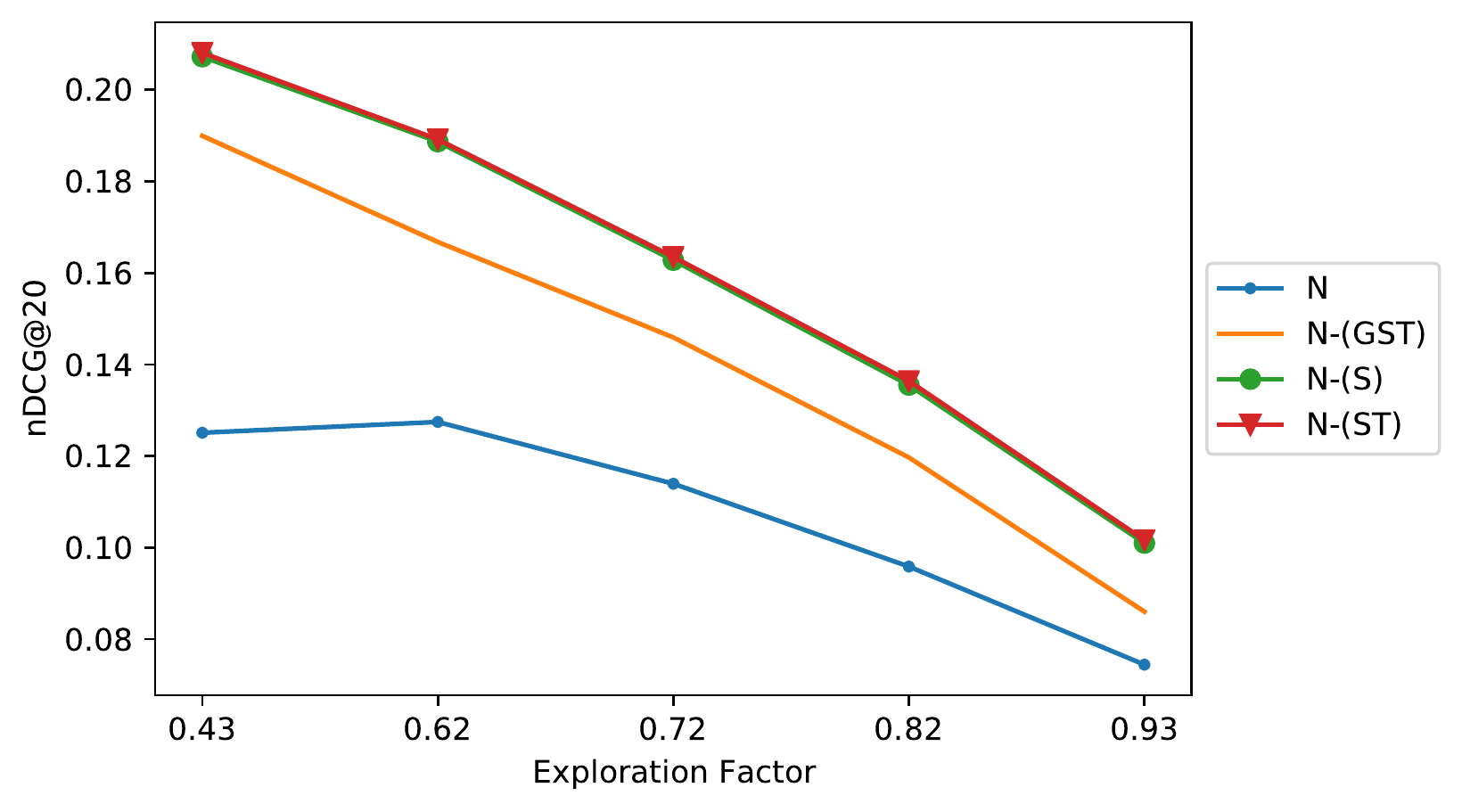}\label{fig:exFactor_NNs_gowalla}}
  \hfill
  \subfloat[Top models with geographical information]{\includegraphics[scale=0.4]{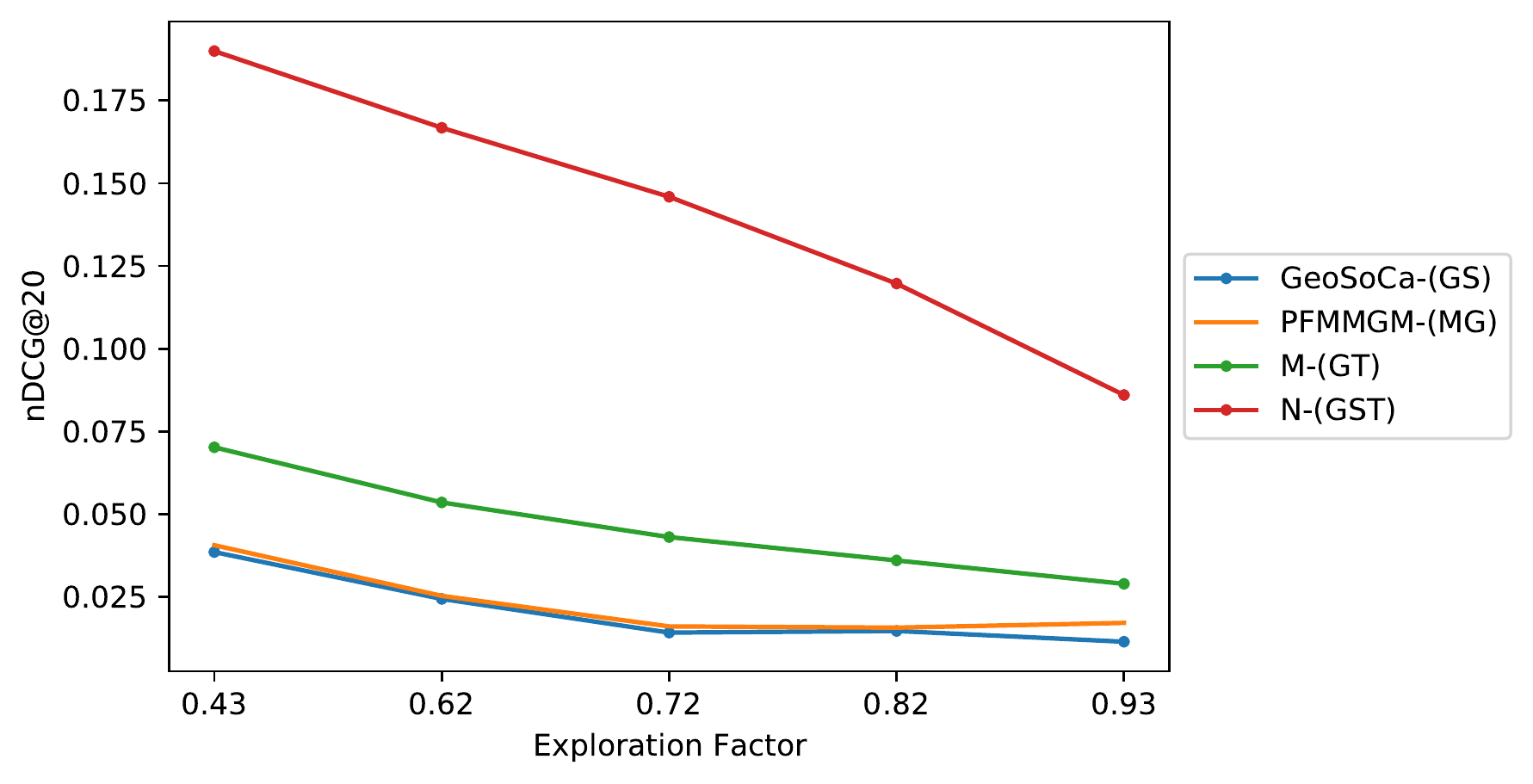}\label{fig:exFactor_Geo_gowalla}}
  \hfill
  \subfloat[Top models with temporal information]{\includegraphics[scale=0.4]{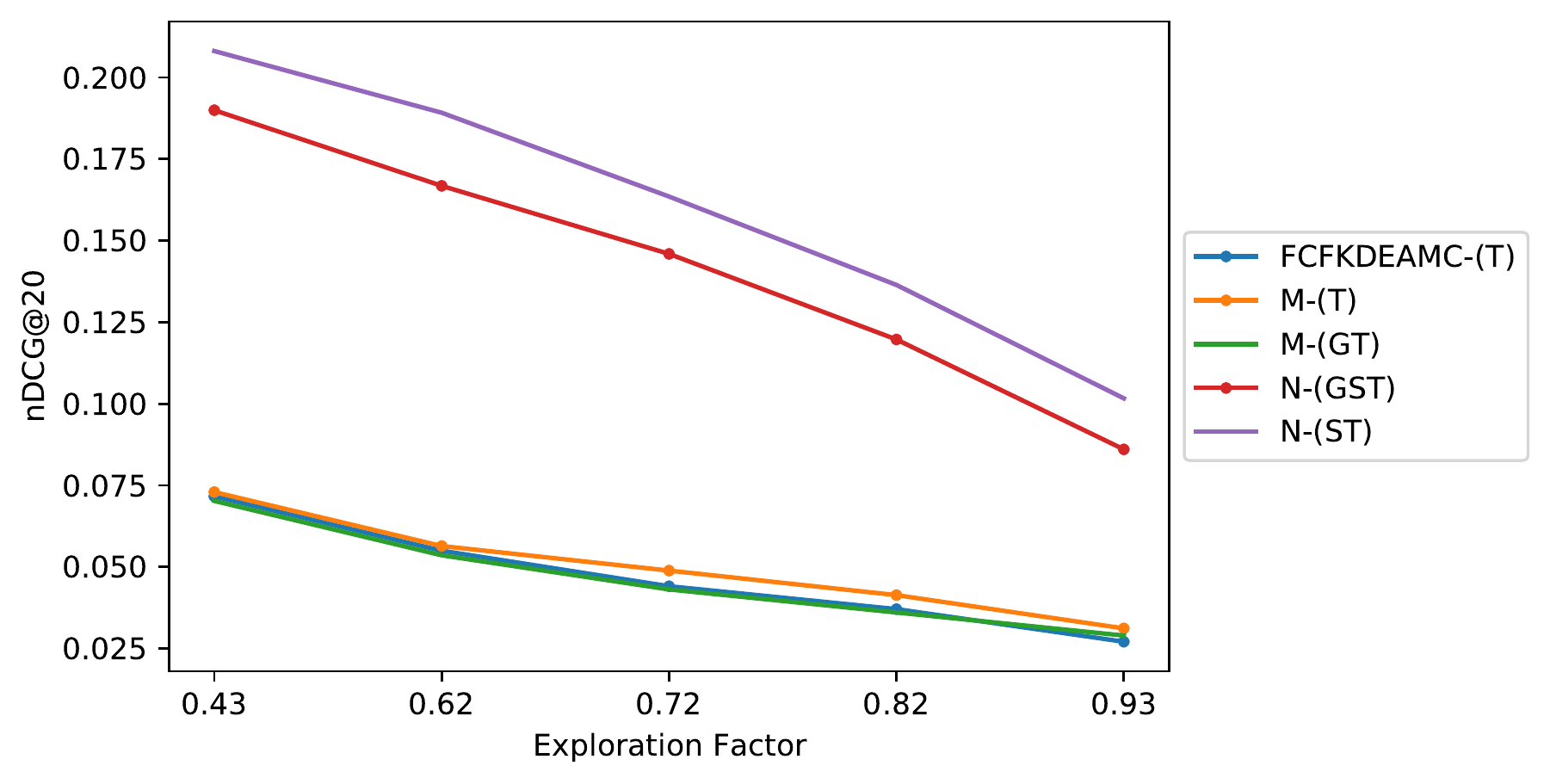}\label{fig:exFactor_Temp_gowalla}}
  \hfill
  \subfloat[Top models with social information]{\includegraphics[scale=0.4]{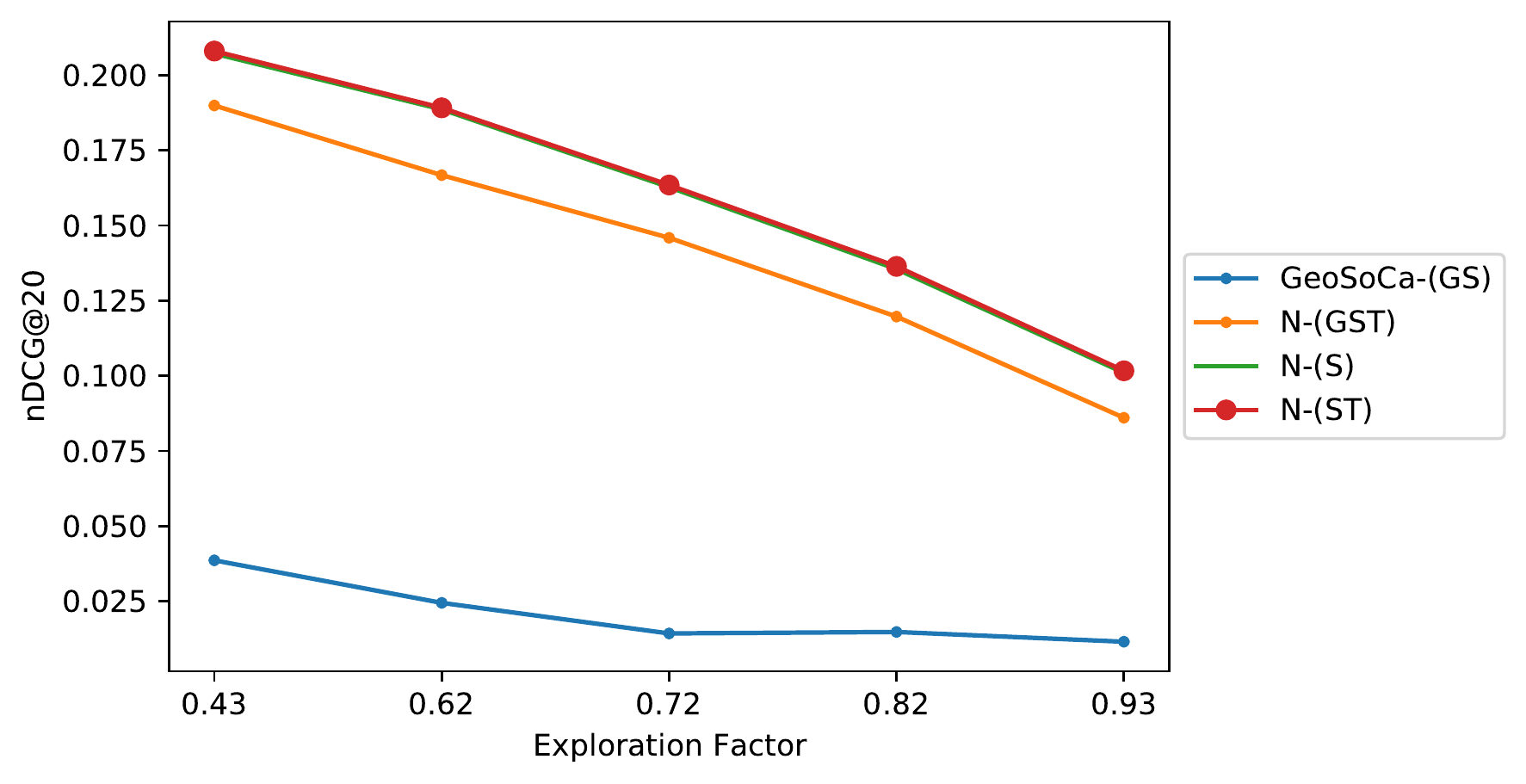}\label{fig:exFactor_So_gowalla}}
  \caption{Analysis of recommendation performance as a function of exploration factor of users on Gowalla.}
  \label{fig:exFactor_gowalla}
\end{figure}

\subsubsection{Impact of Exploration}
In Fig.~\ref{fig:exFactor_gowalla}, we show the results of our experiments on studying the impact of the exploration factor on users' behavior. In the graphs, on the x-axis, we present the exploration factor. The five values presented on the x-axis are extracted from the users' exploration factor that is calculated using Eq.~\ref{eq:EF}.

We see that all models perform worse as users tend to explore more. Higher exploration, in this case, means that users tend to visit different POIs (i.e., visiting more POIs but with less frequency). Therefore, it is harder for models to predict which novel POIs users would visit next. It is interesting to see that neural network-based approaches (i.e., Fig.~\ref{fig:exFactor_NNs_gowalla}) are performing more effectively compared to the rest of the models. This is not in line with what we can usually expect from neural models. They are supposed to learn high-dimensional embeddings of users and POIs, enabling them to better match users with unknown venues. This suggests that the tested neural models cannot learn such representations effectively and instead gain performance improvement mainly on users with more repeated check-ins (i.e., lower exploration value).

Based on our categorization of models presented in Figs.~\ref{fig:exFactor_Geo_gowalla}, \ref{fig:exFactor_Temp_gowalla}, and \ref{fig:exFactor_So_gowalla}, we see the effect of using different contextual information on user groups. While geographical and social models (i.e., Figs.~\ref{fig:exFactor_So_gowalla} and \ref{fig:exFactor_Geo_gowalla}) are affected similarly, we see that temporal models (i.e., Fig.~\ref{fig:exFactor_Temp_gowalla}) are affected differently. This could be due to the existence of some latent relation between exploration and temporal behavior of users that help the models in some cases to perform better. We leave the further investigation of this effect for future work.

Between the contextual models, \textit{T}, \textit{G}, and their combination improve the performance more than the other contextual information and combination in terms of all three evaluation metrics. As we can see from the baselines and \textit{M} models (i.e., Figs.~\ref{fig:exFactor_baselines_gowalla} and \ref{fig:exFactor_MFs_gowalla}), one of the most important factors to improve recommendation accuracy is the temporal factor. We can see in Fig.~\ref{fig:exFactor_gowalla} that when we consider temporal factors, the models' performance improves significantly. On the contrary, in the \textit{N} models (i.e., Fig.~\ref{fig:exFactor_NNs_gowalla}), social information is one of the most effective contextual information. Neural network-based models have better results than matrix factorization ones due to their capability to model non-linear information related to users and item (Figs.~\ref{fig:exFactor_NNs_gowalla} and \ref{fig:exFactor_MFs_gowalla}).

We can also see that a simple combination of the \textit{N} model with contextual models achieves better results than different \textit{M} models. In fact, Fig.~\ref{fig:exFactor_NNs_gowalla} shows that \textit{N-(ST)} gives the best performance among all models. These results indicate that the non-linearity of neural networks obtain users' preference in a better way than linear models such as matrix factorization.

In summary, our main findings related to RQ4 are:
\begin{itemize}
    \item The analysis on the geographical distance indicates the accuracy of recommendation is better when users stay in smaller neighborhoods.
    \item The recommendation task gets more difficult when the users display a higher movement range which causes low accuracy.
    \item Non-linear models are better at modeling the consecutive check-ins behavior of users.
    \item The performance of recommendation decreases as the users' consecutive check-ins become more distant in time.
    \item In general, POI recommendation models perform worse as users tend to explore more.
\end{itemize}

\section{Conclusions and Future Work}
\label{sec:conclusion}
In this paper, we analyzed and extensively evaluated the role of contextual information on POI recommendation systems. We initially provided an extensive survey of POI recommendation, categorized and compared the approaches used so far in modeling contextual information (i.e., geographical, temporal, social, and categorical). Next, with experiments on two benchmark datasets, we analyzed the performance of POI recommendation from different perspectives using well-adopted evaluation metrics in the field of information retrieval. First, we studied how the performance of recommendation is affected by the combinations of contextual information incorporated. To do so, we compared the performance of three prominent baseline methods selected from the literature, confirming that there is indeed a difference in how different combinations of contextual information impact the performance.
Second, we examined the importance of each piece of contextual information and assessed their priorities when incorporated in a different set of models (i.e., linear and non-linear) in POI recommendation. For this goal, we used the two most widely-used linear and non-linear models, namely, matrix factorization and neural networks. Third, we also analyzed if and how the quality of POI recommendation systems is affected by different user behaviors, represented in the geographical distance between check-ins, temporal check-in density, and exploration.

Our findings show that fusing all contextual information into a recommendation model is not the best strategy and one needs first to evaluate the impact of different combinations of proposed contextual information and the method used for fusing them (e.g., some models such as \textit{FCFKDEAMC} achieved better results when the social context is removed). Moreover, our results further show that the impact of geographical and temporal information is more than social and categorical information. More importantly, we showed that different information retrieval evaluation metrics such as precision, recall, and nDCG yield different results for a proposed contextual recommendation model. For instance, using precision, a geographical model outperforms a temporal model, while recall indicates the superiority of the temporal model. However, by looking only at the list of recommended POIs, none of these metrics evaluates if the recommendation lists retain the temporal and geographical context.

Finally, there are many interesting future directions worthy of exploration. Other factors and pieces of information can be incorporated into POI recommendation models, for instance, those extracted from comments, tips, or images available from a POI. Also, for generalization purposes, the analysis of different datasets with sufficient information (e.g., users, POIs, categories) is very important. Evaluating contextual information shows inconsistent results using different metrics. One future direction of work could be proposing different evaluation metrics to test the quality of contextual information in context-aware recommender systems (see, for example \cite{sanchez2019exploiting}). 
Moreover, our study suggests that the usage of contextual information leads to different levels of improvement on the two datasets. This can be due to various factors that are dependent on each dataset's characteristics, such as sparsity. Thus, we plan to extend our study to various datasets of certain characteristics and perform a thorough investigation on the impact of different dataset characteristics on performance.

\vspace{5pt}
\noindent \textbf{Acknowledgments.} 
We thank the anonymous reviewers for the valuable feedback. This work was in part supported by the Swiss State Secretariat for Education, Research and Innovation (SERI) mobility grant between Switzerland and Iran, and in part by the NWO (No. 016.Vidi.189.039 and No. 314-99-301).

\bibliographystyle{ACM-Reference-Format}
\bibliography{sample-base}

\end{document}